\documentclass[12pt]{article}
\usepackage{titling}
\usepackage{amsmath}
\usepackage{slashed}
\usepackage{amssymb}
\usepackage{epsfig}
\usepackage{graphicx}
\usepackage{multirow}
\usepackage{color}
\usepackage{subcaption}
\usepackage{rotating}
\usepackage{hyperref}
\usepackage[margin=1.0in]{geometry}
\usepackage[table]{xcolor}
\usepackage{enumitem}
\usepackage[utf8x]{inputenc}
\usepackage[compress,numbers,sort]{natbib}
\usepackage{authblk}
\usepackage{colortbl}
\usepackage{pdflscape}
\usepackage{color}
\usepackage{float}
\usepackage{siunitx}
\usepackage[compat=1.1.0]{tikz-feynman}
\usepackage{lscape}
\usepackage{mathtools}
\usepackage{amsfonts}
\usepackage{bbold}
\usepackage{soul}
\usepackage{comment}
\usepackage{adjustbox}
\usepackage{cleveref}
\usepackage{subcaption} % Add to preamble

\definecolor{nicered}{rgb}{0.6,0.1,0.1}
\definecolor{nicegreen}{rgb}{0.1,0.5,0.1}
\definecolor{mediumcandyapplered}{rgb}{0.99, 0.12, 0.07}
\definecolor{red}{rgb}{1.0, 0, 0}
\hypersetup{colorlinks,citecolor= nicegreen,linkcolor= nicered}

\renewcommand{\bar}{\overline}

\definecolor{LightCyan}{rgb}{0.88,1,1}
\definecolor{piggypink}{rgb}{0.99, 0.87, 0.9}
\definecolor{applegreen}{rgb}{0.55, 0.71, 0.0}
\definecolor{darkpastelgreen}{rgb}{0.01, 0.75, 0.24}
\definecolor{green-yellow}{rgb}{0.68, 1.0, 0.18}

\newcommand{\beq}{\begin{equation}}
	\newcommand{\eeq}{\end{equation}}
\newcommand{\bea}{\begin{eqnarray}}
	\newcommand{\eea}{\end{eqnarray}}

\DeclareSIUnit\barn{b}

\title{\bf{Precise predictions for double Higgs production in association with a vector boson in Effective Field Theory} 
}

\author[1,2]
{Ramona Gr\"ober \thanks{ramona.groeber@pd.infn.it}}
\author[1,2]
{Micha\l{} Ryczkowski \thanks{michaljakub.ryczkowski@unipd.it}}
\author[1,3]
{Gioia Sacchi \thanks{gioia.sacchi@studenti.unipd.it}}

\affil[1]{\emph{\normalsize Dipartimento di Fisica e Astronomia ``G. Galilei'', Universit\`a di Padova, Via F. Marzolo 8, I-35131, Padova, Italy}}
\affil[2]{\emph{\normalsize Istituto Nazionale di Fisica Nucleare, Sezione di Padova, Via F. Marzolo 8, I-35131, Padova, Italy}}
\affil[3]{\emph{\normalsize Scuola Galileiana di Studi Superiori, Universit\`a degli Studi di Padova, via Marzolo 6,
I-35131 Padova, Italy}}
\makeatletter
\patchcmd{\@maketitle}
  {\null\vskip 2em}
  {\null\vskip 2em%
   \noindent\hfill COMETA-2026-28\par\vspace{1em}}
  {}{}
\makeatother

\date{}

\begin{document}
\maketitle
\begin{abstract}
	\normalsize
We present next-to-next-to-leading order (NNLO) QCD predictions for Higgs boson pair production in association with a weak gauge boson, $pp \to Vhh$ with $V=W^\pm,Z$, in effective field theory descriptions of new physics. We consider both the Standard Model Effective Field Theory (SMEFT), truncated at dimension six, and the Higgs Effective Field Theory (HEFT) at leading order in the chiral expansion. The calculation builds on the factorisation of the quark-induced production process into Drell-Yan production of an off-shell vector boson and its subsequent decay into $Vhh$, supplemented in the $Zhh$ channel by the loop-induced $gg \to Zhh$ contribution that first enters at NNLO QCD. We provide inclusive predictions at $\sqrt{s}=13.6$ and $14.0$ TeV, including scale, PDF, and $\alpha_s$ uncertainties, and express the results in terms of numerical coefficients that allow fast evaluations for arbitrary EFT parameters in the considered ranges. We find that, for $W^\pm hh$ production, QCD corrections largely factorise from the EFT dependence, leading to almost flat $K$-factors. In contrast, $Zhh$ production shows a stronger dependence on the EFT coefficients because of the gluon-induced component. 

\end{abstract}	
	
\clearpage

\section{Introduction} 
\label{sec:introduction}
One of the flagship measurements of the upcoming High-Luminosity Large Hadron Collider (HL-LHC) phase is Higgs boson pair production, as it provides direct access to the trilinear Higgs self-coupling and hence constitutes a first step towards an experimental verification of the Higgs potential \cite{Baglio:2012np, Dolan:2012rv, Djouadi:1999rca, DiMicco:2019ngk}. 
The dominant Higgs pair production mode, gluon fusion, has a cross section that is about three orders of magnitude smaller than that of single-Higgs production, so high luminosities are required for its observation. 
While gluon fusion may remain the only channel with sensitivity to the trilinear Higgs self-coupling close to its Standard Model (SM) value, other Higgs pair production mechanisms are also of considerable interest. 
In particular, they can provide information on whether electroweak symmetry breaking is realised linearly or non-linearly, for instance by measuring whether the couplings of one Higgs boson and of two Higgs bosons to vector bosons follow the pattern implied by the \(SU(2)_L\) doublet nature of the scalar sector. 
This has motivated experimental searches for subleading Higgs pair production channels such as vector-boson fusion \cite{CMS:2026nuu}, \(t\bar t hh\) production \cite{ATLAS:2026dvm}, and associated production with a \(W^\pm\) or \(Z\) boson \cite{CMS:2024fkb, ATLAS:2022fpx}.

In this work we concentrate on Higgs boson pair production in association with a weak gauge boson, \(Vhh\) with \(V=W^\pm,Z\), which is the fourth-largest Higgs pair production mode. 
At a centre-of-mass energy of \(\sqrt{s}=13.6\,\text{TeV}\), its SM cross section amounts to \(0.35\,\text{fb}\), \(0.19\,\text{fb}\), and \(0.40\,\text{fb}\) for \(W^+hh\), \(W^-hh\), and \(Zhh\) production, respectively, at next-to-next-to-leading order (NNLO) in QCD \cite{Grober:2026ece}. Fully differential NNLO QCD corrections are given in Refs.~\cite{Li:2016nrr, Li:2017lbf}. Current experimental constraints read $183\times \sigma_{\rm SM}$ \cite{ATLAS:2022fpx} and $294\times \sigma_{\rm SM}$ \cite{CMS:2024fkb}.

 While the process involves nearly the same couplings as Higgs boson pair production in vector boson fusion, which has a sizeably larger cross section, $Vhh$ production can probe separately whether the couplings of two Higgs bosons to $W$ or $Z$ bosons follow the same pattern. Vector-boson fusion Higgs pair production constrains only a combination of these couplings. Related HEFT studies of Higgs pair production in vector boson fusion at NLO QCD, including Higgs-strahlung contributions, have been presented in Ref.~\cite{Braun:2025hvr}.\footnote{See also Ref.~\cite{Jaeger:2025isz} for the vector boson fusion part only.}

In recent years, effective field theory (EFT) descriptions of new physics have become increasingly popular, motivated by the absence of clear evidence for light new particles at collider experiments.
In this paper, we study \(Vhh\) production in two EFT frameworks: the Standard Model Effective Field Theory (SMEFT) \cite{BUCHMULLER1986621, dim6smeft} and the Higgs Effective Field Theory (HEFT) \cite{Feruglio:1992wf, Grinstein:2007iv, Buchalla:2012qq, Alonso:2012px, Brivio:2013pma, Brivio:2016fzo}. 
In the absence of new light degrees of freedom, both EFTs provide model-independent descriptions of heavy new physics. 
They differ, however, in their treatment of the Higgs boson. 
In SMEFT, the Higgs field is assumed to belong to an \(SU(2)_L\) doublet, whereas in HEFT the Goldstone bosons are encoded in a Goldstone matrix while the Higgs boson transforms as a singlet. 
Correspondingly, the organisation of the two expansions is different: SMEFT is ordered in inverse powers of a high new-physics scale \(\Lambda\), while HEFT is organised according to chiral counting, involving powers of \(\Lambda\) and \(4\pi\)~\cite{Brivio:2025yrr}. 
To exploit the discriminatory power of \(Vhh\) production at the HL-LHC, precise theory predictions are essential. 
Here we provide NNLO QCD predictions in both EFT frameworks, building on the NNLO QCD description of Drell--Yan production \cite{Hamberg:1990np, Brein:2003wg}, in analogy to single Higgs production in association with a vector boson, $Vh$, \cite{Harlander:2002wh}. 
For \(Zhh\) production, additional NNLO contributions from heavy-quark loop-induced triangle, box, and pentagon topologies have to be included. Since this process involves for this specific partonic channel only the leading order contribution, it will turn out to be the driver for deviations from flat $K$-factors and also of the theory uncertainty.

The paper is organised as follows. 
In Sec.~\ref{sec:EFT} we review the two EFT frameworks. 
In Sec.~\ref{sec:cxn} we discuss the computation of the NNLO QCD cross section. 
Our numerical results are presented in Sec.~\ref{sec:results}. 
We conclude in Sec.~\ref{sec:conclusion}.

\section{Effective Field Theory Descriptions of New Physics \label{sec:EFT}}
In this section, we describe the effective field theory frameworks used in our calculation. In order to introduce our notation we will start with the SM Lagrangian, invariant under a local $SU(3)_c\times SU(2)_L\times U(1)_Y$ symmetry with gauge fields $G^A_\mu, W^I_\mu, B_\mu$,
\begin{align}
\mathcal{L}_{\mathrm{SM}} =& - \frac{1}{4}G_{\mu\nu}^A G^{A\mu\nu}- \frac{1}{4}W_{\mu\nu}^I W^{I\mu\nu}-\frac{1}{4}B_{\mu\nu}B^{\mu\nu} + (D_\mu \varphi)^\dagger (D^\mu \varphi) -V(\varphi)
\nonumber\\ &+\sum_{\psi} \bar{\psi} i\slashed{D}\psi-\left[\bar{q}_L Y_d \varphi d_R+ \bar{q}_L Y_u \tilde \varphi u_R + \bar{l}_L Y_e \varphi e_R+\text{h.c.}\right]\,.
\label{eq:LagSM_SMEFT}
\end{align}
The index $\psi$ runs over the 5 fermion fields of the SM,
\begin{align}
 q_L &= \begin{pmatrix}
u_L \\
d_L
\end{pmatrix}\, ,
  \quad l_L= \begin{pmatrix}
\nu_L \\
e_L
\end{pmatrix}\, , \quad u_R\,, \quad  d_R\,, \quad e_R\,.
\end{align}
Each of the fermion fields carries flavor indices which are implicitly contracted in Eq.~\eqref{eq:LagSM_SMEFT}. 
%The Yukawa couplings $Y_u,Y_d,Y_e$ are $3\times3$ complex matrices in flavor space, that have to be diagonalized through proper fermion field redefinitions to obtain the mass (physical) basis.
The Yukawa couplings $Y_u$, $Y_d$, and $Y_e$ are $3\times3$ complex matrices in flavor space; appropriate fermion field redefinitions are required to diagonalize them and define the physical mass basis.
The field strength tensors are of form
\begin{equation}
  \begin{aligned}
    W_{\mu \nu}^I & =\partial_\mu W_\nu^I-\partial_\nu W_\mu^I-g
    \epsilon^{I J K} W_\mu^J W_\nu^K\, , \\
    B_{\mu \nu} & =\partial_\mu B_\nu-\partial_\nu B_\mu\, .
\end{aligned}
\end{equation}
The Higgs potential is given by
\begin{equation}
    V(\varphi) = -\frac{m_h^2}{2} \varphi^\dag \varphi + \frac{1}{2}\lambda (\varphi^\dag \varphi)^2\,, \quad \quad \lambda = \frac{m_h^2}{v^2}\, .
\end{equation}
The Higgs doublet $\varphi$ can acquire a vacuum expectation value $v$ such that,
\begin{equation}
\varphi =\frac{1}{\sqrt{2}}\begin{pmatrix}
\pi_1+i \pi_2 \\ v+h +i \pi_0
\end{pmatrix}\,,
\end{equation}
where $h$ is the Higgs boson and the $\pi_i$ are Goldstone degrees of freedom. We have also introduced the dual Higgs doublet,
\begin{align}
\tilde \varphi = i\sigma_2 \varphi^*,
\end{align}
where $\sigma_2$ is the second Pauli matrix.
The covariant derivative acting on the Higgs field is defined as,
\begin{equation}
D_{\mu}\varphi=\left(\partial_{\mu} + i g W^I\frac{\sigma^I}{2} + i \frac{g'}{2} B_{\mu} \right)\varphi \, ,
\end{equation}
with $\sigma^I$ denoting the Pauli matrices.

\subsection{SMEFT}
The Standard Model Effective Field Theory provides a model-independent framework to parametrize the effects of heavy and decoupling beyond-the-Standard-Model physics by adding all possible higher-dimensional operators to the SM Lagrangian that respect the gauge symmetries of the SM. The particle content remains the one of the SM, with all fields transforming under the gauge symmetries in exactly the same way as in the SM. The SMEFT expansion is organized via a power counting in $1/\Lambda$ with a Lagrangian of the form,
\begin{equation}\mathcal{L}_{\mathrm{SMEFT}}=\mathcal{L}_{\mathrm{SM}}+
\sum_i
\frac{C_i}{\Lambda^{d_i-4}}\mathcal{O}_i \, ,
  \label{eq:SMEFTLagrGeneral}
\end{equation}
where $\mathcal{L}_{\mathrm{SM}}$ denotes the SM Lagrangian and $\mathcal{O}_i$ are all higher-dimensional operators of dimension $d_i$ invariant under the SM symmetries. The $C_i$ are dimensionless Wilson coefficients, suppressed by powers of the new physics scale $\Lambda$. The operators $\mathcal{O}_i$ build a basis when all redundant operators are removed by making use of integration by parts, Fierz identities and field redefinitions. A complete, non-redundant basis has been defined for the first time in Ref.~\cite{dim6smeft}, usually referred to as the \textit{Warsaw} basis. We adopt this basis in the following.

In this work we neglect light quark masses and their Yukawa interactions. We adopt a minimal-flavour-violation-inspired flavour setup \cite{DAmbrosio:2002vsn}, 
in which chirality-flipping Higgs-quark operators are aligned with the SM Yukawa matrices. As a result, the corresponding interactions involving first- and second-generation quarks are Yukawa suppressed and are neglected, while possible third-generation effects are retained where relevant. In the same spirit, the right-handed charged-current operator
$\mathcal{O}_{\varphi ud}=
i(\tilde{\varphi}^{\dagger}D_{\mu}\varphi)
(\bar{u}_{p}\gamma^{\mu}d_{r})
$ is suppressed by two Yukawa insertions in minimal flavour violation, such that we do not include it in the discussion. Furthermore, we neglect CP-violating operators.
We then are left with the operators in Table~\ref{tab:operators}. Since dipole operators do not interfere with the SM results, they arise at higher order in the EFT expansion in the limit of massless initial state quarks.
\begin{table}[H]
\centering
\begin{tabular}{||c|c||c|c||}\hline\hline\rule{0pt}{3ex}
  $\mathcal{O}_\varphi$   &$\left(\varphi^\dagger \varphi\right)^3$ &$\mathcal{O}_{\varphi\square}$ &$\left(\varphi^\dagger \varphi\right)\square\left(\varphi^\dagger \varphi\right)$\\[5pt]\hline\rule{0pt}{3ex}$\mathcal{O}_{\varphi D}$&$\left(\varphi^\dagger D^\mu \varphi\right)^*\left(\varphi^\dagger D_\mu \varphi\right)$&
  $\mathcal{O}_{\varphi W}$&$\varphi^\dagger\varphi\, W^a_{\mu\nu}W^{a\mu\nu}$
     \\[5pt]\hline\rule{0pt}{3ex} 
     $\mathcal{O}_{\varphi B}$&$\varphi^\dagger\varphi\, B_{\mu\nu}B^{\mu\nu}$ & $\mathcal{O}_{\varphi WB}$ & $\varphi^\dagger\sigma^a\,\varphi\, W^a_{\mu\nu}B^{\mu\nu}$
        \\[5pt]\hline\rule{0pt}{3ex}$\mathcal{O}_{ u\varphi}$&$\left(\varphi^\dagger \varphi\right)\left(\bar{q}_p\widetilde{\varphi} u_r\right)$&$\mathcal{O}_{ d\varphi}$ & $\left(\varphi^\dagger \varphi\right)\left(\bar{q}_p \varphi d_r\right)$
     \\[5pt]\hline\rule{0pt}{3ex} 
     $\mathcal{O}_{\varphi q}^{(1)}$&$\left(\varphi^\dagger i \overleftrightarrow{D}_\mu\varphi\right)\left(\bar{q}_p\gamma^\mu q_r\right)$ &$\mathcal{O}_{\varphi q}^{(3)}$&$\left(\varphi^\dagger i \overleftrightarrow{D^a_{\mu}}\varphi\right)\left(\bar{q}_p\,\sigma^a\,\gamma^\mu q_r\right)$ 
      \\[5pt]\hline\rule{0pt}{3ex} 
     $\mathcal{O}_{\varphi u}$&$\left(\varphi^\dagger i \overleftrightarrow{D}_\mu\varphi\right)\left(\bar{u}_p\gamma^\mu u_r\right)$ & $\mathcal{O}_{\varphi d}$&$\left(\varphi^\dagger i \overleftrightarrow{D}_\mu\varphi\right)\left(\bar{d}_p\gamma^\mu d_r\right)$ 
  \\[5pt]\hline\hline
\end{tabular}
\caption{SMEFT operators that contribute to $Vhh$ production, with $\varphi^{\dagger}\overleftrightarrow{D}_\mu \varphi=  i \varphi^\dagger (D_\mu \varphi ) - i (D_\mu \varphi )^\dagger \varphi  $ and $\varphi^{\dagger}\overleftrightarrow{D}^a_\mu \varphi= i \varphi^\dagger \sigma^a (D_\mu \varphi ) - i (D_\mu \varphi )^\dagger \sigma^a \varphi$. 
}
\label{tab:operators}
\end{table}

Throughout this work, we distinguish between the operators involving first- and third-generation up-type quarks, denoting the former by $\mathcal{O}_{u\varphi}$, $\mathcal{O}_{\varphi u}$, and $\mathcal{O}_{\varphi q}^{(1,3)}$, and the latter by $\mathcal{O}_{t\varphi}$, $\mathcal{O}_{\varphi t}$, and $\mathcal{O}_{\varphi Q}^{(1,3)}$.

\subsection{HEFT}
The HEFT loosens the assumption of SMEFT that the Higgs boson comes together with the three Goldstone bosons in an $SU(2)_L$ doublet. This comes at the price of a loss of perturbative unitarity at scales above $\Lambda \gtrsim 4\pi v$. As a consequence, the power counting in HEFT is different from SMEFT \cite{Buchalla:2013eza, Brivio:2025yrr}, and one can power count inverse orders of the scale $\Lambda$ and $4\pi$ adopting naive dimensional analysis \cite{MANOHAR1984189,  Cohen:1997rt, Luty:1997fk, Gavela:2016bzc}. 
In the end, Ref.~\cite{Brivio:2025yrr} proposes a power counting prescription that counts, at the matrix element level, the loop order, the number of external legs, and the chiral dimension of each vertex. The chiral dimension of each operator is determined by counting the derivatives, couplings and fermion bilinears. Scalar self-interactions count as two orders. 

Altogether, while the HEFT power counting should count loops, this is only true for electroweak loops. The HEFT expansion can stay separate from the perturbative expansion in the strong coupling constant~\cite{Brivio:2025yrr}.
Since we focus on higher order corrections in QCD, from the point of view of the HEFT power counting this means that we count
\begin{equation}
N_{\chi}-N_{g_s}\,, \label{eq:powercounting}
\end{equation}
where $N_{\chi}$ is the chiral dimension and $N_{g_s}$ the number of insertions of the strong coupling constant. This counting allows us to keep the perturbative QCD expansion and the EFT expansion separate. In this work, while we go up to NNLO in the QCD expansion, we remain at LO in the HEFT expansion.

We write the Goldstone matrix as
\begin{equation}
\mathbf{U}=\exp\left( \frac{i \pi_a \sigma^a}{v}\right)\, ,
\end{equation}
where $\pi^a$ are the Goldstone bosons and $\sigma^a$ the Pauli matrices. The matrix $\mathbf{U}$ transforms under a global $SU(2)_L\times SU(2)_R$ as bidoublet.
Its covariant derivative is given by
\begin{align}
    D_{\mu} {\bf U} = \partial_{\mu} {\bf U} + \frac{i g}{2} W_{\mu}^I\, \sigma^I {\bf U} + \frac{i g'}{2} B_{\mu}\, {\bf U} \sigma^3\,.
\end{align}
For convenience, it is useful to define the objects
\begin{align}
{\bf V}_{\mu} &= (D_{\mu} {\bf U}) {\bf U}^{\dagger}\,,
    &
 {\bf T} &= {\bf U }\sigma^3{\bf U}^\dagger   \,,
\end{align}
which both transform in the adjoint representation of $SU(2)_L$. ${\bf V}_\mu$ is a singlet under $SU(2)_R$, and ${\bf T}$ breaks $SU(2)_R$ and hence constitutes an explicit breaking of the custodial symmetry.  The LO HEFT Lagrangian allows for generic deviations from the SM, and can be written as
\begin{equation}
    \begin{aligned}
\mathcal{L}_{\text{LO}}^{\text{HEFT}} &= 
- \frac{1}{4}G_{\mu\nu}^A G^{A\,\mu\nu}
- \frac{1}{4}  W^I_{\mu\nu} W^{I\mu\nu}
- \frac{1}{4} B_{\mu\nu} B^{\mu\nu}
\\
&+ \frac{1}{2} \partial_\mu h \partial^\mu h - V(h)
- \frac{v^2}{4} \text{Tr}\left( \mathbf{V}_\mu \mathbf{V}^\mu \right) \mathcal{F}_C(h) \\
&+ i \bar{q}_L \slashed{D} q_L + i \bar{q}_R \slashed{D} q_R
+ i \bar{l}_L \slashed{D} l_L + i \bar{l}_R \slashed{D} l_R \\
&- \frac{v}{\sqrt{2}} \left( \bar{q}_L \mathbf{U} \mathcal{Y}_q(h) q_R + \text{h.c.} \right)
- \frac{v}{\sqrt{2}} \left( \bar{l}_L \mathbf{U} \mathcal{Y}_l(h) l_R + \text{h.c.} \right),
    \end{aligned}
    \label{HEFT:LO:Lagrangian}
\end{equation}
with the SM fermions grouped conveniently into the doublets of global $SU(2)_{L/R}$ symmetries, so in addition to the usual $SU(2)_L$ doublets we define
\begin{equation}
q_R = \begin{pmatrix} u_R \\ d_R \end{pmatrix}\, , \quad
l_R = \begin{pmatrix} 0 \\ e_R \end{pmatrix}\, .
\end{equation}

The relevant parts of Higgs potential $V(h)$, functional $\mathcal{F}_C(h)$, and the Yukawa terms are given by
\begin{equation}
\begin{aligned}
V(h)& = \frac{1}{2} m_h^2 h^2 
             + \kappa_{3\lambda} \frac{m_h^2}{2v} h^3 + v^4\sum_{n=4} a_{V,n} \left(\frac{h}{v}\right)^n \,, \quad \mathcal{F}_C(h)=1+ 2 a_{C,1}\left(\frac{h}{v}\right) + 
\sum_{n=2}a_{C,n}\left(\frac{h}{v}\right)^n\,, \\
\mathcal{Y}_q(h)& = \text{diag}\left( \sum_{n=0} Y_u^{(n)} \frac{h^n}{v^n}, \sum_{n=0} Y_d^{(n)} \frac{h^n}{v^n} \right), \quad
\mathcal{Y}_l(h) = \text{diag}\left( 0, \sum_{n=0} Y_l^{(n)} \frac{h^n}{v^n} \right).
\end{aligned}
\label{eq:VFY_expansion}
\end{equation}

%The Yukawa functionals  $\mathcal{Y}_{D/Q/L}$ are obtained in analogy to $\mathcal{Y}_{U}$.

In this work, we focus on the multiplicative factors modifying Higgs-vector and Higgs-top vertices relevant for this process, namely $hhh$, $VVh$, $VVhh$ and $\bar{t}th$, denoted by
\begin{equation}
\kappa_{\lambda}\,, \quad \kappa_{V}=a_{C,1}\,, \quad \kappa_{2V}=a_{C,2}\,, \quad \kappa_t = Y_t^{(1)}\, .\label{eq:relHEFT1}
\end{equation}
The $V$ stands generically for $V=W^{\pm}, Z$.
 We leave the inclusion of operators from the NLO HEFT Lagrangian to future work, but refer to Refs.~\cite{Brivio:2025sib, Herrero:2022krh, Domenech:2025gmn} for studies of those operators in Higgs pair production in gluon fusion and vector boson fusion.
From the power counting rule employed in Eq.~\eqref{eq:powercounting}, a custodial violating operator
\begin{equation}
    \mathcal{O}_T= - \frac{v^2}{4}\text{Tr}\left(\bf{T V_{\mu}} \right)^2\mathcal{F}_T(h) \quad\text{with} \quad \mathcal{F}_T(h)=\sum_{n=0}^{\infty}a_{T,n}\left(\frac{h}{v} \right)^n\, ,
\end{equation}
can be added to the LO Lagrangian. 
Such operator is often moved to the NLO HEFT basis as the effect is constrained by electroweak precision tests, generically attributing operators with appearance of ${\bf T}$ to higher orders in the EFT expansion. 
In the mass basis, and in the unitary gauge, this operator can be written as
\begin{equation}
    \mathcal{O}_T = -\frac{1}{4}(g^2+g'^2) v^2 \left( a_{T,0} + \left(\frac{h}{v}\right) a_{T,1} + \left(\frac{h}{v}\right)^2 a_{T,2}\right) Z_\mu Z^\mu\, .
\end{equation}
This term can be a source of the splitting between $\kappa_W$ and $\kappa_Z$ factors (and similarly for $\kappa_{2W}$ and $\kappa_{2Z}$) in the following way: 
\begin{equation}
\begin{aligned}
    \kappa_{W}&=a_{C,1}\,, \quad \kappa_{Z}=a_{C,1} + a_{T,1}\, , \\
    \kappa_{2W}&=a_{C,2}\,, \quad \kappa_{2Z}=a_{C,2} + 2 a_{T,2}\, . \label{eq:relHEFT2}
\end{aligned}
\end{equation}
The coefficient $a_{T,0}$ is strongly constrained ($a_{T,0} \lesssim 0.2$) due to its violation of custodial symmetry and its contribution to the $T$ parameter. While it is reasonable to assume that other coefficients $a_{T,i>0}$ are of a similar order, this assumption can be tested experimentally by independently measuring $\kappa_{W/2W}$ and $\kappa_{Z/2Z}$. For $\kappa_V$, such measurements can be performed in $Vh$ production or Higgs decays. For $\kappa_{2V}$, $Vhh$ production currently offers the most promising avenue for probing the splitting between $\kappa_{2W}$ and $\kappa_{2Z}$. Hence, we retain $\mathcal{O}_T$ to preserve generality and to explicitly account for this splitting.

\section{Cross section at NNLO QCD\label{sec:cxn}}
In this section, we will discuss how the results in the various EFTs are obtained at NNLO QCD and how the theoretical uncertainty is computed. We will start by reviewing the SM results, for which we rely on Ref.~\cite{Baglio:2012np}.
\subsection{SM}
\label{subsec:SM}
\begin{figure}[t!]
 \hspace*{-1.1cm}   \includegraphics[width=1.1\linewidth]{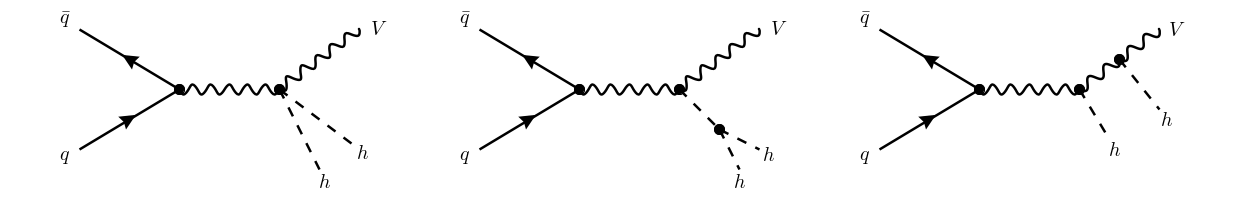}
    \caption{Representative diagrams contributing to quark-initiated $Vhh$ production in unitary gauge. In Feynman gauge, the corresponding Goldstone-boson exchange diagrams are included as well.}
    \label{fig:VHH:Diagrams}
\end{figure}
The LO results were reproduced using \texttt{FeynArts} and \texttt{FormCalc} \cite{Hahn:2000kx,Hahn:2008fyh} to distinguish contribution of each diagram.
We show the LO diagrams in Fig.~\ref{fig:VHH:Diagrams}. We adopt the Feynman gauge, where we have in addition to the vector bosons, also Goldstone bosons are exchanged.
Defining
\begin{equation}
    \begin{aligned}
        \mu_V=\frac{m^2_V}{s} \, ,\qquad \mu_h=\frac{m^2_h}{s} \,  , \qquad
        x_1=\frac{2E_{h1}}{\sqrt{s}} \,  , \qquad x_2=\frac{2E_{h2}}{\sqrt{s}} \,,
    \end{aligned}
\end{equation}
where $s$ is the partonic centre-of-mass energy squared, $E_{hi}$ (with $i=1,2$) the energy of the Higgs bosons, $m_V$ the gauge boson mass and $m_h$ the Higgs boson mass, the differential cross section can be written as
\begin{equation}
    \begin{aligned}
        \frac{d^2\hat{\sigma}^{LO}_{Vhh}}{dx_1dx_2}&=\frac{G_F^3\;m_V^6\left(\alpha_q^2+\upsilon^2_q\right)}{1152\;\sqrt{2}\;\pi^3\;s\left(1-\mu_V\right)}\left[\frac{1}{8}f_0\left(A+B+C+D\right)^2+\right.
        \\
        &\quad+\frac{1}{4\mu_V\left(1-x_1+\mu_h-\mu_V\right)}\times
        \left(\frac{f_1}{1-x_1+\mu_h-\mu_V}\right. +
        \\&\quad\left.\left.+\frac{f_2}{1-x_2+\mu_h-\mu_V}\quad+2\mu_Vf_3\left(A+B+C+D\right)\right)+\left\{x_1 \leftrightarrow x_2\right\}\right]\,.
    \end{aligned}
    \label{eq:sigmalo}
\end{equation}
The reduced couplings of quarks to vector bosons are $\alpha_q=\upsilon_q=\sqrt{2}$ if $V=W$, while if $V=Z$ $\alpha_u=1$, $\upsilon_u=1-\frac{8}{3}\sin^2\theta_W$,  $ \alpha_d=-1$ and $\upsilon_d=-1+\frac{4}{3}\sin^2\theta_W$. Furthermore, we have introduced the following abbreviations
\begin{equation}
    \begin{aligned}
        A&=\frac{3\;m_h^2}{m^2_V}\;\frac{1}{x_1+x_2-1+\mu_V-\mu_h}\, ,&\qquad &B=\frac{2}{1-x_1+\mu_h-\mu_V} \, , \\ C&=\frac{2}{1-x_2+\mu_h-\mu_V} \, , &\qquad &D=\frac{1}{\mu_V} \, .
    \end{aligned}
\end{equation}
The coefficients $f_i$ are
\begin{equation}    
\begin{aligned}
        f_0&=\mu_V\left[\left(2-x_1-x_2\right)^2+8\mu_V\right] \, ,
        \\f_1&=x_1^2\left(\mu_V-1+x_1\right)^2-4\mu_h\left(1-x_1\right)\left(1-x_1-\mu_Vx_1-3\mu_V\right)+\\
        &\quad+\mu_V\left(\mu_V-4\mu_h\right)\left(1-4\mu_h\right)-\mu_V^2 \, , 
        \\f_2&=\left(2\mu_V+x_1+x_2\right)\left[\mu_V\left(x_1+x_2-1+\mu_V-8\mu_h\right)-\left(1-x_1\right)\left(1-x_2\right)\left(1+\mu_V\right)\right]+
        \\&\quad+\left(1-x_1\right)^2\left(1-x_2\right)^2+\left(1-x_1\right)\left(1-x_2\right)\left[\mu_V^2+1+4\mu_h\left(1+\mu_V\right)\right]+
        \\&\quad+4\mu_h\mu_V\left(1+\mu_V+4\mu_h\right)+\mu_V^2 \,  ,
        \\f_3&=x_1\left(x_1-1\right)\left(\mu_V+x_1-1\right)-\left(1-x_2\right)\left(2-x_1\right)\left(1-x_1+\mu_V\right)+
        \\&\quad+2\mu_V\left(\mu_V+1-4\mu_h\right) \,  .
 \end{aligned}   
\end{equation}
The functions $f_1$, $f_2$ and $f_3$ originate from the momentum-dependent tensor structures of the diagrams with two $VVh$ vertices. In unitary gauge these terms are associated with the longitudinal part of the internal vector-boson propagator, while in
Feynman gauge the same contribution is reproduced by the corresponding Goldstone-boson exchange diagrams. 
In order to obtain the cross section, we convolute the result with the parton distribution functions $f_i(x,\mu_F)$ with $i=q,\bar{q},g$, momentum transfer $x$, and dependence on the factorisation scale $\mu_F$
\begin{equation}
    \sigma_{pp\rightarrow Vhh}=\int_{\tau_0}^1d\tau\sum_{i,j}\frac{d\mathcal{L}^{ij}}{d\tau}\int_{\tau_0/\tau}^1dz\;\hat\sigma^{LO}_{Vhh}\left(s=z\tau\hat{s}\right)\Delta_{ij}\left(z;\;ij\rightarrow V^*\right) \, ,
\end{equation}
where the luminosity function is given by
\begin{equation}
    \begin{aligned}
        \frac{d\mathcal{L}^{ij}}{d\tau}=\int_{\tau}^1\frac{dy}{y}f_i\left(y;\mu_F\right)f_{j}\left(\frac{\tau}{y};\mu_F\right) \, ,
    \end{aligned}
    \label{lum}
\end{equation}
and $\tau_0=(2m_h+m_V)^2/\hat{s}$.
At LO we have
\begin{equation}
\Delta_{ij}^{LO}=\delta\left(1-z\right)\delta_{i,q}\delta_{j,\bar{q}/\bar{q}'}\, .
\end{equation}
In order to obtain the NNLO QCD-corrected cross section, we observe that the process can be written as $pp\to V^*$ with subsequent decay $V^*\to Vhh$. The QCD corrections affect only the $pp\to V^*$ production which is a Drell-Yan process, and have been computed at NNLO QCD in Ref.~\cite{Harlander:2002wh, Hamberg:1990np} and more recently also at $\text{N}^3\text{LO}$ QCD in Ref.~\cite{Baglio:2022wzu}. We will remain at NNLO QCD which is sufficient given the current low precision of the measurement. The procedure is in analogy to $Vh$ production, for which the NNLO QCD corrections have been computed in Ref.~\cite{Brein:2003wg}. We are using the results of Ref.~\cite{Baglio:2012np}, whose cross section predictions have been recently updated for the Report 5 of the LHC Higgs Working Group in Ref.~\cite{Grober:2026ece}. 
\begin{figure}[t!]
    \centering    
    \includegraphics[width=16.2cm]{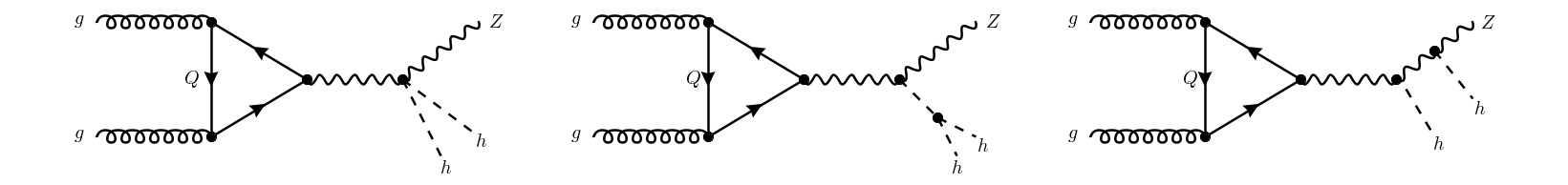}\\
    \includegraphics[width=16.2cm]{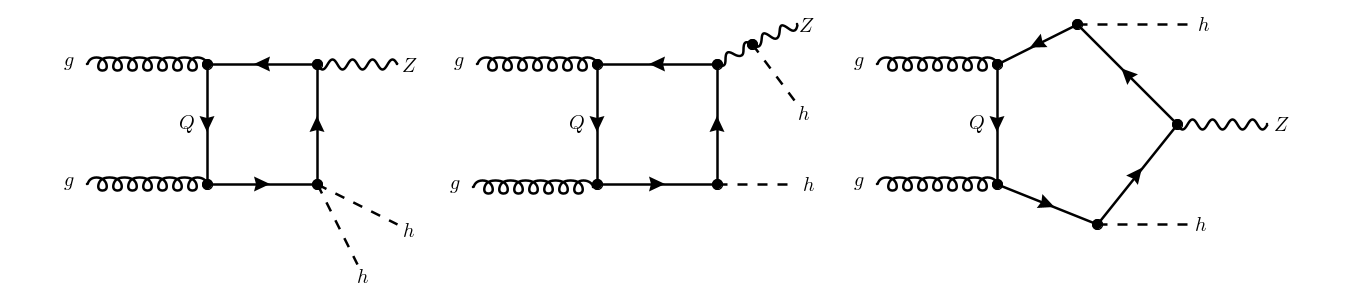}\\
    \caption{Illustrative diagrams for the gluon initiated process.}
    \label{fig:gluin}
\end{figure}
The cross section at NNLO QCD can be written as
\begin{equation}
\sigma_{ppVhh}^{NNLO}=\sigma_{ppVhh}^{LO}+\Delta\sigma_{q\bar{q}/\bar{q}'}+\Delta\sigma_{qg}+\Delta\sigma_{qq'}+\Delta\sigma_{qq}+\Delta\sigma_{gg}+\delta_{V,Z}\Delta \sigma_{ggZhh}\,,
    \label{sigmaquark}
\end{equation}
where the index in $\Delta \sigma$ indicates the initial state and the last piece, $\sigma_{ggZhh}$,  is specific to the $Zhh$ process and not a Drell-Yan type correction. The Feynman diagrams for this sub-process, mediated by heavy quarks, are shown in Fig.~\ref{fig:gluin}.
The contribution to the cross section is given by
\begin{equation}
\Delta \sigma_{ggZhh}=\int_{\tau_0}^1d\tau \frac{d\mathcal{L}} {d\tau}\hat{\sigma}_{ggZhh} \, ,
\end{equation}
where the hat indicates the partonic cross section.
The $gg\to Zhh$ sub-process contributes for the first time at NNLO QCD, even though it arises at one-loop order, as it does not interfere with the quark initiated sub-process at the considered order. This is in analogy to single Higgs production $Zh$, where $gg\to Zh$, even though being a NNLO QCD correction, turns out to be the driver of the remaining theoretical uncertainty. This has triggered the computation of NLO QCD corrections to $gg\to Zh$, see Refs.~\cite{Chen:2022rua, Degrassi:2022mro,  CampilloAveleira:2025rbh, Davies:2026uxl}. Due to their complexity, providing them for $gg\to Zhh$ is beyond the scope of this work. We will though study the EFT effects of the gluon-induced $Zhh$ production in detail.
The Drell-Yan type NNLO QCD corrections read
\begin{align}
 \sigma_{ppVhh}^{LO}& = \sum_{q,q'}\int_{\tau_0}^1d\tau\;\frac{d\mathcal{L}^{q \bar{ q}/\bar{ q}'}}{d\tau} \hat\sigma_{Vhh}^{LO}\left(\tau \hat{s}\right) \, ,
 \\ 
 \Delta\sigma_{q \bar{ q}/\bar{ q}'}&=\sum_{q,q'}\left(\frac{\alpha_s(\mu_R)}{\pi}\right)\int_{\tau_0}^1d\tau\;\frac{d\mathcal{L}^{q \bar{ q}/\bar{ q}'}}{d\tau}
 \\ \nonumber
 &\quad\times\int_{\tau_0/\tau}^1dz\;\hat\sigma_{Vhh}^{LO}\left(z\tau \hat{s}\right)\left(\Delta_{q\bar{q}}^{(1)}(z)+\left(\frac{\alpha_s(\mu_R)}{\pi}\right)\Delta_{q\bar{q}}^{(2)}(z)\right) \, ,
 \\ \allowdisplaybreaks
 \Delta\sigma_{q g}&=\sum_{i=q,\bar{q}}\left(\frac{\alpha_s(\mu_R)}{\pi}\right)\int_{\tau_0}^1d\tau\;\frac{d\mathcal{L}^{ig}}{d\tau}
 \\ \nonumber
 &\quad\times\int_{\tau_0/\tau}^1dz\;\hat\sigma_{Vhh}^{LO}\left(z\tau \hat{s}\right)\left(\Delta_{qg}^{(1)}(z)+\left(\frac{\alpha_s(\mu_R)}{\pi}\right)\Delta_{qg}^{(2)}(z)\right) \, ,
 \\\displaybreak
\Delta\sigma_{qq'}&=\sum_{i=q,\bar{q},j=q',\bar{q}'}\left(\frac{\alpha_s(\mu_R)}{\pi}\right)^2\int_{\tau_0}^1d\tau\; \frac{d\mathcal{L}^{ij}}{d\tau}\int_{\tau_0/\tau}^1dz\;\hat\sigma_{Vhh}^{LO}\left(z\tau \hat{s}\right)\Delta_{qq'}^{(2)}(z) \, ,
 \\ \allowdisplaybreaks
 \Delta\sigma_{qq}&=\sum_{i=q,\bar{q}}\left(\frac{\alpha_s(\mu_R)}{\pi}\right)^2\int_{\tau_0}^1d\tau\; \frac{d\mathcal{L}^{ii}}{d\tau}\int_{\tau_0/\tau}^1dz\;\hat\sigma_{Vhh}^{LO}\left(z\tau \hat{s}\right)\Delta_{qq}^{(2)}(z) \, ,
 \\
 \Delta\sigma_{gg}&=\left(\frac{\alpha_s(\mu_R)}{\pi}\right)^2\int_{\tau_0}^1d\tau\; \frac{d\mathcal{L}^{gg}}{d\tau}\int_{\tau_0/\tau}^1dz\;\hat\sigma_{Vhh}^{LO}\left(z\tau \hat{s}\right)\Delta_{gg}^{(2)}(z) \,.
 \end{align}
Here the different $\Delta^{(i=1,2)}(z)$ are respectively, NLO and NNLO corrections. They can be found in appendix B of Ref.~\cite{Hamberg:1990np} (where $\mu_F$ is indicated as $M$ and $\mu_R$ as $R$), although to be correctly inserted in our expressions, they must first be rescaled by a factor of $\left(\pi/\alpha_s\right)^i$.

\subsection{Uncertainty definitions for inclusive cross sections}
\label{subsec:uncrt}

Let us introduce the dimensionless scale ratios
\begin{equation}
    x_F = \frac{\mu_F}{m_{Vhh}}\,, \qquad x_R = \frac{\mu_R}{m_{Vhh}}\,,
\end{equation}
where $m_{Vhh}$ is the invariant mass of the $Vhh$ system, with $\mu_F$ and $\mu_R$ as the factorisation and renormalisation scales. We denote by $\sigma^{(n)}(x_F,x_R)$ the inclusive cross section (at LO or NNLO) obtained with PDF member $n$ at scales $\mu_F = x_F\, m_{Vhh}$ and $\mu_R = x_R\, m_{Vhh}$. The default (central) scale choice thus corresponds to $\sigma^{(n)}(1,1) \equiv \sigma^{(n)}(\mu_F=m_{Vhh},\, \mu_R=m_{Vhh})$.

The PDF uncertainty for the PDF set \texttt{PDF4LHC21\_40}~\cite{Cridge:2021qjj,PDF4LHCWorkingGroup:2022cjn} (40 replicas) is obtained from
\begin{equation}
\Delta_{\mathrm{PDF}} = \sqrt{\sum_{n=1}^{40}\bigl(\sigma^{(n)}(1,1)-\sigma^{(0)}(1,1)\bigr)^2}\,,
\end{equation}
where $\sigma^{(n)}(1,1)$ is the cross section of the $n$th replica of the PDF, computed at the central scale choice. The $\alpha_s$ uncertainty uses the dedicated PDF members $n=41$ and $n=42$ at the default scale,
\begin{equation}
\Delta_{\alpha_s} = \frac{\sigma^{(42)}(1,1)-\sigma^{(41)}(1,1)}{2}\,.
\end{equation}
The PDF and $\alpha_s$ uncertainties are combined in quadrature,
\begin{equation}
\Delta_{\mathrm{PDF}+\alpha_s} = \sqrt{\Delta_{\mathrm{PDF}}^2+\Delta_{\alpha_s}^2}\,.
\end{equation}
For the central PDF ($n=0$), we additionally evaluate the seven standard scale points
\begin{equation}
\mathcal{S} = \bigl\{(x_F,x_R)\bigr\} = \bigl\{(0.5,0.5),\,(1,0.5),\,(0.5,1),\,(1,1),\,(2,1),\,(1,2),\,(2,2)\bigr\}\,,
\end{equation}
restricted to $1/2 \le \mu_F/\mu_R \le 2$, following the usual fixed-order seven-point prescription. The scale uncertainties are then defined as
\begin{align}
\Delta^{\mathrm{scale},+} &= \max_{(x_F,x_R)\in\mathcal{S}}\, \sigma^{(0)}(x_F,x_R) - \sigma^{(0)}(1,1)\,,\\
\Delta^{\mathrm{scale},-} &= \sigma^{(0)}(1,1) - \min_{(x_F,x_R)\in\mathcal{S}}\, \sigma^{(0)}(x_F,x_R)\,.
\end{align}

\subsection{HEFT corrections}
\label{subsec:HEFT:param}

In the case of HEFT corrections, we use multiplicative factors which correspond to the LO HEFT Lagrangian, as discussed in Eq.~\eqref{HEFT:LO:Lagrangian}, using the relations in Eqs.~\eqref{eq:relHEFT1} and~\eqref{eq:relHEFT2}, and a power counting prescription that does not count orders of $g_s$ \cite{Brivio:2025yrr}. This approach corresponds to the coupling modifiers framework (also known as $\kappa$-framework) used in various experimental analyses~\cite{ATLAS:2025tkn, ATLAS:2024ish}. These modification factors can be introduced into Eq.~\eqref{eq:sigmalo} as follows
\begin{equation}
\begin{aligned}
A\to A_{HEFT}=&\kappa_V \times \kappa_\lambda \times A\\
B\to B_{HEFT}=& \kappa_V^2 \times B\,,\\
C\to C_{HEFT}=&  \kappa_V^2 \times C\,,\\
D\to D_{HEFT}=&  \kappa_{2V} \times D\,,\\
f_1\to f_1^{HEFT}=& \kappa_V^4 \times f_1\,,\\
f_2\to f_2^{HEFT}=&\kappa_V^4 \times f_2\,,\\
f_3\to f_3^{HEFT}=&\kappa_V^2 \times f_3\,.
\end{aligned}
\end{equation}

This type of corrections allows us to conveniently parametrise the $pp\rightarrow W^\pm hh$ EFT cross section prediction as
\begin{align} \sigma_{\textrm{LO}}^{(W,\text{HEFT})}&=A^{W}_{1,\textrm{LO}}\, \kappa_{W}^4+A^W_{2,\textrm{LO}}\,\kappa_{W}^3\times \kappa_\lambda+A^W_{3,\textrm{LO}}\,\kappa_{W}^2\times \kappa_{2W}+
    \\&\quad +A^W_{4,\textrm{LO}}\,\kappa_{W}^2\times \kappa_\lambda^2+A^W_{5,\textrm{LO}}\,\kappa_{W}\times \kappa_{\lambda}\times \kappa_{2W}+A^W_{6,\textrm{LO}}\, \kappa_{2W}^2 \, , \nonumber \\ 
      \label{intheftw} \sigma_{\textrm{NNLO}}^{(W,\text{HEFT})}&=A^{W}_{1,\textrm{NNLO}}\, \kappa_{W}^4+A^{W}_{2,\textrm{NNLO}}\,\kappa_{W}^3\times \kappa_\lambda+A^{W}_{3,\textrm{NNLO}}\,\kappa_{W}^2\times \kappa_{2W}+
    \\&\quad +A^{W}_{4,\textrm{NNLO}}\,\kappa_{W}^2\times \kappa_\lambda^2+A^{W}_{5,\textrm{NNLO}}\,\kappa_{W}\times \kappa_\lambda\times \kappa_{2W}+A^{W}_{6,\textrm{NNLO}}\, \kappa_{2W}^2 \, , \nonumber
\end{align}
where the point $(\kappa_\lambda,\kappa_W,\kappa_{2W})=(1,1,1)$ corresponds to the SM cross section. Using a sufficient number of simulated data points, we interpolate the coefficients $A_i^W$ (with different $A_i^W$ coefficients for $W^+$ and $W^-$). and obtain a closed-form expression for the cross section as a function of the coupling modifiers. Numerical results for the $A_i^W$ coefficients are given in Sec.~\ref{subsec:HEFT}.

For the $Z$ boson, the LO cross section follows an analogous relation to the one for $W$. At NNLO, as discussed in Sec.~\ref{subsec:SM}, we must additionally include the contributions that enter only via the gluon-initiated amplitude squared. The complete expressions are thus
\begin{align}    
\sigma_{\textrm{LO}}^{(Z,\text{HEFT})}&=A^{Z}_{1,\textrm{LO}}\, \kappa_{Z}^4+A^{Z}_{2,\textrm{LO}}\,\kappa_{Z}^3\times \kappa_\lambda+A^{Z}_{3,\textrm{LO}}\,\kappa_{Z}^2\times \kappa_{2Z}+
    \\&\quad +A^{Z}_{4,\textrm{LO}}\,\kappa_{Z}^2\times \kappa_\lambda^2+A^{Z}_{5,\textrm{LO}}\,\kappa_{Z}\times \kappa_\lambda\times \kappa_{2Z}+A^{Z}_{6,\textrm{LO}}\, \kappa_{2Z}^2 \, , \nonumber \\ 
\label{intheftz} \sigma_{\textrm{NNLO}}^{(Z,\text{HEFT})} &=
A^Z_{1,\text{NNLO}} \kappa_{Z}^4 +
A^Z_{2,\text{NNLO}} \kappa_{Z}^3 \kappa_\lambda +
A^Z_{3,\text{NNLO}} \kappa_{Z}^2 \kappa_{2Z} +
A^Z_{4,\text{NNLO}} \kappa_{Z}^2 \kappa_\lambda^2 \\
&\quad +
A^Z_{5,\text{NNLO}} \kappa_{Z} \kappa_\lambda \kappa_{2Z} +
A^Z_{6,\text{NNLO}} \kappa_{2Z}^2 +
A^Z_{7,\text{NNLO}} \kappa_{2Z} \kappa_t \kappa_\lambda +
A^Z_{8,\text{NNLO}} \kappa_{2Z} \kappa_Z \kappa_t  \nonumber\\
&\quad +
A^Z_{9,\text{NNLO}} \kappa_{2Z} \kappa_t^2 +
A^Z_{10,\text{NNLO}} \kappa_Z^3 \kappa_t +
A^Z_{11,\text{NNLO}} \kappa_Z^2 \kappa_t^2 +
A^Z_{12,\text{NNLO}} \kappa_Z^2 \kappa_\lambda \kappa_t \nonumber \\
&\quad +
A^Z_{13,\text{NNLO}} \kappa_Z \kappa_\lambda^2 \kappa_t +
A^Z_{14,\text{NNLO}} \kappa_Z \kappa_\lambda \kappa_t^2 +
A^Z_{15,\text{NNLO}} \kappa_t^2 \kappa_\lambda^2 +
A^Z_{16,\text{NNLO}} \kappa_\lambda \kappa_t^3  \nonumber \\
&\quad +
A^Z_{17,\text{NNLO}} \kappa_Z \kappa_t^3 +
A^Z_{18,\text{NNLO}} \kappa_t^4 \, . \nonumber
\end{align}
The interpolation to obtain the numerical coefficients $A_i^Z$, though computationally heavier, follows the same procedure as for $A_i^{W}$. The corresponding results are presented again in Sec.~\ref{subsec:HEFT}.

\subsection{SMEFT corrections}
\label{sec:SMEFT:corr}
In order to calculate relevant SMEFT vertex corrections, we exploited \texttt{FeynRules}~\cite{Alloul:2013bka} and \texttt{SmeftFR}~\cite{Dedes:2023zws}, with output passed to \texttt{FeynArts}~\cite{Hahn:2000kx} and \texttt{FormCalc}~\cite{Hahn:2008fyh}.

In this work, we only consider SMEFT operators that affect the electroweak vertices, so apart from $gg\to Zhh$ production, we only need to modify $\hat{\sigma}^{LO}_{Vhh}$. In principle, the operator $\mathcal{O}_G=G_{\mu\nu}G^{\mu\alpha}G_{\alpha}^{\nu}$, which modifies the trilinear gluon self-coupling, would also enter the NNLO QCD corrections; however, being loop-induced, it effectively contributes only at three-loop order. Furthermore, the chromomagnetic operators $\mathcal{O}_{uG}=(\bar{q}_L\sigma_{\mu\nu}T^a\tilde{\varphi}u_R)G^{a,\mu\nu}$ and $\mathcal{O}_{dG}=(\bar{q}_L\sigma_{\mu\nu}T^a\varphi d_R)G^{a,\mu\nu}$ would contribute at one-loop level. Since this operator is loop-generated, it enters formally at NNLO. However, it also changes the helicity structure of the amplitude, such that it does not interfere with the LO diagrams, and its contribution is therefore beyond the considered order. In the $gg\to Zh$ diagrams, the chromomagnetic operator, being loop-generated in weakly interacting theories~\cite{Arzt:1994gp}, also leads to a contribution beyond the considered order when adopting a loop counting \cite{Buchalla:2022vjp}.

We use as input parameters $G_F,\; m_W, \; m_Z$ and $m_h$. 
The trilinear Higgs self-coupling gets modified with respect to the SM by the operators $\mathcal{O}_{\varphi}$, $\mathcal{O}_{\varphi D}$ and $\mathcal{O}_{\varphi\Box}$. The former simply rescales the SM value, but the latter introduce a momentum dependent contribution.
In our notation, the modification with respect to the SM value is $C_{hhh}=g^{hhh}_{SMEFT}/g^{hhh}_{SM}-1$
\begin{equation}
    \begin{aligned}
        C_{hhh}&=\frac{1}{\Lambda^2}\left(-\frac{1}{G_F^2m_h^2}C_\varphi+\frac{\sqrt{2}}{6\,G_Fm_h^2}\left(7\mu_h s+6s\left( -1+x_1+x_2+\mu_V\right)\right)C_{\varphi\square}-\right.
        \\&\quad\left.-\frac{\sqrt{2}}{24\,G_Fm_h^2}\left(7\mu_h s-6s\left( -1+x_1+x_2+\mu_V\right)\right)C_{\varphi D}\right)+\mathcal{O}\left(\frac{1}{\Lambda^4} \right).
    \end{aligned}
\end{equation}
For the contributions of $\mathcal{O}_{\varphi D}$ and $\mathcal{O}_{\varphi \Box}$ to $VVh$ and $VVhh$ couplings, one generates multiplicative factors of form

\begin{align}
C_{WWh}=& \frac{1}{\sqrt{2}G_F}\frac{1}{\Lambda^2}\left(C_{\varphi\Box}-\frac{1}{4}C_{\varphi D} \right)\,, \label{eq:CWWh}\\
C_{ZZh}=& \frac{1}{\sqrt{2}G_F}\frac{1}{\Lambda^2}\left(C_{\varphi\Box}+\frac{1}{4}C_{\varphi D} \right)\, , \\
C_{WWhh}=& \frac{\sqrt{2}}{G_F}\frac{1}{\Lambda^2}\left(C_{\varphi\Box}-\frac{1}{4}C_{\varphi D} \right)\, , \\
C_{ZZhh}=& \frac{\sqrt{2}}{G_F}\frac{1}{\Lambda^2}\biggl(C_{\varphi\Box}+C_{\varphi D} \biggr)\, .
\label{eq:CZZhh}
\end{align}
\newpage
Operators of type $\psi^2 \varphi^2 D$ and $X^2 \varphi^2$ generate two distinct types of contributions. The first is a correction to the $\bar{q} q V$ vertices, of the form
\begin{align} \label{eq:CqqW}
C_{\bar q q W}
&= \frac{1}{\sqrt{2}G_F}\,
\frac{C_{\varphi q}^{(3)}}{\Lambda^2}\,,
\\ 
C_{\bar u u Z}
&=
\frac{3m_Z^2(m_Z^2-4m_W^2)}
{\sqrt{2}G_F\,D_u}
\left(
\frac{C_{\varphi q}^{(1)}}{\Lambda^2}
-\frac{C_{\varphi q}^{(3)}}{\Lambda^2}
\right)
+
\frac{12m_Z^2(m_Z^2-m_W^2)}
{\sqrt{2}G_F\,D_u}
\frac{C_{\varphi u}}{\Lambda^2}
\nonumber\\
&\quad
+
\frac{4m_W^4-m_Z^4}
{2\sqrt{2}G_F(4m_W^4-5m_W^2m_Z^2+m_Z^4)}
\left(\frac{C_{\varphi D}}{\Lambda^2}
+\frac{4m_W\sqrt{m_Z^2-m_W^2}}{m_Z^2}
\frac{C_{\varphi WB}}{\Lambda^2} \right)\,, 
\\ 
C_{\bar d d Z}
&=
\frac{3m_Z^2(m_Z^2+2m_W^2)}
{\sqrt{2}G_F\,D_d}
\left(
\frac{C_{\varphi q}^{(1)}}{\Lambda^2}
+\frac{C_{\varphi q}^{(3)}}{\Lambda^2}
\right)
-
\frac{6m_Z^2(m_Z^2-m_W^2)}
{\sqrt{2}G_F\,D_d}
\frac{C_{\varphi d}}{\Lambda^2}
\nonumber\\
&\quad
-
\frac{2m_W^4+m_Z^4}
{2\sqrt{2}G_F(-2m_W^4+m_W^2m_Z^2+m_Z^4)} \left( \frac{C_{\varphi D}}{\Lambda^2}
+\frac{4m_W\sqrt{m_Z^2-m_W^2}}{m_Z^2}
\frac{C_{\varphi WB}}{\Lambda^2}\right)\, .
\label{eq:CddZ}
\end{align}
with
\begin{equation}
D_u=32m_W^4-40m_W^2m_Z^2+17m_Z^4 \, ,\qquad
D_d=8m_W^4-4m_W^2m_Z^2+5m_Z^4\,,
\end{equation}
while the second type of contribution leads to new terms arising due to novel interaction vertices, and will be addressed later.
The modification factors of Eqs.~\eqref{eq:CWWh}--\eqref{eq:CZZhh} and Eqs.~\eqref{eq:CqqW}--\eqref{eq:CddZ} can be introduced into Eq.~\eqref{eq:sigmalo} as follows
\begin{equation}
\begin{aligned}
A\to A_{SMEFT}=&A\left(1+ C_{hhh} + C_{VVh} + 2 C_{\bar{q} q V} \right)\,,\\
B\to B_{SMEFT}=& B \left(1+ 2 C_{VVh} + 2 C_{\bar{q} q V}\right) \,,\\
C\to C_{SMEFT}=& C \left(1+2 C_{VVh}+ 2 C_{\bar{q} q V}\right)\,,\\
D\to D_{SMEFT}=& D (1+C_{VVhh}+ 2 C_{\bar{q} q V}) \,, \\
f_1\to f_1^{SMEFT}=& f_1 (1+4 C_{VVh}+ 4 C_{VVh}^2+ 2 C_{\bar{q} q V}) \,,\\
f_2\to f_2^{SMEFT}=& f_2 (1+ 4 C_{VVh}+ 4 C_{VVh}^2+ 2 C_{\bar{q} q V}) \,,\\
f_3\to f_3^{SMEFT}=& f_3 (1+2 C_{VVh}+ 2 C_{\bar{q} q V}) \,,\label{eq:SMEFTf3}
\end{aligned}
\end{equation}
whereas $f_0$ remains unchanged. Inserted in this way, they lead to $\mathcal{O}(1/\Lambda^4)$ terms in the matrix element squared. Staying strictly at the $\mathcal{O}(1/\Lambda^2)$ order in the matrix element squared, the combinations that appear in Eq.~\eqref{eq:sigmalo} need to be further expanded.

\begin{figure}[t!]
\centering
 \includegraphics[width=0.6\linewidth]{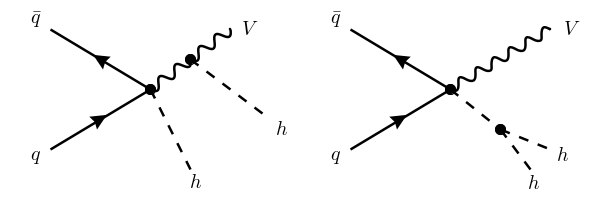}
    \caption{Example diagrams featuring genuine SMEFT contributions to quark-initiated $Vhh$ production.}
    \label{fig:VHH:Diagrams:SMEFT}
\end{figure}

Instead some of the contributions from operators of type $\psi^2 \varphi^2 D$ and $X^2 \varphi^2$, cannot just be written in terms of the $f_i$ coefficients of the SM, but one has to introduce new ones, which is due to the appearance of new Lorentz structures in the vertices, and new diagram topologies (see Fig.~\ref{fig:VHH:Diagrams:SMEFT}). Their contribution to the cross section is given by
\begin{equation}
\frac{d^2\hat\sigma_{\rm SMEFT}}{dx_1dx_2}=\frac{d^2\hat\sigma_{\rm LO}}{dx_1dx_2}+\frac{G_F^3 m_V^6}{1152\sqrt{2}\,\pi^3\,s\,(1-\mu_V)^2}\left[\frac{C_{Vq}}{\Lambda^2}\,
\mathcal{F}_{Vq}(x_1,x_2;\mu_h,\mu_V)\right]+\frac{d^2\hat\sigma_{X^2\varphi^2}}{dx_1dx_2}(V) \, , \label{eq:sigmaSMEFT}
\end{equation}
where $\hat\sigma_{\rm LO}$ has to be understood as the contribution with the replacements in Eq.~\eqref{eq:SMEFTf3} expanded to $\mathcal{O}(1/\Lambda^2)$.
We begin by discussing the contribution of $C_{Vq}$. The form factor $\mathcal{F}_{Vq}$ is given by
\begin{equation}
\mathcal{F}_{Vq}
=\mathcal{S}_{12}\!\left[f^1_{Vq}+f^2_{Vq}+f^5_{Vq}\right]
+f^3_{Vq}+f^4_{Vq} \, ,
\qquad
\mathcal{S}_{12}F(x_1,x_2)\equiv F(x_1,x_2)+F(x_2,x_1) \, .
\end{equation}
The five independent kinematic functions can then be written as
\begin{equation}
\begin{aligned}
f^1_{Vq}
&=\frac{(\mu_V-1)\,N_1}
        {2\sqrt{2}\,G_F\,\mu_V^2\,\Delta_2}
  \left(2-\frac{4\mu_V}{\Delta_1}
       -\frac{4\mu_V}{\Delta_2}
       +\frac{6\mu_h}{\Delta_{12}}\right),
\\[2mm]
f^2_{Vq}
&=\frac{(\mu_V-1)\,R}
        {\sqrt{2}\,G_F\,\mu_V\,\Delta_1}
  \left(2-\frac{2\mu_V}{\Delta_1}
       -\frac{2\mu_V}{\Delta_2}
       +\frac{6\mu_h}{\Delta_{12}}\right),
\\[2mm]
f^3_{Vq}
&=-\frac{3\mu_h(\mu_V-1)\,R}
        {2\sqrt{2}\,G_F\,\mu_V^2\,\Delta_{12}}
  \left(2+\frac{3\mu_h}{\Delta_{12}}
       +\frac{\Delta_{12}}{3\mu_h}\right),
\\[2mm]
f^4_{Vq}
&=-\frac{(\mu_V-1)\,P\,Q}
        {\sqrt{2}\,G_F\,\mu_V^2\,\Delta_1\Delta_2},
\\[2mm]
f^5_{Vq}
&=-\frac{(\mu_V-1)(x_1^2-4\mu_h)
        \left[(x_1-1+\mu_V)^2-4\mu_h\mu_V\right]}
        {2\sqrt{2}\,G_F\,\mu_V^2\,\Delta_1^2},
\end{aligned}
\end{equation}
where
\begin{equation}
\begin{aligned}
N_1&=(x_2-1)\left[2+(x_1+x_2)(x_2-2)\right]
     +\left[x_2^2-8\mu_h-x_1(x_2-2)\right]\mu_V
     +2\mu_V^2 \, ,
\\
P&=x_1+x_2-x_1x_2-1+(x_1+x_2-4\mu_h)\mu_V+\mu_V^2 \, ,
\\
Q&=-x_1(x_2-2)+2(x_2-1-2\mu_h+\mu_V),\\
R&=(2-x_1-x_2)^2+8\mu_V \,,
\end{aligned}
\end{equation}
and
\begin{equation}
\Delta_i=x_i-1-\mu_h+\mu_V\, ,\qquad
\Delta_{12}=x_1+x_2-1-\mu_h+\mu_V\, .
\end{equation}
The Wilson-coefficient combinations $C_{Vq}$ appearing in Eq.~\eqref{eq:sigmaSMEFT} are
\begin{equation}
\begin{aligned}
C_{Wq}
&=9 C_{\varphi q}^{(3)},\\
C_{Zu}
&=\frac{m_Z^2}{3 \left(32m_W^4-40m_W^2m_Z^2+17m_Z^4\right)}
\left[
-4(9C_{\varphi q}^{(1)}-9C_{\varphi q}^{(3)}+C_{\varphi u})m_W^2
\right. \\  & \hspace{42mm}
+(9C_{\varphi q}^{(1)}-9C_{\varphi q}^{(3)}+4C_{\varphi u})m_Z^2
  \bigg]\,,
\\
C_{Zd}
&=\frac{m_Z^2}
        {3\left(32m_W^4-40m_W^2m_Z^2+17m_Z^4\right)}
  \bigg[
  2\left(C_{\varphi q}^{(1)}+C_{\varphi q}^{(3)}+C_{\varphi d}\right)m_W^2
\\
&\hspace{42mm}
  +\left(C_{\varphi q}^{(1)}+C_{\varphi q}^{(3)}
  -2C_{\varphi d}\right)m_Z^2
  \bigg]\,.
\end{aligned}
\end{equation}
Now we turn to the contributions of the operators of type $X^2\varphi^2$. Their contribution to the LO cross section is
\begin{align}
\frac{d^2\hat\sigma_{X^2\varphi^2}}{dx_1dx_2}(W)
&=\frac{G_F^3 m_V^6}{1152\sqrt{2}\,\pi^3\,s\,(1-\mu_V)^2} \frac{C_{X^2\varphi^2}(W)}{\Lambda^2}\,
\mathcal{F}_{X^2\varphi^2}(x_1,x_2;\mu_h,\mu_V) \, ,
\\
\frac{d^2\hat\sigma_{X^2\varphi^2}}{dx_1dx_2}(Z)
&=\frac{G_F^3 m_V^6}{1152\sqrt{2}\,\pi^3\,s\,(1-\mu_V)^2} \left[\frac{C_{X^2\varphi^2}(Z)}{\Lambda^2}\,
\mathcal{F}_{X^2\varphi^2}(x_1,x_2;\mu_h,\mu_V) \right.
\\ &  \left. +\frac{C^{Z{\rm -only}}_{X^2\varphi^2}}{\Lambda^2}\,
\mathcal{F}^{Z{\rm -only}}_{X^2\varphi^2}(x_1,x_2;\mu_h,\mu_V)\right] \, . \nonumber
\end{align}
The Wilson-coefficient combinations are
\begin{equation}
\begin{aligned}
C_{X^2\varphi^2}(W)&=C_{\varphi W} \, ,
\\
C_{X^2\varphi^2}(Z)
&=\frac{m_W^2}{m_Z^4}
\left[
m_W^2 C_{\varphi W}
+(m_Z^2-m_W^2)C_{\varphi B}
-m_W\sqrt{m_Z^2-m_W^2}\,C_{\varphi WB}
\right] \, ,
\\
C^{Z{\rm -only}}_{X^2\varphi^2}
&=C_{\varphi B}-C_{\varphi W}
+\frac{2m_W^2-m_Z^2}{2m_W\sqrt{m_Z^2-m_W^2}}\,C_{\varphi WB}\, .
\end{aligned}
\end{equation}
For the contribution from diagrams allowed both for $V=W^{\pm},Z$ we find
\begin{equation}
\mathcal{F}_{X^2\varphi^2}
=f^1_{X^2\varphi^2}
+\mathcal{S}_{12}\!\left[
f^2_{X^2\varphi^2}
+f^3_{X^2\varphi^2}
+f^4_{X^2\varphi^2}
+f^5_{X^2\varphi^2}
\right],
\end{equation}
with
\begin{equation}
\begin{aligned}
f^1_{X^2\varphi^2}
&=-\frac{18\sqrt{2}\,(x_1+x_2-2)\mu_h}
        {G_F\mu_V\Delta_{12}}
\left(1+\frac{\Delta_{12}}{6\mu_h}
        +\frac{3\mu_h}{2\Delta_{12}}\right),
\\[2mm]
f^2_{X^2\varphi^2}
&=-\frac{\sqrt{2}\,N_2}
        {G_F\mu_V\Delta_2}
\left(1+\frac{3\mu_h}{\Delta_{12}}\right),
\\[2mm]
f^3_{X^2\varphi^2}
&=\frac{\sqrt{2}\,N_3}
        {G_F\mu_V\Delta_2}
\left(1+\frac{x_2-1+(x_2-3)\mu_V}{\Delta_2}
        +\frac{3\mu_h}{\Delta_{12}}\right),
\\[2mm]
f^4_{X^2\varphi^2}
&=\frac{2\sqrt{2}\,N_4}{G_F\Delta_2^2} \, ,
\\[2mm]
f^5_{X^2\varphi^2}
&=\frac{N_5}{G_F\mu_V\Delta_1\Delta_2} \, .
\end{aligned}
\end{equation}
Here,
\begin{equation}
\begin{aligned}
N_2&=(x_2-1)(x_1+x_2-2)
    +\left[30+x_1(x_2-9)+(x_2-19)x_2\right]\mu_V
    +4\mu_V^2 \, ,
\\
N_3&=2x_1(x_2-1-\mu_h)
    +(x_2-2)(x_2-1-2\mu_h+\mu_V) \, ,
\\
N_4&=(x_2-1)(x_1+x_2-2)
    +\left[18+x_1(x_2-3)+(x_2-13)x_2\right]\mu_V
    +4\mu_V^2 \, ,
\\
N_5&=\sqrt{2}\bigg\{
(x_2-1)\left[x_2-x_1x_2+2\mu_h(x_1+x_2-2)\right]
\\
&\quad
-\left[6-6x_1-8x_2+4x_1x_2+x_2^2+x_1x_2^2
-2\left(-6+x_1+(x_1-1)x_2+x_2^2\right)\mu_h\right]\mu_V
\\
&\quad
+\left[4-x_1(x_2-2)+x_2-8\mu_h\right]\mu_V^2
+2\mu_V^3
\bigg\} \, .
\end{aligned}
\end{equation}
For the \(Zhh\)-only diagrams, one finds
\begin{equation}
\mathcal{F}^{Z{\rm -only}}_{X^2\varphi^2}
=f^{0,Z}_{X^2\varphi^2}
+f^{1,Z}_{X^2\varphi^2}
+\mathcal{S}_{12}\!\left[
f^{2,Z}_{X^2\varphi^2}
+f^{3,Z}_{X^2\varphi^2}
+f^{4,Z}_{X^2\varphi^2}
\right] \, .
\end{equation}
It is useful to set
\begin{equation}
D_Z=32m_W^4-40m_W^2m_Z^2+17m_Z^4 \, ,
\qquad
\mathcal{N}_Z=
\frac{4\sqrt{2}\,m_W^2(\mu_V-1)(m_W^2-m_Z^2)(8m_W^2-5m_Z^2)}
     {G_Fs\,\mu_V^2D_Z} \, .
\end{equation}
Then
\begin{equation}
\begin{aligned}
f^{0,Z}_{X^2\varphi^2}
&=3\mathcal{N}_Z(x_1+x_2-2)
\left(1-\frac{2\mu_V}{\Delta_1}
        -\frac{2\mu_V}{\Delta_2}\right),
\\[2mm]
f^{1,Z}_{X^2\varphi^2}
&=\frac{9\mathcal{N}_Z\mu_h(x_1+x_2-2)}{\Delta_{12}}
\left(2-\frac{2\mu_V}{\Delta_1}
        -\frac{2\mu_V}{\Delta_2}
        +\frac{3\mu_h}{\Delta_{12}}\right),
\\[2mm]
f^{2,Z}_{X^2\varphi^2}
&=-\frac{\mathcal{N}_Z\,N_3}{\Delta_2}
\left(1+\frac{x_2-1-\mu_V}{\Delta_2}
        +\frac{3\mu_h}{\Delta_{12}}\right),
\\[2mm]
f^{3,Z}_{X^2\varphi^2}
&=-\frac{\mathcal{N}_Z\,N_6}{\Delta_2}
\left(1-\frac{2\mu_V}{\Delta_1}
        -\frac{2\mu_V}{\Delta_2}
        +\frac{3\mu_h}{\Delta_{12}}\right),
\\[2mm]
f^{4,Z}_{X^2\varphi^2}
&=\frac{\mathcal{N}_Z\,N_7}{\Delta_1\Delta_2} \, ,
\end{aligned}
\end{equation}
where
\begin{equation}
\begin{aligned}
N_6&=(1-x_2)(x_1+x_2-2)+\left[x_1+5(x_2-2)\right]\mu_V \, ,
\\
N_7&=(x_2-1)\left[(x_1-1)x_2-2(x_1+x_2-2)\mu_h\right]
\\
&\quad
+\left[4-4x_1-6x_2+3x_1x_2+x_2^2
      -2(x_1+x_2-6)\mu_h\right]\mu_V
+(x_2-4)\mu_V^2 \, .
\end{aligned}
\end{equation}

While for the operators of type $X^2\varphi^2$, $\varphi^4$ and $\varphi^4D^2$ the factorisation of the QCD corrections from the electroweak EFT part is clear, we comment on the operators $O_{u \varphi}$, $O_{\varphi q}^{(1,3)}$, $O_{\varphi u}$ and $O_{\varphi d}$. They also induce contact interactions between the quark current and the Higgs/gauge fields. Since these interactions are still built from color-singlet quark currents with the same QCD quantum numbers and Lorentz structure as the $q\bar{q}V$ couplings in the SM Drell-Yan current, the initial-state QCD corrections factorise in the same way. We therefore apply the standard Drell-Yan coefficient functions to the full SMEFT hard function.

As for the HEFT we can express the SMEFT cross section in terms of some numerical coefficients, $B_i$, and the Wilson coefficients. At linear order in $1/\Lambda^2$ one can write it as 
\begin{equation}
\begin{aligned}
 \sigma_{\textrm{LO}}^{(W,\text{SMEFT})}=
\sigma_{\textrm{LO}}^{W,\textrm{SM}}+B_{1,\rm LO}^W \frac{C_{\varphi}}{\Lambda^2}+B_{2,\rm LO}^W \frac{C_{\varphi \Box}}{\Lambda^2}+ B_{3,\rm LO}^W \frac{C_{\varphi D}}{\Lambda^2}+ B_{4,\rm LO}^W \frac{C_{\varphi q}^{(3)}}{\Lambda^2}+ B_{5,\rm LO}^W \frac{C_{\varphi W}}{\Lambda^2} \, ,
\label{eq:sigmaWhhSMEFT}
\end{aligned}
\end{equation}
for $W^{\pm}hh$ (with different $B_i$ coefficient depending on whether it is a positively charged $W$ or a negatively charged one).
The NNLO coefficients are defined in analogy to the LO one replacing in the labels ``LO'' with ``NNLO''. 
The $Zhh$ cross section can be parametrised at LO and NNLO as follows
\begin{align} \label{eq:sigmaZhhSMEFTLO}
 \sigma_{\textrm{LO}}^{(Z,\text{SMEFT})}&=
\sigma_{\textrm{LO}}^{Z,\textrm{SM}}+B_{1,\rm LO}^Z \frac{C_{\varphi}}{\Lambda^2}+B_{2,\rm LO}^Z \frac{C_{\varphi \Box}}{\Lambda^2}+ B_{3,\rm LO}^Z \frac{C_{\varphi D}}{\Lambda^2}+ B_{4,\rm LO}^Z \frac{C_{\varphi q}^{(3)}}{\Lambda^2}+ B_{5,\rm LO}^Z \frac{C_{\varphi W}}{\Lambda^2} \\ &+B_{6,\rm LO}^Z \frac{C_{\varphi q}^{(1)}}{\Lambda^2}  +B_{7,\rm LO}^Z \frac{C_{\varphi u}}{\Lambda^2}+B_{8,\rm LO}^Z \frac{C_{\varphi d}}{\Lambda^2}+ B_{9,\rm LO}^Z \frac{C_{\varphi B}}{\Lambda^2}+ B_{10,\rm LO}^Z \frac{C_{\varphi WB}}{\Lambda^2} \nonumber\,,  \\
 \sigma_{\textrm{NNLO}}^{(Z,\text{SMEFT})}&=
\sigma_{\textrm{NNLO}}^{Z,\textrm{SM}}+B_{1,\rm NNLO}^Z \frac{C_{\varphi}}{\Lambda^2}+B_{2,\rm NNLO}^Z \frac{C_{\varphi \Box}}{\Lambda^2}+ B_{3,\rm NNLO}^Z \frac{C_{\varphi D}}{\Lambda^2}+ B_{4,\rm NNLO}^Z \frac{C_{\varphi q}^{(3)}}{\Lambda^2} \\ &+ B_{5,\rm NNLO}^Z \frac{C_{\varphi W}}{\Lambda^2} +B_{6,\rm NNLO}^Z \frac{C_{\varphi q}^{(1)}}{\Lambda^2}  +B_{7,\rm NNLO}^Z \frac{C_{\varphi u}}{\Lambda^2}+B_{8,\rm NNLO}^Z \frac{C_{\varphi d}}{\Lambda^2} \nonumber\\ &+ B_{9,\rm NNLO}^Z \frac{C_{\varphi B}}{\Lambda^2}+ B_{10,\rm NNLO}^Z \frac{C_{\varphi WB}}{\Lambda^2}+B_{11,\rm NNLO}^Z \frac{C_{t \varphi}}{\Lambda^2} +B_{12,\rm NNLO}^Z \frac{C_{\varphi t}+C_{\varphi Q}^{(3)}-C_{\varphi Q}^{(1)}}{\Lambda^2}\,. \nonumber
\label{eq:sigmaZhhSMEFT}
\end{align}
The last two terms are specific to the gluon induced channel $gg\to Zhh$ arising at the NNLO. We neglect the bottom Yukawa couplings in this formula. The coefficients $C_{t\varphi}$, $C_{\varphi t}$, $C_{\varphi Q}^{(1)}$ and $C_{\varphi Q}^{(3)}$ indicate the modifications of the $Z$ couplings to top quarks, which we consider different from the first and second generation ones. Only the indicated combination of $C_{\varphi t}$, $C_{\varphi Q}^{(1)}$ and $C_{\varphi Q}^{(3)}$ enters, from modifications of the box and pentagon diagrams, similar to $gg\to Zh$ production, where it was shown in Refs.~\cite{Rossia:2023hen, inprep} that the effects of the respective operators in the triangle type diagrams exactly cancel. 
\section{Results \label{sec:results}}
In order to obtain the numerical results in this section we have used the following values of input parameters
\begin{align}
                  m_t   &=  172.5\;\text{GeV} \, ,       & \quad \quad \quad m_b &= 4.9 \;\text{GeV} \, ,  \nonumber \\
                m_Z &=  91.1876 \;\text{GeV}\, ,      &
                m_W &=  80.379 \;\text{GeV}\, ,\\      
                G_F    &= 1.16637\times 10^{-5}~\mbox{GeV}^{-2}\, , & \quad \quad \quad
                \nonumber 
                m_h &=  125.0\;\text{GeV}\,. 
\end{align}
The parton distribution functions (PDF) \texttt{PDF4LHC21\_40}~\cite{Cridge:2021qjj,PDF4LHCWorkingGroup:2022cjn} with $\alpha_s(m_{Z})= 0.118$ are utilised through the \textsc{LHAPDF} framework~\cite{Buckley:2014ana} and the central scale choice for renormalisation and factorisation scale is $\mu_R=\mu_F=m_{Vhh}$, the invariant mass of the $Vhh$ system.
The numerical values of the $A$ (HEFT) and $B$ (SMEFT) coefficients, along with their uncertainties, used to obtain our results can be found in the \href{https://github.com/Ryczek/VHH-NNLO}{GitHub repository} associated with this publication.

\subsection{HEFT}
\label{subsec:HEFT}

In this section, we present the results of the HEFT analysis for the $Vhh$ cross sections.
The cross sections are expressed in terms of the $A_i$ coefficients, as defined in the parametrisation of Sec.~\ref{subsec:HEFT:param}.
The central values are summarized in Tables~\ref{tab:A_wphh_combined},~\ref{tab:A_wmhh_combined} and~\ref{tab:zhh_coeff_combined}.
These tables enable the computation of predictions for any set of LO HEFT effective couplings: $\kappa_V$, $\kappa_\lambda$, $\kappa_{2V}$, and $\kappa_t$.
The central values of these coefficients are determined by solving a system of linear equations derived from Eqs.~\eqref{intheftw} and~\eqref{intheftz}, evaluated at a sufficient number of simulated phase space points.

Current experimental limits on the HEFT coefficients used in this work, aligned with the results of Refs.~\cite{CMS:2022dwd, ATLAS:2022vkf, CMS:2026nuu}, are collected in Table~\ref{tab:HEFT:coeff}.

\begin{table}[t]
\begin{center}
\resizebox{\textwidth}{!}{%
\begin{tabular}{|c|c|c|c|c|c|c|}
\hline\rule{0pt}{3ex}
Coefficient & $\kappa_{W}$ & $\kappa_{Z}$ & $\kappa_{2W}$ & $\kappa_{2Z}$ & $\kappa_\lambda$ & $\kappa_t$ \\[3pt]\hline\rule{0pt}{3ex}
Uncertainty Interval & $[0.85, 1.20]$ & $[0.90, 1.20]$ & $[0.70, 1.30]$ & $[0.70, 1.30]$ & $[-0.70, 6.10]$ & $[0.80, 1.20]$ \\[3pt]\hline
\end{tabular}}
\end{center}
\caption{Exclusion limits for the HEFT coefficients at the $95\%$ CL, assuming universal $\kappa_{2V}$ as constrained in Higgs pair production in vector boson fusion. 
}
\label{tab:HEFT:coeff}
\end{table}

\begin{figure}[htbp]
\centering
\includegraphics[width=0.45\textwidth]{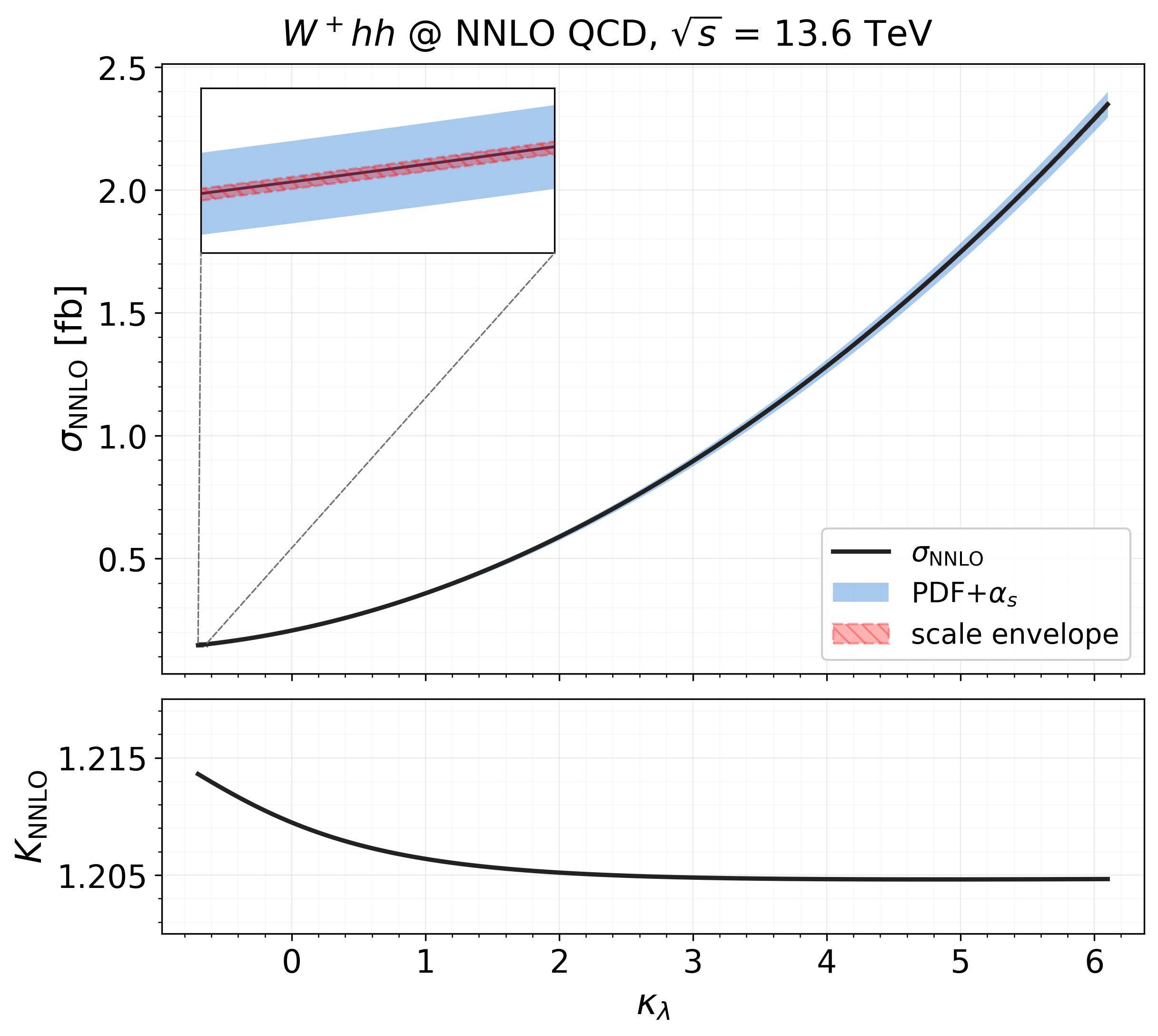}
\includegraphics[width=0.45\textwidth]{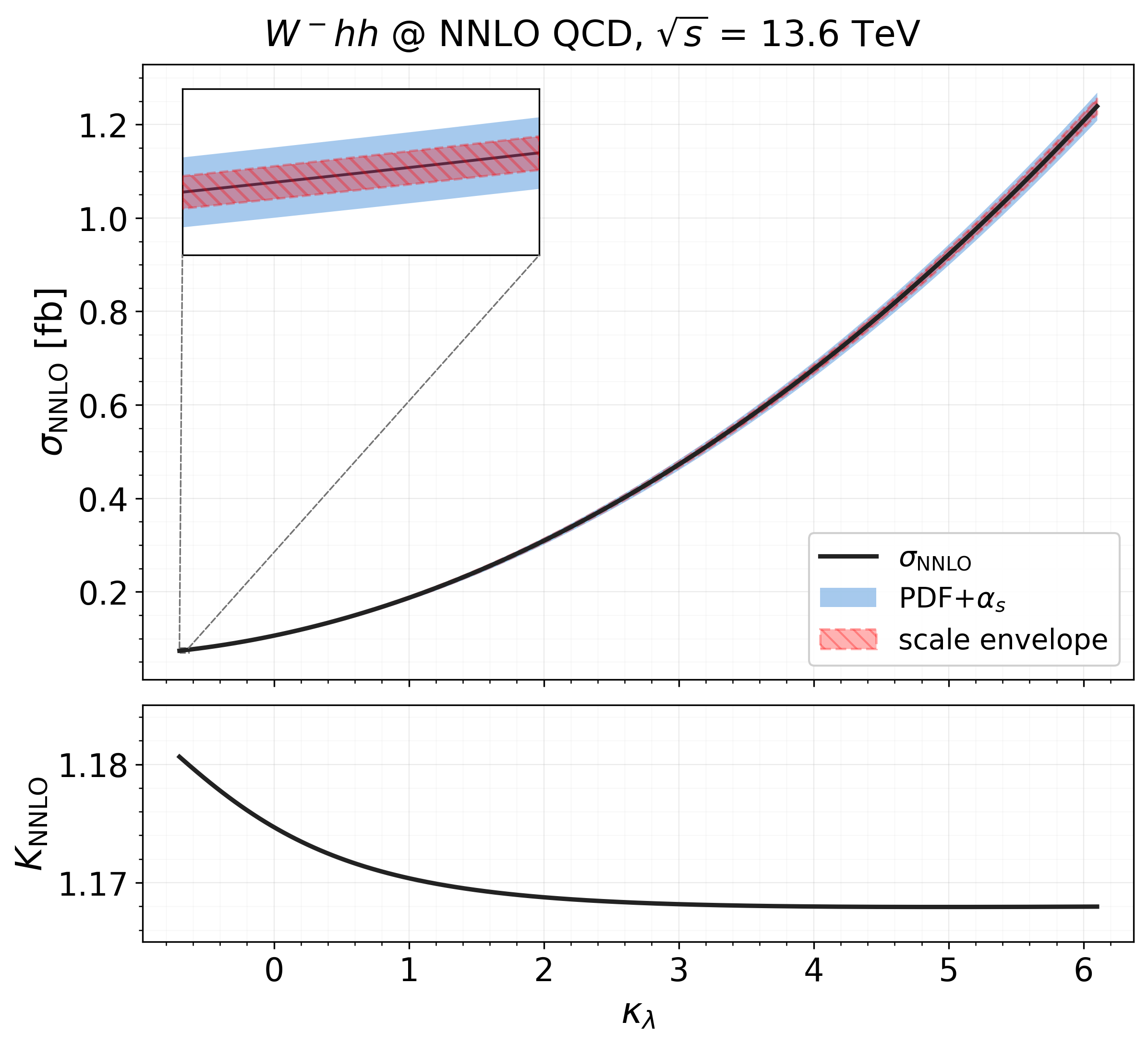}
\includegraphics[width=0.45\textwidth]{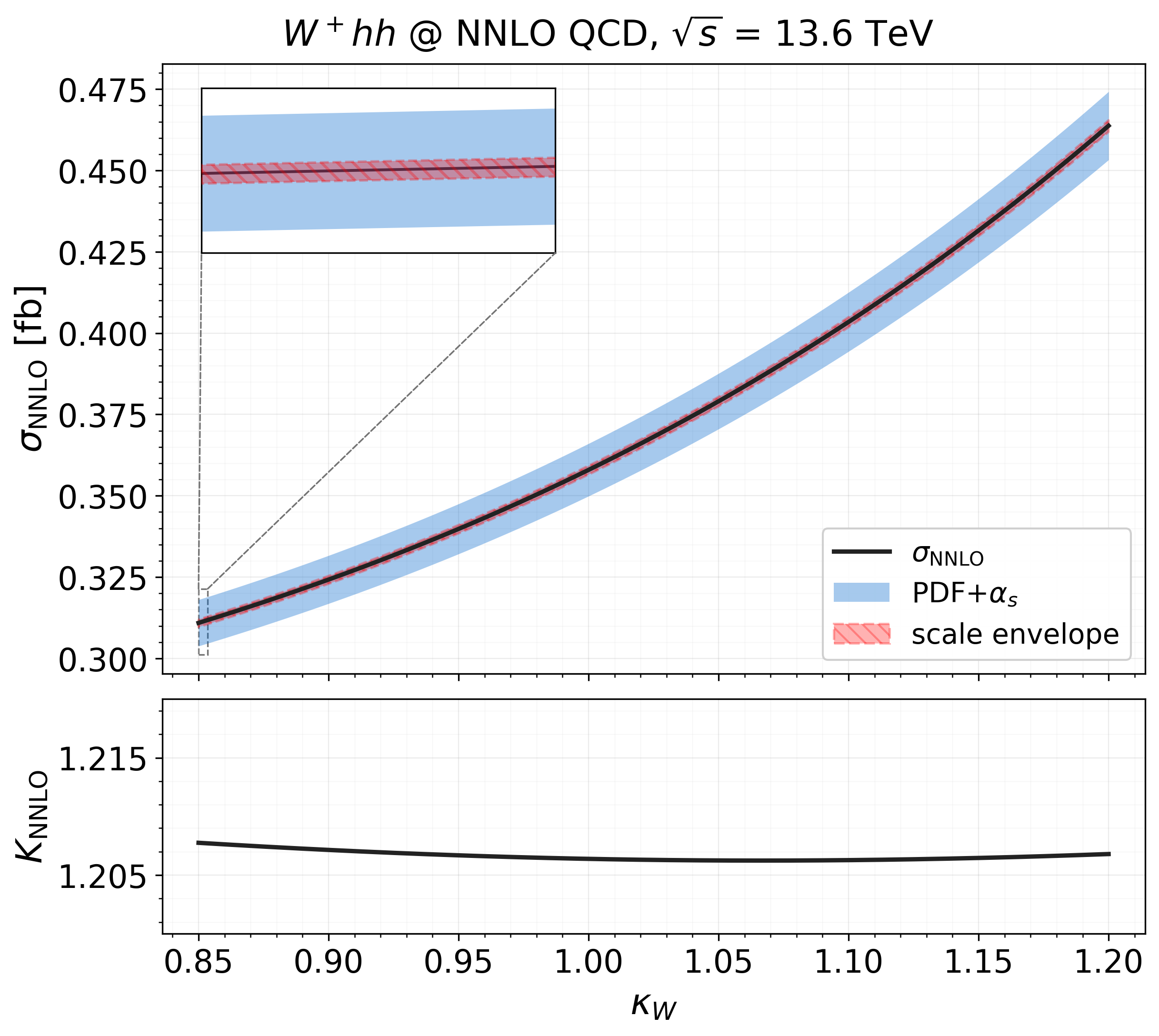}
\includegraphics[width=0.45\textwidth]{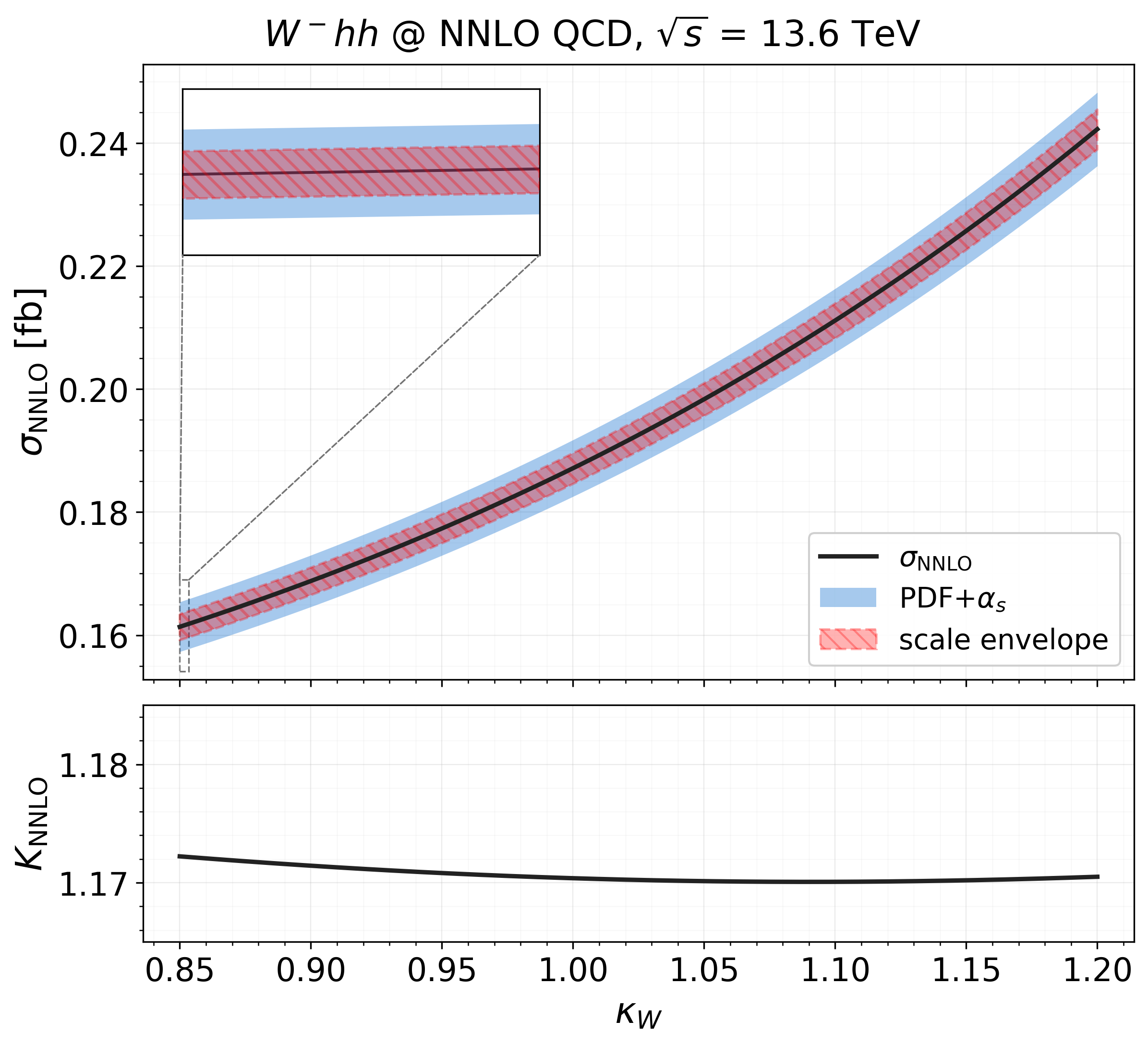}
\includegraphics[width=0.45\textwidth]{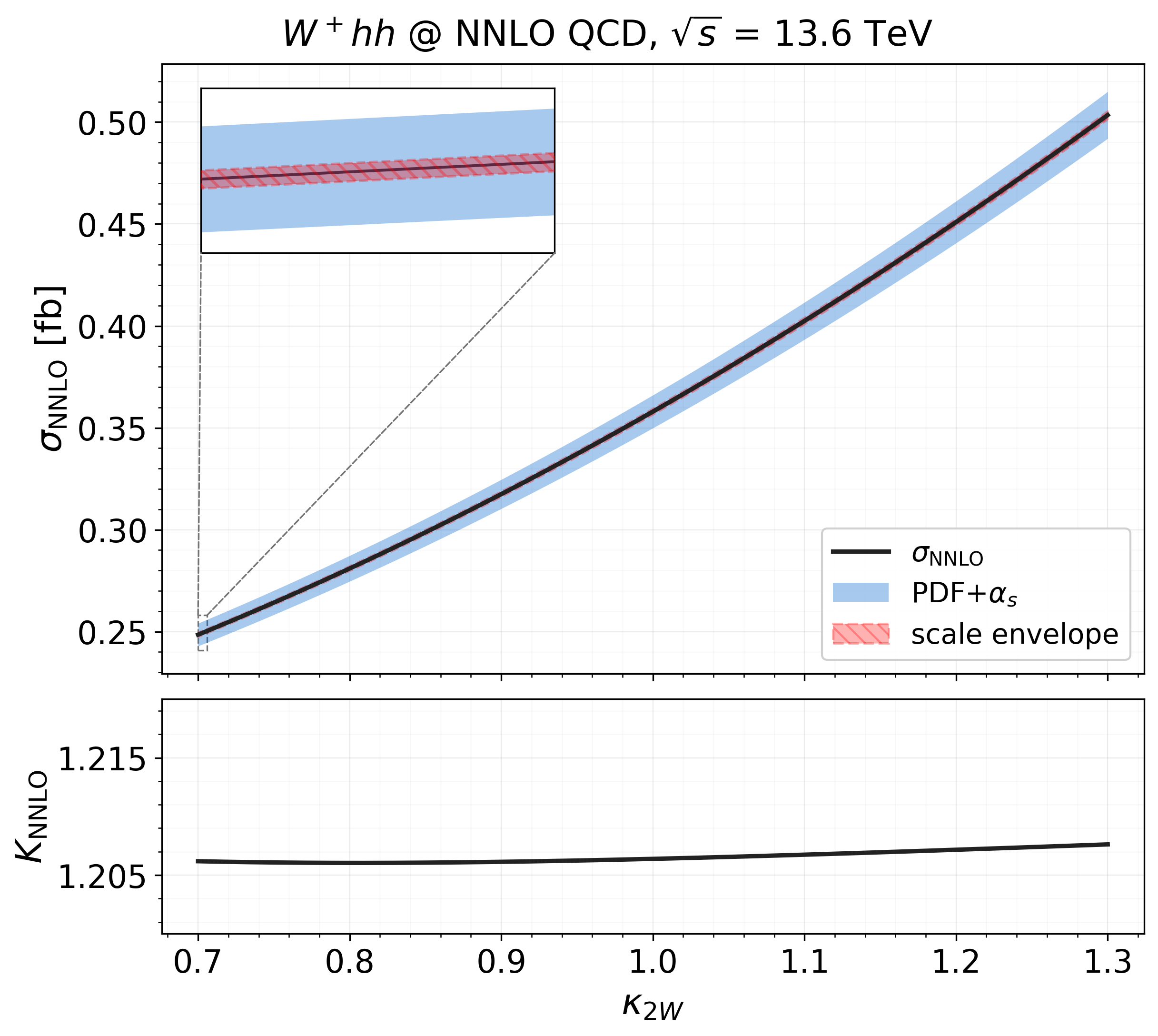}
\includegraphics[width=0.45\textwidth]{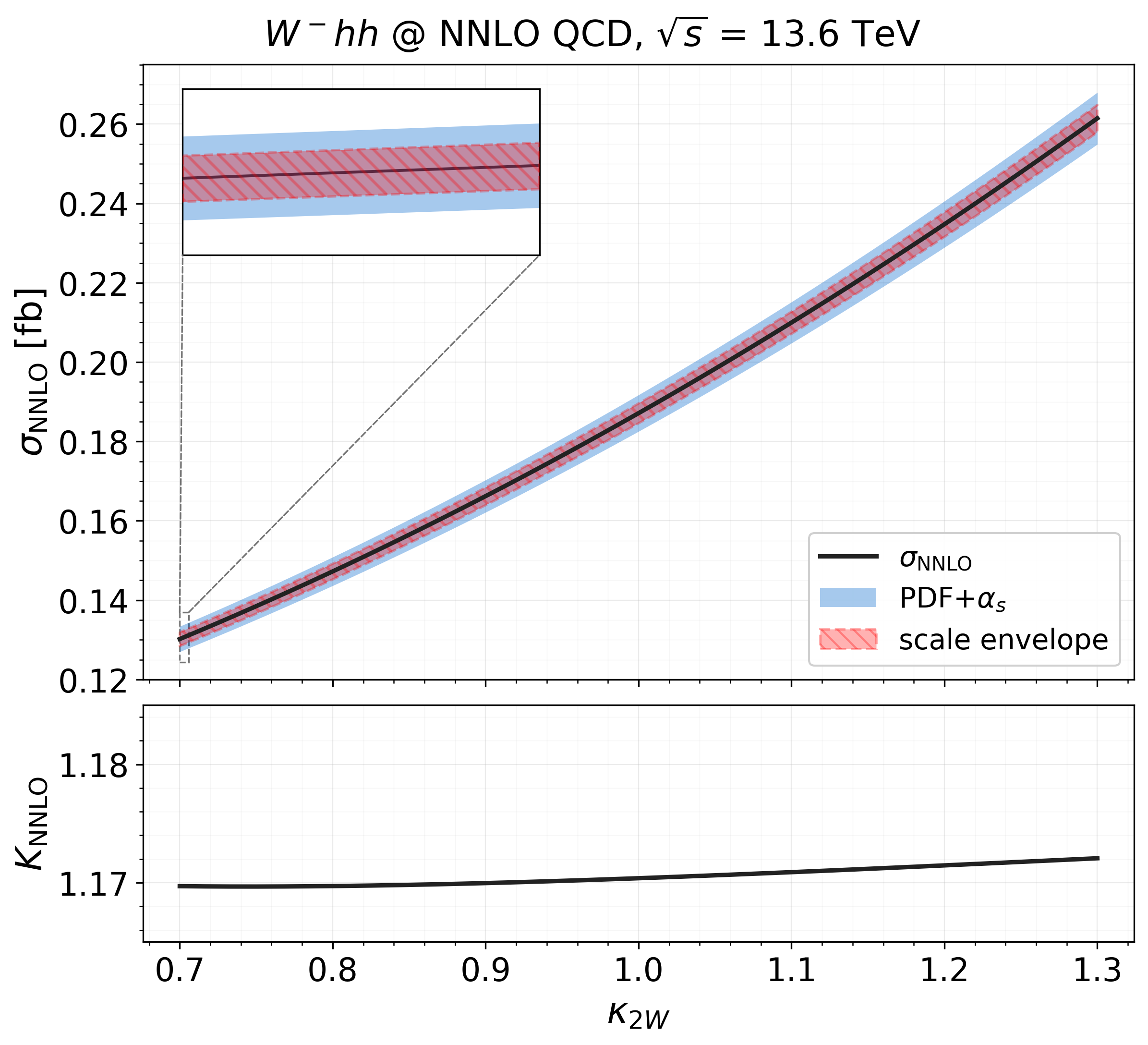}
\caption{$W^+hh$ (left) and $W^-hh$ (right) cross section at NNLO QCD and $K$-factor as a function of various $\kappa$ coefficients at $\sqrt{s}=13.6$~TeV. All other $\kappa$ coefficients are set to their SM values.}
\label{fig:req_scans_wplus_136}
\end{figure}

\begin{figure}[htbp]
\centering
\includegraphics[width=0.45\textwidth]{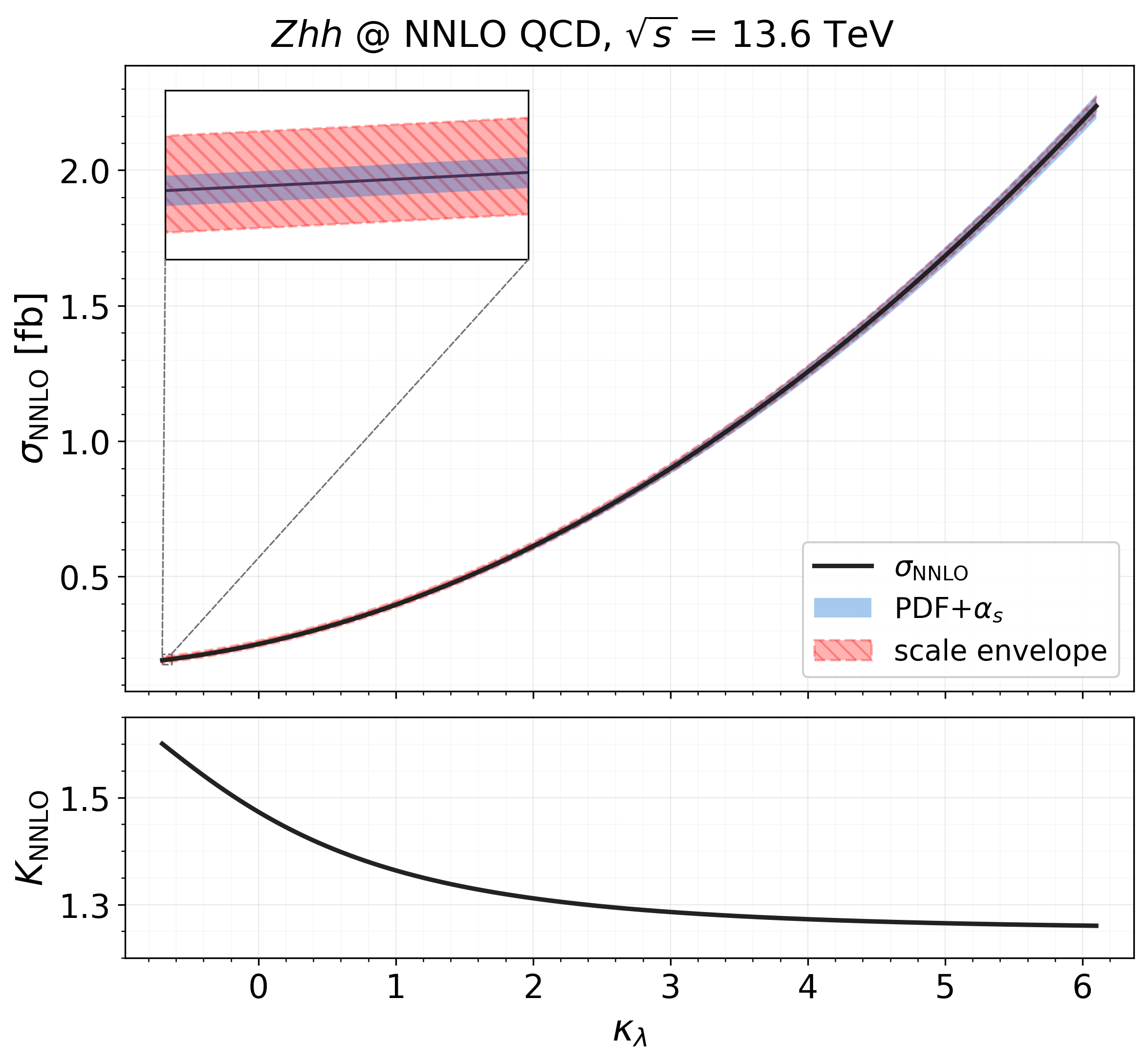}
\includegraphics[width=0.45\textwidth]{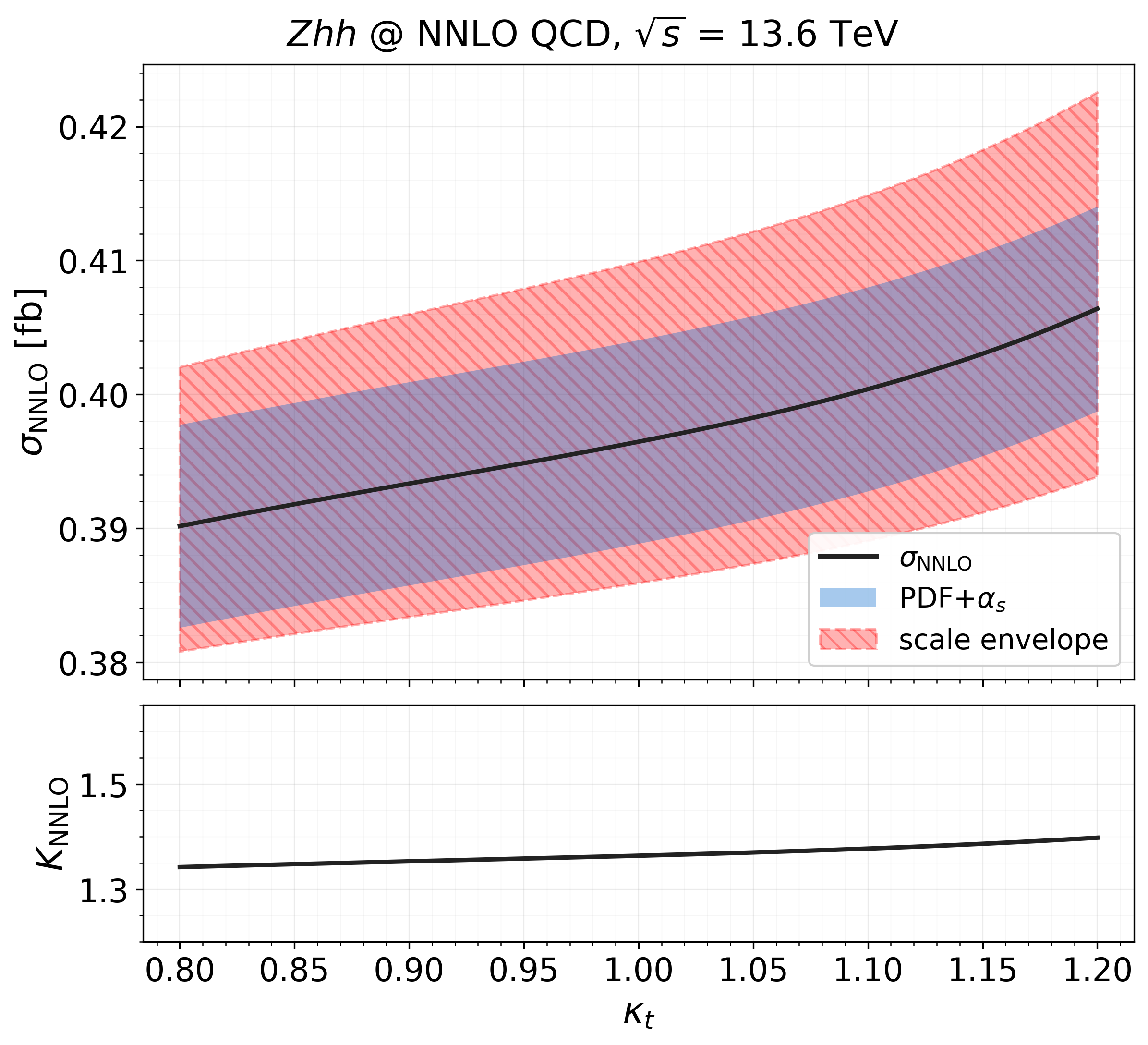}
\includegraphics[width=0.45\textwidth]{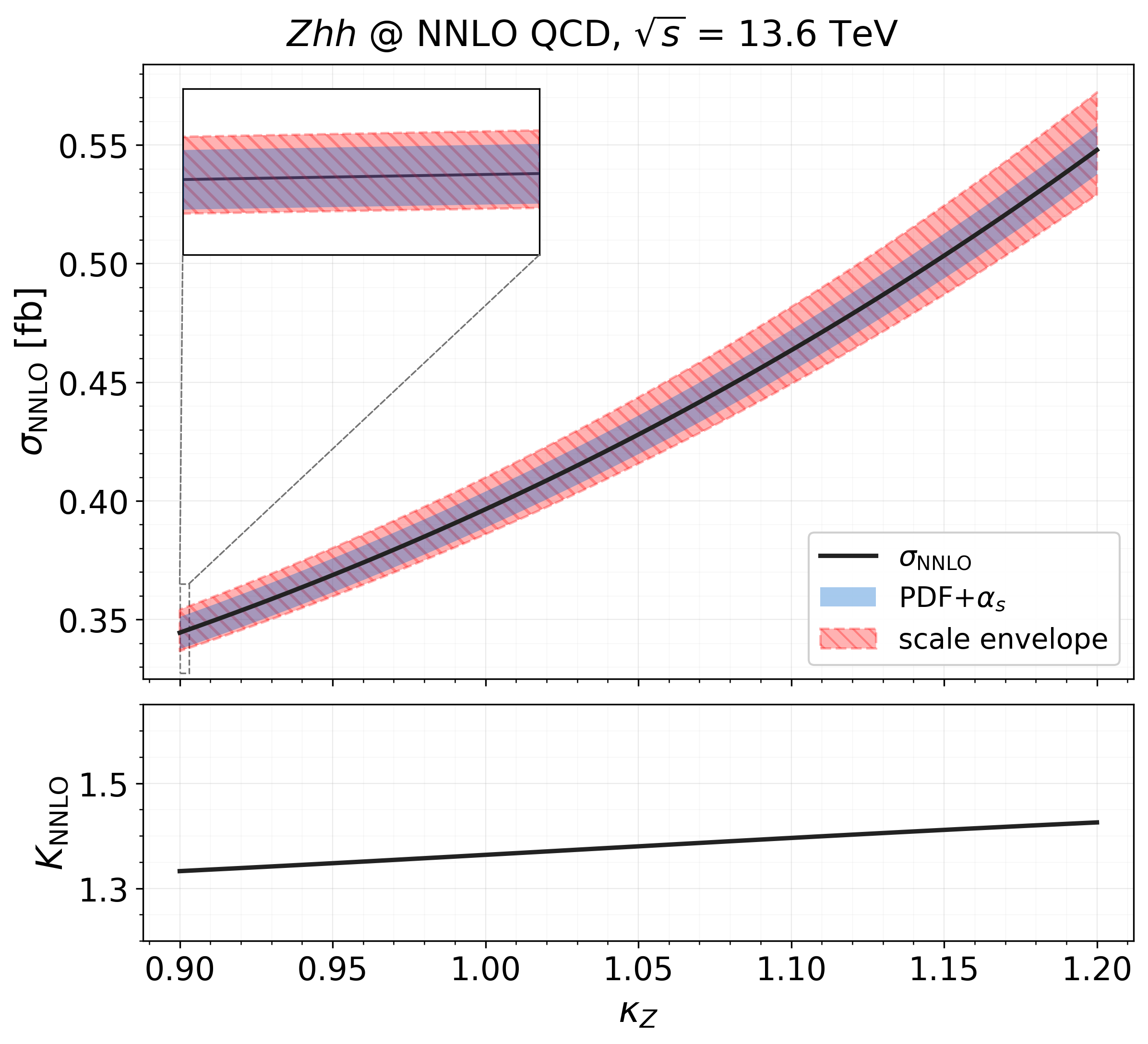}
\includegraphics[width=0.45\textwidth]{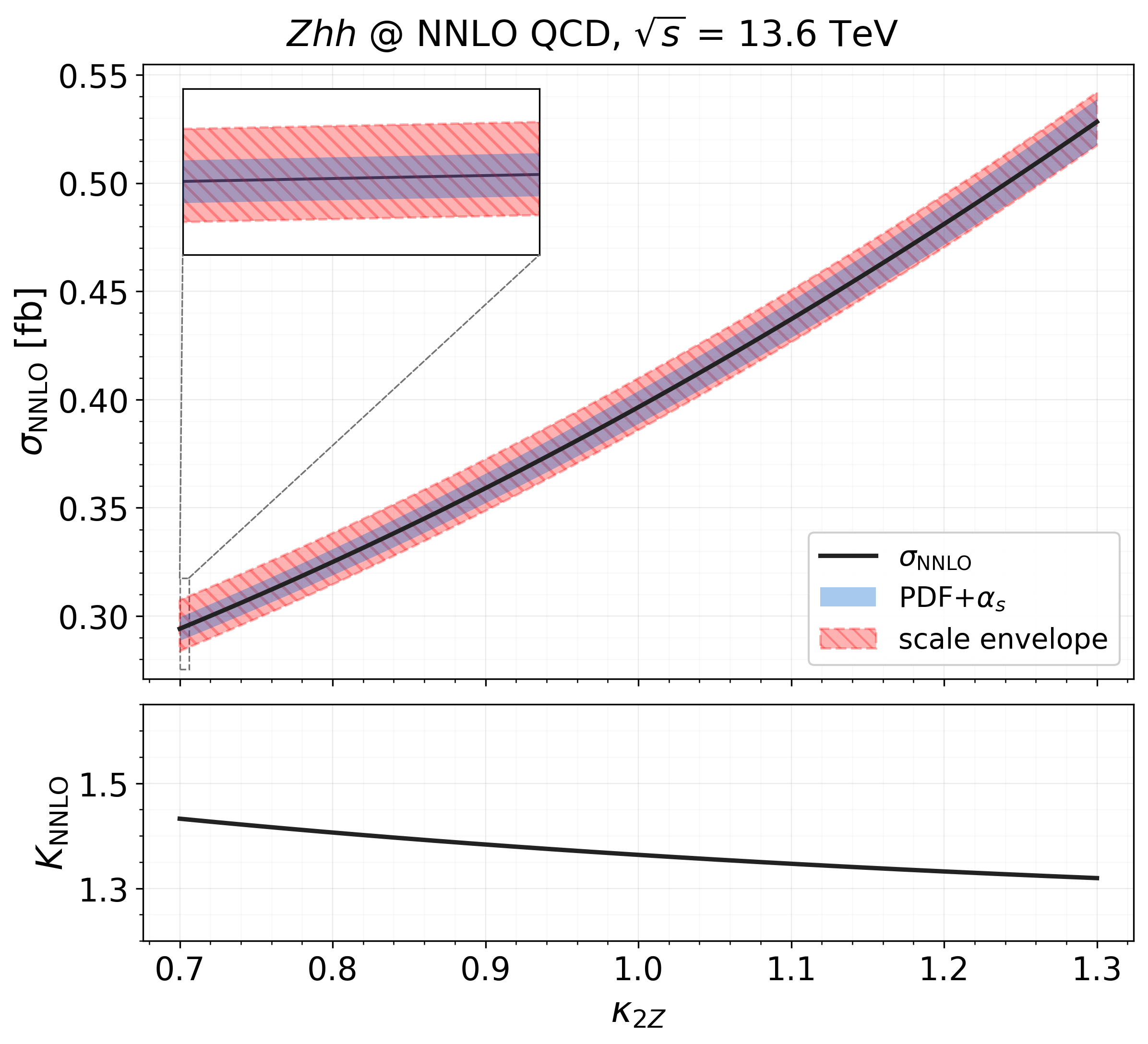}
\caption{$Zhh$ cross section at NNLO QCD and $K$-factor as a function of various $\kappa$ coefficients at $\sqrt{s}=13.6$~TeV. All other $\kappa$ coefficients are set to their SM values.}
\label{fig:req_scans_z_1}
\end{figure}

Example predictions, obtained by scanning the allowed parameter space with one non-SM $\kappa$ at a time, are shown in Figs.~\ref{fig:req_scans_wplus_136} and \ref{fig:req_scans_z_1}. Additionally, predictions for values of the $\kappa$'s taken close to their exclusion limits are given in Tables~\ref{tab:wplushh_HEFT},~\ref{tab:wminushh_HEFT},~\ref{tab:zhh_HEFT_kl_kt} and~\ref{tab:zhh_HEFT_kz_k2z}. The quoted uncertainties are the scale uncertainties (first) and the PDF+$\alpha_s$ uncertainty (second). All results are consistent with the findings of Refs.~\cite{Baglio:2012np,Grober:2026ece} regarding both the SM predictions and the variations of the trilinear Higgs self-coupling.

As one can note, the predicted theoretical uncertainties are below $4\%$ for $W^{\pm}hh$, while for $Zhh$ production they can be above $5\%$ due to the gluon-initiated contribution. Moreover, the maximal cross-section enhancement for the HL-LHC can reach $\mathcal{O}(6)$, for $\kappa_\lambda = 6.1$ in $Zhh$, $W^+hh$, and $W^-hh$.
Furthermore, we can also observe that the maximal impact of NNLO QCD corrections, encoded in the $K$-factor
\begin{equation}
K_{\text{NNLO}}=\sigma_{\rm NNLO}/\sigma_{\rm LO}\, ,
\end{equation}
 is of $1.6$ for $Zhh$ and $1.2$ for $W^\pm hh$. 
 
 The dependence of the $K$-factor on the HEFT coefficients is rather flat for $W^\pm hh$, as the NNLO QCD corrections largely factorise from the HEFT LO cross section. Additionally, the theory uncertainty remains approximately constant across different values of the HEFT coefficients. We also note the lower uncertainty for $W^+ hh$ compared to $W^- hh$. Since the two processes receive different relative contributions from the $q\bar{q}$ and $qg$ initial states due to the different PDFs, which nearly cancel~\cite{Grober:2026ece}, differences in the theory uncertainty do not come as a surprise. 

 For the $Zhh$ process, the situation is slightly different due to the gluon-induced contributions appearing at NNLO QCD.
As can be seen in Fig.~\ref{fig:req_scans_z_1}, the $K$-factor shows the strongest dependence on $\kappa_{\lambda}$, varying between $1.25$ and $1.6$, given the still rather large range of allowed $\kappa_{\lambda}$ values.

While HL-LHC projections for these processes are not yet available, the current experimental constraints of $183 \times \sigma_\mathrm{SM}$~\cite{ATLAS:2022fpx} and $294 \times \sigma_\mathrm{SM}$~\cite{CMS:2024fkb} are an order of magnitude larger than the obtained maximal theoretical enhancements. This ensures that the precision of our predictions is sufficient for the HL-LHC upgrade.

\begin{table}[htbp]
\centering
  \renewcommand{\arraystretch}{1.2}%
\begin{tabular}{c c c c c}
\hline
 & \multicolumn{2}{c}{$\sqrt{s}=13.6$~TeV [fb]} & \multicolumn{2}{c}{$\sqrt{s}=14.0$~TeV [fb]} \\
\cline{2-5}
 & $A^{W^+}_{i,\mathrm{LO}}$ & $A^{W^+}_{i,\mathrm{NNLO}}$ & $A^{W^+}_{i,\mathrm{LO}}$ & $A^{W^+}_{i,\mathrm{NNLO}}$ \\
\hline
$A^{W^+}_{1}$ & $0.1053$ & $0.1297$ & $0.1115$ & $0.1372$ \\
$A^{W^+}_{2}$ & $-0.0288$ & $-0.0357$ & $-0.0306$ & $-0.0379$ \\
$A^{W^+}_{3}$ & $-0.0984$ & $-0.1235$ & $-0.1053$ & $-0.1319$ \\
$A^{W^+}_{4}$ & $0.0324$ & $0.0391$ & $0.0340$ & $0.0409$ \\
$A^{W^+}_{5}$ & $0.1225$ & $0.1482$ & $0.1284$ & $0.1552$ \\
$A^{W^+}_{6}$ & $0.1637$ & $0.2001$ & $0.1726$ & $0.2107$ \\
\hline
\end{tabular}
\caption{Central values of $A^{W^+}_i$ coefficients for $W^+hh$ at $\sqrt{s}=13.6$ and $14.0$~TeV.}
\label{tab:A_wphh_combined}
\end{table}

\begin{table}[htbp]
\centering
  \renewcommand{\arraystretch}{1.2}%
\begin{tabular}{c c c c c}
\hline
 & \multicolumn{2}{c}{$\sqrt{s}=13.6$~TeV [fb]} & \multicolumn{2}{c}{$\sqrt{s}=14.0$~TeV [fb]} \\
\cline{2-5}
 & $A^{W^-}_{i,\mathrm{LO}}$ & $A^{W^-}_{i,\mathrm{NNLO}}$ & $A^{W^-}_{i,\mathrm{LO}}$ & $A^{W^-}_{i,\mathrm{NNLO}}$ \\
\hline
$A^{W^-}_{1}$ & $0.0497$ & $0.0598$ & $0.0530$ & $0.0637$ \\
$A^{W^-}_{2}$ & $-0.0125$ & $-0.0154$ & $-0.0134$ & $-0.0165$ \\
$A^{W^-}_{3}$ & $-0.0404$ & $-0.0503$ & $-0.0437$ & $-0.0542$ \\
$A^{W^-}_{4}$ & $0.0176$ & $0.0205$ & $0.0185$ & $0.0216$ \\
$A^{W^-}_{5}$ & $0.0645$ & $0.0759$ & $0.0681$ & $0.0800$ \\
$A^{W^-}_{6}$ & $0.0811$ & $0.0966$ & $0.0861$ & $0.1024$ \\
\hline
\end{tabular}
\caption{Central values of $A^{W^-}_i$ coefficients for $W^-hh$ at $\sqrt{s}=13.6$ and $14.0$~TeV.}
\label{tab:A_wmhh_combined}
\end{table}

\begin{table}[htbp]
\centering
  \renewcommand{\arraystretch}{1.2}%
\begin{tabular}{c c c c c}
\hline
 & \multicolumn{2}{c}{$\sqrt{s}=13.6$~TeV [fb]} & \multicolumn{2}{c}{$\sqrt{s}=14.0$~TeV [fb]} \\
\cline{2-5}
 & $A_{i,\mathrm{LO}}^Z$ & $A_{i,\mathrm{NNLO}}^Z$ & $A_{i,\mathrm{LO}}^Z$ & $A_{i,\mathrm{NNLO}}^Z$ \\
\hline
$A^Z_{1}$ & $0.0844$ & $0.1132$ & $0.0896$ & $0.1203$ \\
$A^Z_{2}$ & $-0.0119$ & $0.0085$ & $-0.0128$ & $0.0092$ \\
$A^Z_{3}$ & $-0.0487$ & $-0.0637$ & $-0.0529$ & $-0.0690$ \\
$A^Z_{4}$ & $0.0280$ & $0.0547$ & $0.0294$ & $0.0581$ \\
$A^Z_{5}$ & $0.1041$ & $0.1236$ & $0.1096$ & $0.1299$ \\
$A^Z_{6}$ & $0.1348$ & $0.1639$ & $0.1427$ & $0.1732$ \\
$A^Z_{7}$ & --- & $0.0011$ & --- & $0.0011$ \\
$A^Z_{8}$ & --- & $0.0051$ & --- & $0.0055$ \\
$A^Z_{9}$ & --- & $-0.0037$ & --- & $-0.0041$ \\
$A^Z_{10}$ & --- & $-0.0798$ & --- & $-0.0860$ \\
$A^Z_{11}$ & --- & $0.4028$ & --- & $0.4349$ \\
$A^Z_{12}$ & --- & $-0.1107$ & --- & $-0.1194$ \\
$A^Z_{13}$ & --- & $-0.0324$ & --- & $-0.0351$ \\
$A^Z_{14}$ & --- & $0.1531$ & --- & $0.1654$ \\
$A^Z_{15}$ & --- & $0.0130$ & --- & $0.0141$ \\
$A^Z_{16}$ & --- & $-0.0656$ & --- & $-0.0710$ \\
$A^Z_{17}$ & --- & $-0.4633$ & --- & $-0.5014$ \\
$A^Z_{18}$ & --- & $0.1767$ & --- & $0.1916$ \\
\hline
\end{tabular}
\caption{Central values of $A^Z_i$ coefficients for $Zhh$ at $\sqrt{s}=13.6$ and $14.0$~TeV.}
\label{tab:zhh_coeff_combined}
\end{table}

\begin{table}[htbp!]
\centering
\resizebox{\textwidth}{!}{%
  \renewcommand{\arraystretch}{1.45}%
  \setlength{\tabcolsep}{7pt}%
  \begin{tabular}{c @{\hspace{10pt}}c@{\hspace{10pt}}c@{\hspace{10pt}}c@{\hspace{10pt}}c@{\hspace{10pt}}c@{\hspace{10pt}}c@{\hspace{10pt}}c}
    \hline
    \multirow{2}{*}{$\sqrt{s}$} & \multicolumn{7}{c}{$\sigma_{NNLO}$} \\
    \cline{2-8}
     & SM & $\kappa_\lambda=-0.7$ & $\kappa_\lambda=6.1$ & $\kappa_W=0.85$ & $\kappa_W=1.2$ & $\kappa_{2W}=0.7$ & $\kappa_{2W}=1.3$ \\
    \hline
    $13.6$ & $0.358^{+0.36\%}_{-0.40\%} \pm 2.3\%$ & $0.147^{+0.30\%}_{-0.42\%} \pm 2.5\%$ & $2.348^{+0.40\%}_{-0.41\%} \pm 2.2\%$ & $0.311^{+0.34\%}_{-0.40\%} \pm 2.3\%$ & $0.464^{+0.36\%}_{-0.40\%} \pm 2.3\%$ & $0.248^{+0.37\%}_{-0.40\%} \pm 2.3\%$ & $0.503^{+0.34\%}_{-0.40\%} \pm 2.3\%$ \\
    $14.0$ & 
    $0.374^{+0.37\%}_{-0.40\%} \pm 2.2\%$ & 
    $0.154^{+0.30\%}_{-0.41\%} \pm 2.4\%$ & 
    $2.453^{+0.41\%}_{-0.40\%} \pm 2.2\%$ & 
    $0.325^{+0.35\%}_{-0.40\%} \pm 2.3\%$ & 
    $0.485^{+0.37\%}_{-0.40\%} \pm 2.2\%$ & 
    $0.260^{+0.38\%}_{-0.40\%} \pm 2.2\%$ & 
    $0.527^{+0.35\%}_{-0.40\%} \pm 2.3\%$ \\
    \hline
  \end{tabular}%
}
\caption{Cross-section values $\sigma$ [fb] at NNLO QCD in HEFT for $W^+hh$ at $\sqrt{s} = 13.6$, $14.0$ TeV, for the SM and various $\kappa$ values at their exclusion boundaries. All other $\kappa$ parameters are set to their SM values.}
\label{tab:wplushh_HEFT}
\end{table}

\begin{table}[htbp]
\centering
\resizebox{\textwidth}{!}{%
  \renewcommand{\arraystretch}{1.45}%
  \setlength{\tabcolsep}{7pt}%
  \begin{tabular}{c@{\hspace{10pt}}c@{\hspace{10pt}}c@{\hspace{10pt}}c@{\hspace{10pt}}c@{\hspace{10pt}}c@{\hspace{10pt}}c@{\hspace{10pt}}c}
    \hline
    \multirow{2}{*}{$\sqrt{s}$} & \multicolumn{7}{c}{$\sigma_{NNLO}$} \\
    \cline{2-8}
     & SM & $\kappa_\lambda=-0.7$ & $\kappa_\lambda=6.1$ & $\kappa_W=0.85$ & $\kappa_W=1.2$ & $\kappa_{2W}=0.7$ & $\kappa_{2W}=1.3$ \\
    \hline
    $13.6$ & 
    $0.187^{+1.31\%}_{-1.36\%} \pm 2.5\%$ & 
    $0.074^{+1.24\%}_{-1.29\%} \pm 2.7\%$ & 
    $1.239^{+1.31\%}_{-1.37\%} \pm 2.4\%$ & 
    $0.161^{+1.29\%}_{-1.35\%} \pm 2.5\%$ & 
    $0.242^{+1.31\%}_{-1.37\%} \pm 2.5\%$ & 
    $0.130^{+1.32\%}_{-1.38\%} \pm 2.5\%$ & 
    $0.261^{+1.29\%}_{-1.35\%} \pm 2.5\%$ \\
    $14.0$ & 
    $0.197^{+1.33\%}_{-1.38\%} \pm 2.4\%$ & 
    $0.078^{+1.26\%}_{-1.30\%} \pm 2.7\%$ & 
    $1.303^{+1.33\%}_{-1.39\%} \pm 2.4\%$ & 
    $0.170^{+1.31\%}_{-1.36\%} \pm 2.5\%$ & 
    $0.255^{+1.33\%}_{-1.39\%} \pm 2.4\%$ & 
    $0.137^{+1.34\%}_{-1.39\%} \pm 2.4\%$ & 
    $0.275^{+1.31\%}_{-1.37\%} \pm 2.5\%$ \\
    \hline
  \end{tabular}%
}
\caption{Cross-section values $\sigma$ [fb] at NNLO QCD in HEFT for $W^-hh$ at $\sqrt{s} = 13.6$, $14.0$ TeV, for the SM and various $\kappa$ values at their exclusion boundaries. All other $\kappa$ parameters are set to their SM values.}
\label{tab:wminushh_HEFT}
\end{table}

\begin{table}[htbp]
\centering
\resizebox{\textwidth}{!}{%
  \renewcommand{\arraystretch}{1.45}%
  \setlength{\tabcolsep}{7pt}%
  \begin{tabular}{c@{\hspace{10pt}}c@{\hspace{10pt}}c@{\hspace{10pt}}c@{\hspace{10pt}}c@{\hspace{10pt}}c}
    \hline
    \multirow{2}{*}{$\sqrt{s}$} & \multicolumn{5}{c}{$\sigma_{NNLO}$} \\
    \cline{2-6}
     & SM & $\kappa_\lambda=-0.7$ & $\kappa_\lambda=6.1$ & $\kappa_t=0.8$ & $\kappa_t=1.2$ \\
    \hline
    $13.6$ & 
    $0.396^{+3.39\%}_{-2.66\%} \pm 1.9\%$ & 
    $0.191^{+6.70\%}_{-5.15\%} \pm 1.9\%$ &
    $2.24^{+1.50\%}_{-1.26\%} \pm 2.0\%$ & 
    $0.390^{+3.04\%}_{-2.40\%} \pm 1.9\%$ &
    $0.406^{+3.97\%}_{-3.09\%} \pm 1.9\%$ \\
    $14.0$ & 
    $0.417^{+3.42\%}_{-2.69\%} \pm 1.9\%$ &
    $0.203^{+6.74\%}_{-5.18\%} \pm 1.8\%$ & 
    $2.35^{+1.51\%}_{-1.27\%} \pm 2.0\%$ & 
    $0.411^{+3.08\%}_{-2.43\%} \pm 1.9\%$ & 
    $0.428^{+4.02\%}_{-3.12\%} \pm 1.9\%$ \\
    \hline
  \end{tabular}%
}
\caption{Cross-section values $\sigma$ [fb] at NNLO QCD in HEFT for $Zhh$ at $\sqrt{s} = 13.6$, $14.0$ TeV, for the SM and $\kappa_\lambda$, $\kappa_{t}$ at their exclusion boundaries. All other $\kappa$ parameters are set to their SM values.}
\label{tab:zhh_HEFT_kl_kt}
\end{table}

\begin{table}[htbp]
\centering
\resizebox{\textwidth}{!}{%
  \renewcommand{\arraystretch}{1.45}%
  \setlength{\tabcolsep}{7pt}%
  \begin{tabular}{c@{\hspace{10pt}}c@{\hspace{10pt}}c@{\hspace{10pt}}c@{\hspace{10pt}}c@{\hspace{10pt}}c}
    \hline
    \multirow{2}{*}{$\sqrt{s}$} & \multicolumn{5}{c}{$\sigma_{NNLO}$} \\
    \cline{2-6}
     & SM & $\kappa_Z=0.9$ & $\kappa_Z=1.2$ & $\kappa_{2Z}=0.7$ & $\kappa_{2Z}=1.3$ \\
    \hline
    $13.6$ & 
    $0.396^{+3.39\%}_{-2.66\%} \pm 1.9\%$ & 
    $0.344^{+2.83\%}_{-2.25\%} \pm 2.0\%$ &
    $0.548^{+4.44\%}_{-3.44\%} \pm 1.9\%$ & 
    $0.294^{+4.53\%}_{-3.50\%} \pm 1.8\%$ & 
    $0.528^{+2.58\%}_{-2.07\%} \pm 2.0\%$ \\
    $14.0$ & 
    $0.417^{+3.42\%}_{-2.69\%} \pm 1.9\%$ & 
    $0.362^{+2.86\%}_{-2.27\%} \pm 1.9\%$ & 
    $0.577^{+4.49\%}_{-3.47\%} \pm 1.8\%$ & 
    $0.310^{+4.57\%}_{-3.54\%} \pm 1.8\%$ & 
    $0.556^{+2.60\%}_{-2.09\%} \pm 2.0\%$ \\
    \hline
  \end{tabular}%
}
\caption{Cross-section values $\sigma$ [fb] at NNLO QCD in HEFT for $Zhh$ at $\sqrt{s} = 13.6$, $14.0$ TeV, for the SM and $\kappa_Z$, $\kappa_{2Z}$ at their exclusion boundaries. All other $\kappa$ coefficients are set to their SM values.}
\label{tab:zhh_HEFT_kz_k2z}
\end{table}

\subsection{SMEFT}

\begin{figure}[htbp]
\centering
\includegraphics[width=0.45\textwidth]{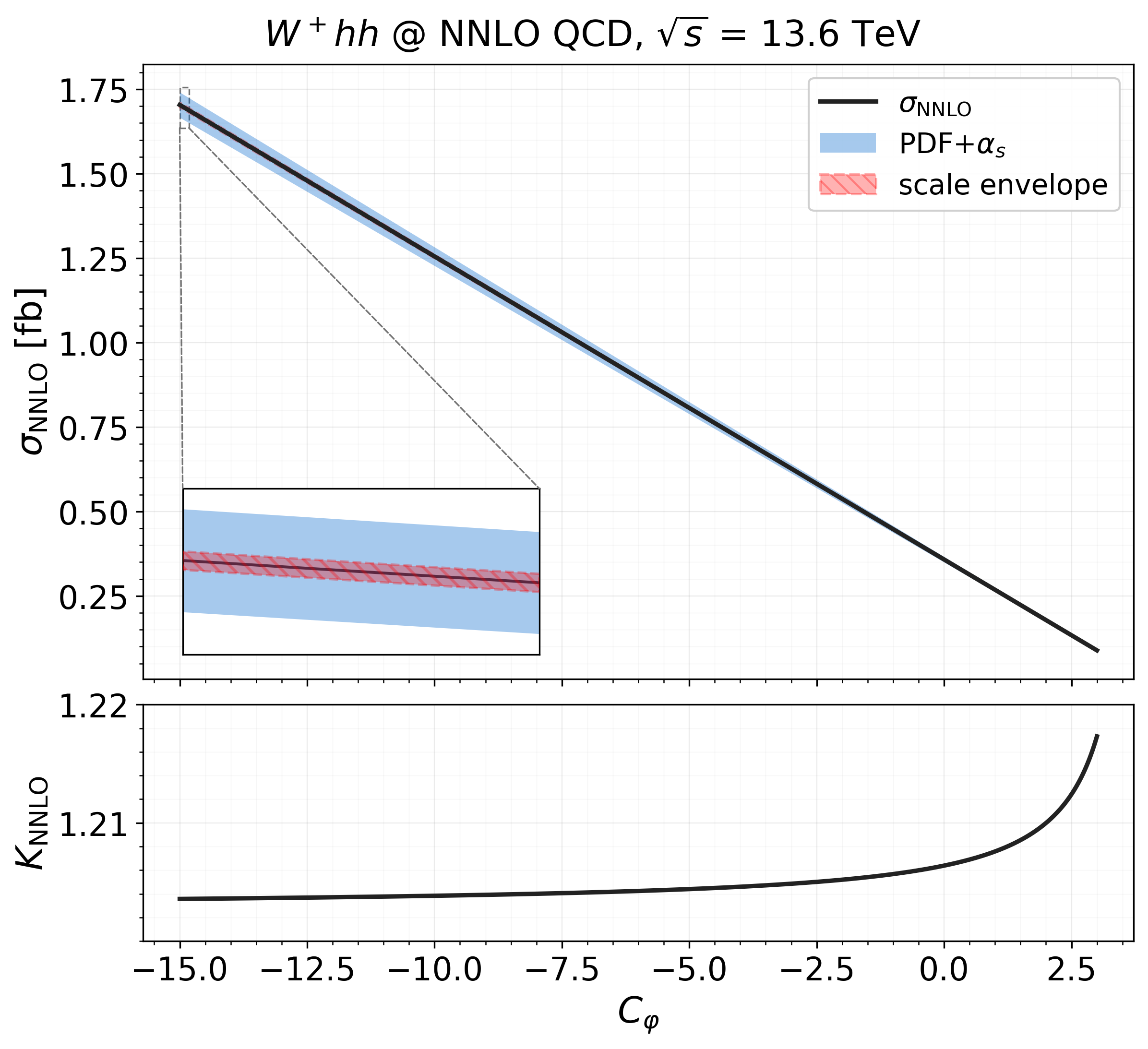}
\includegraphics[width=0.45\textwidth]{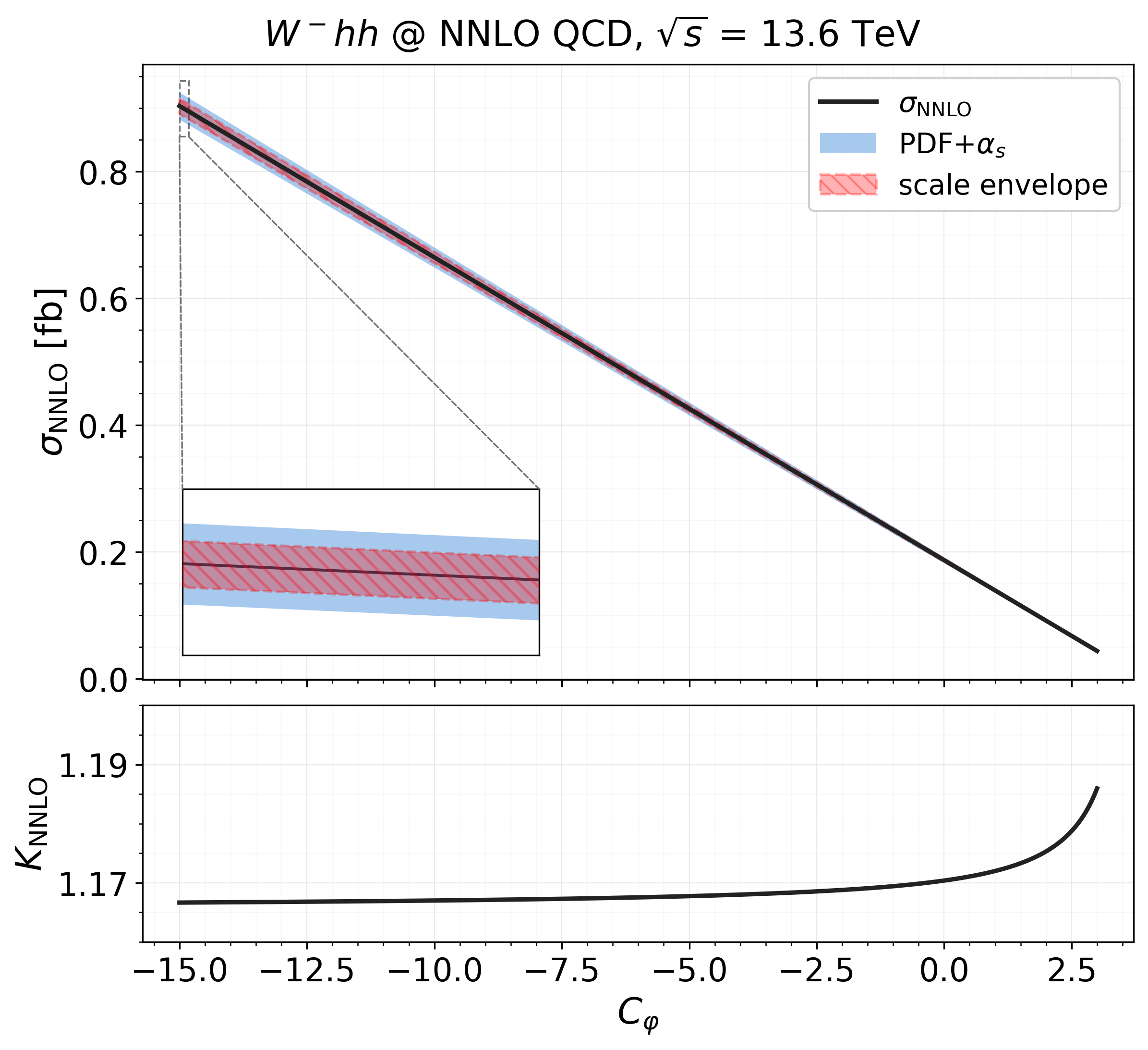}
\includegraphics[width=0.45\textwidth]{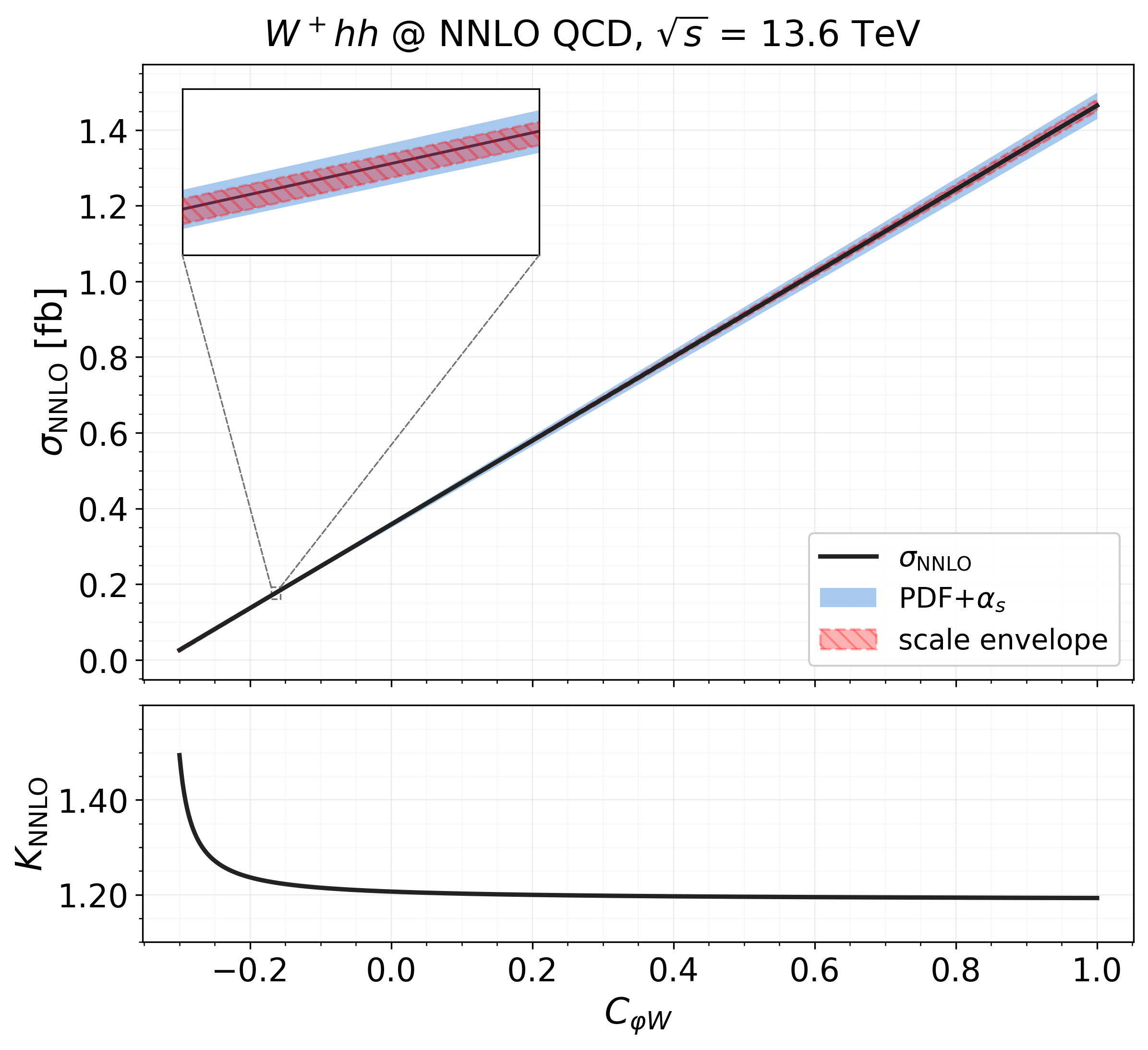}
\includegraphics[width=0.45\textwidth]{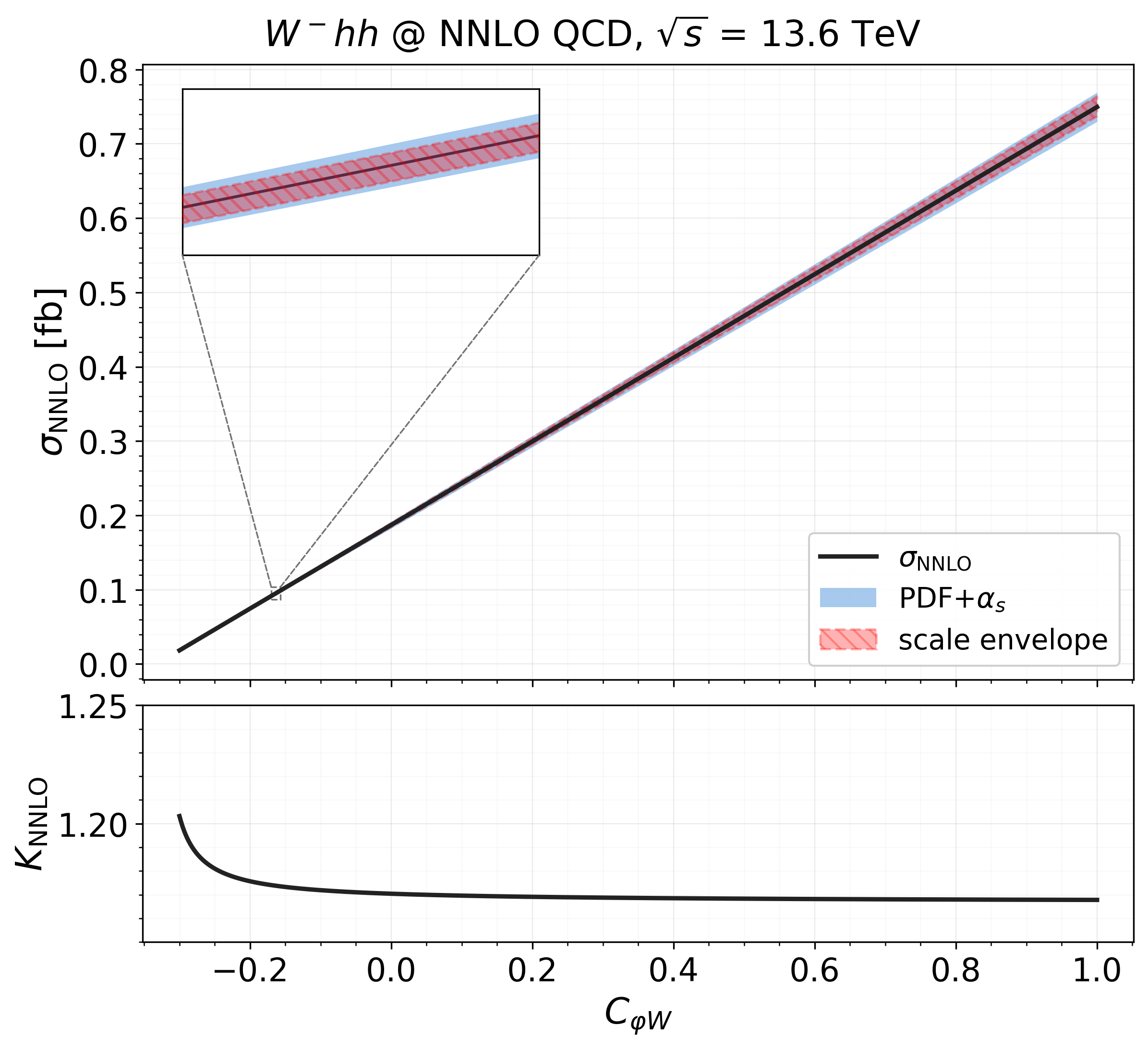}
\caption{Cross section and $K$-factor of  $W^+ hh$ (left) and $W^- hh$ (right) as function of $C_{\varphi}$ (top) and $C_{\varphi W}$ (bottom) at $\sqrt{s}=13.6$~TeV.}
\label{fig:smeft_phi_phiW:WHH}
\end{figure}

\begin{figure}[htbp]
\centering
\includegraphics[width=0.45\textwidth]{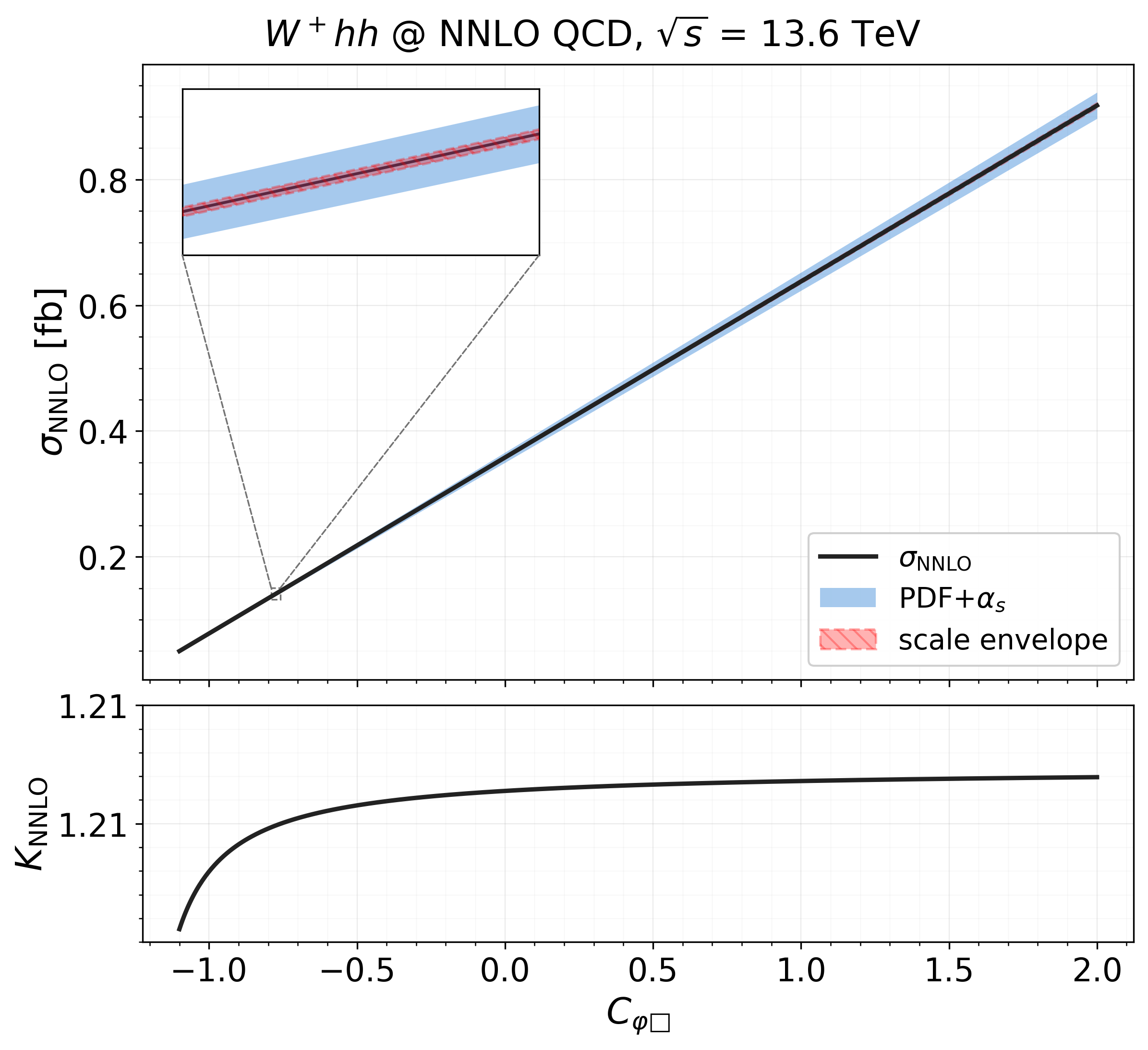}
\includegraphics[width=0.45\textwidth]{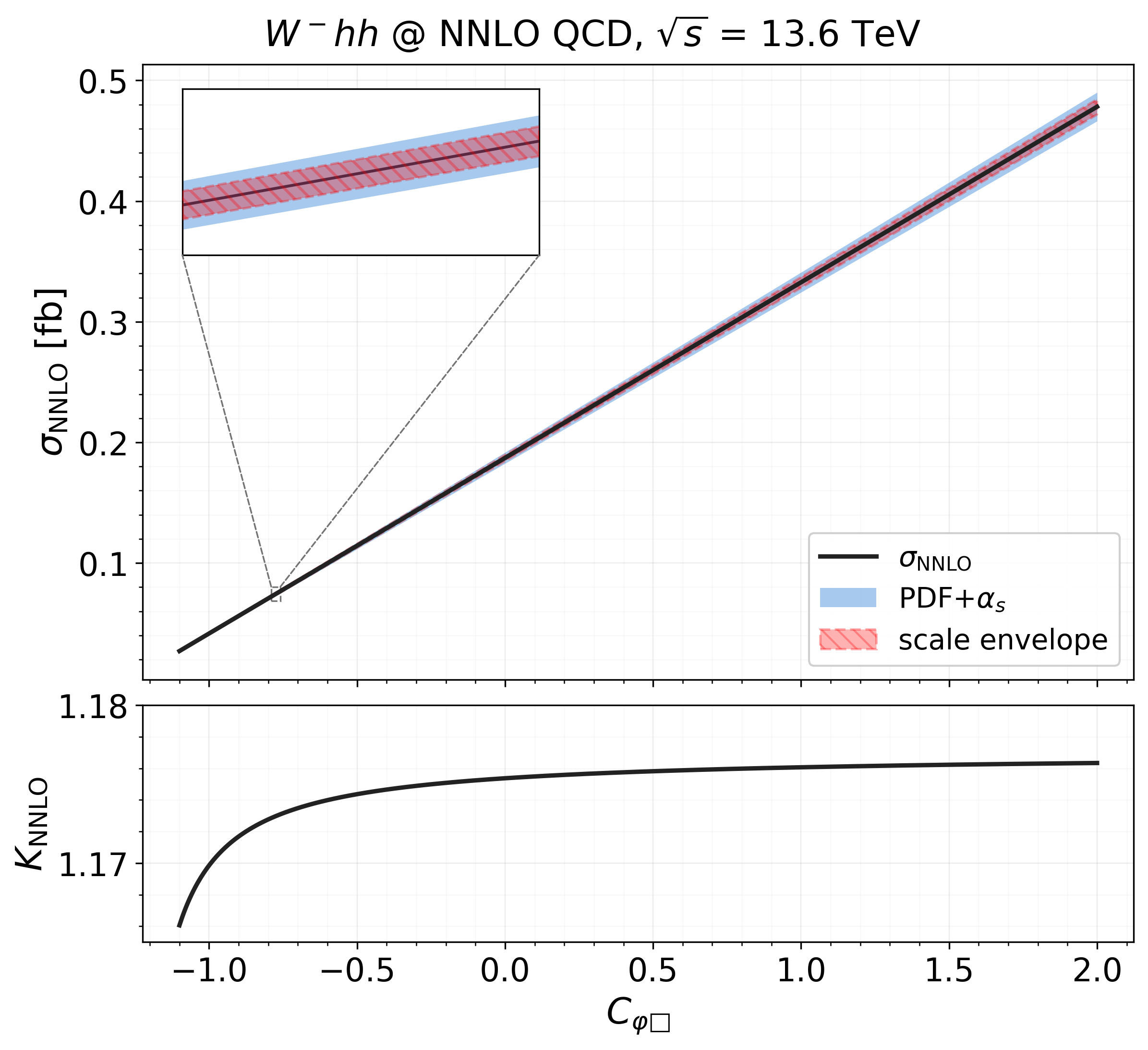}
\includegraphics[width=0.45\textwidth]{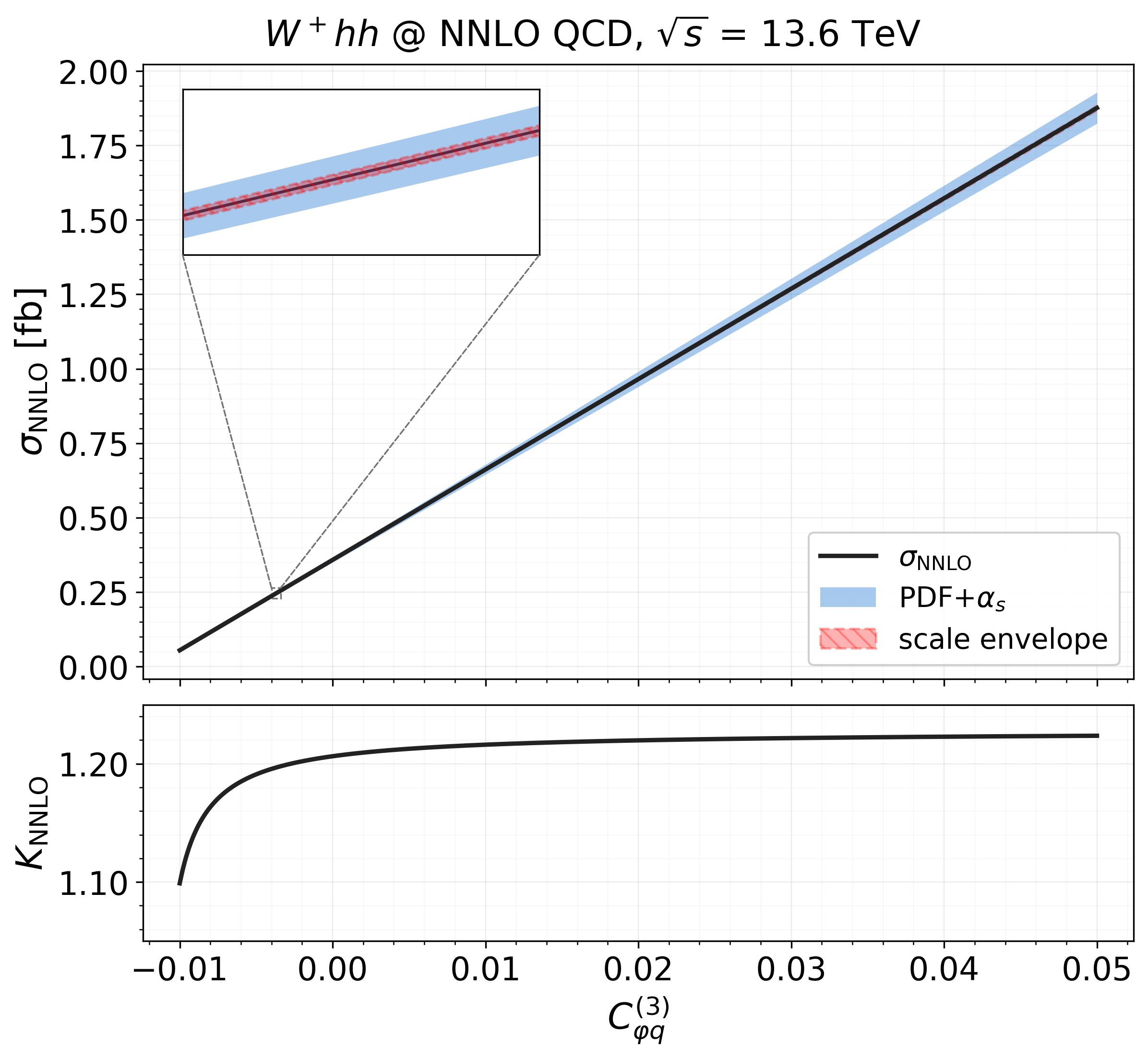}
\includegraphics[width=0.45\textwidth]{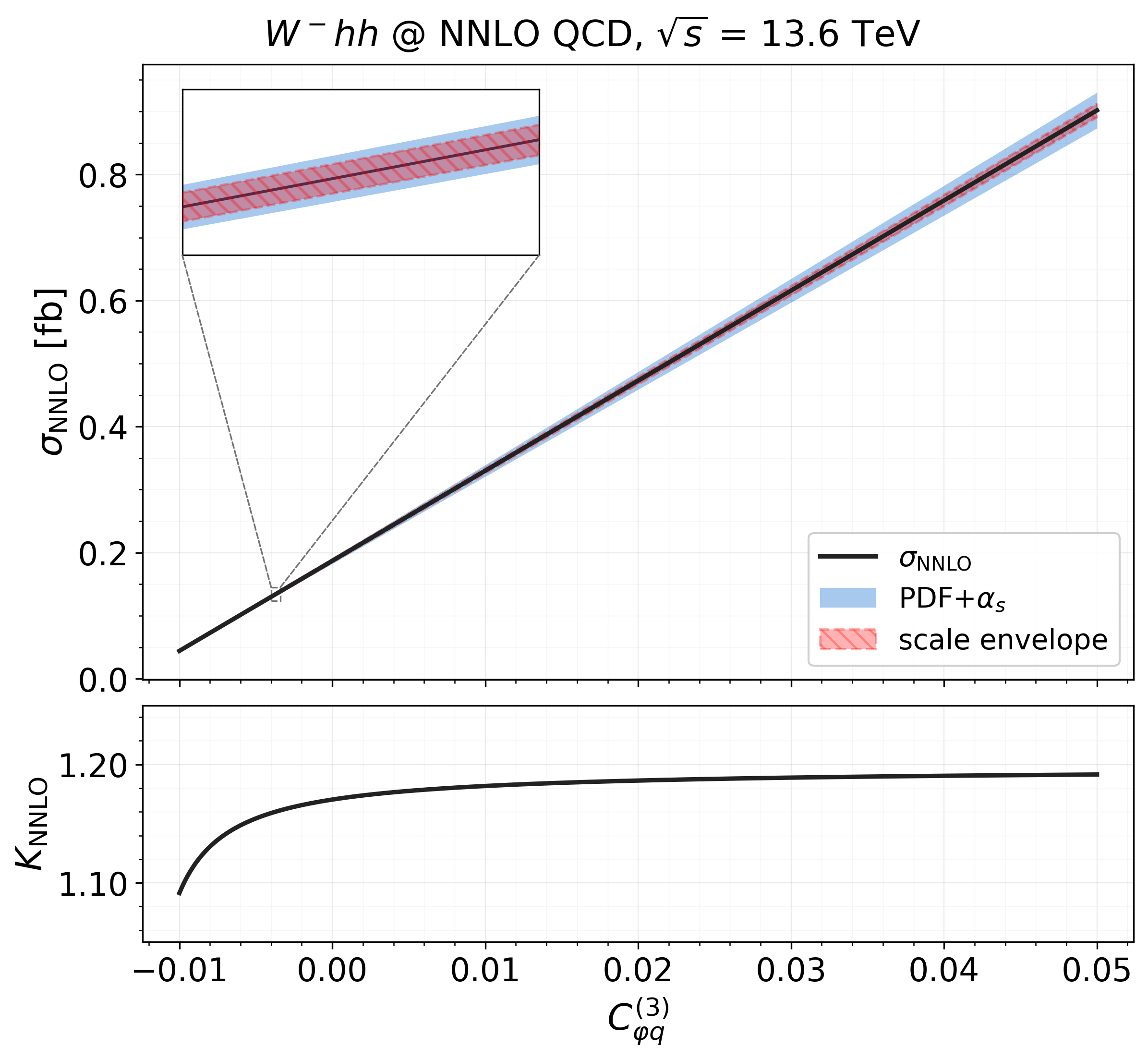}
\caption{Cross section and $K$-factor of  $W^+ hh$ (left) and $W^- hh$ (right) as function of $C_{\varphi \Box}$ (top) and $C_{\varphi q}^{(3)}$ (bottom) at $\sqrt{s}=13.6$~TeV.}
\label{fig:smeft_phiBox_phiq:WHH}
\end{figure}

\begin{figure}[htb!]
\centering
\includegraphics[width=0.45\textwidth]{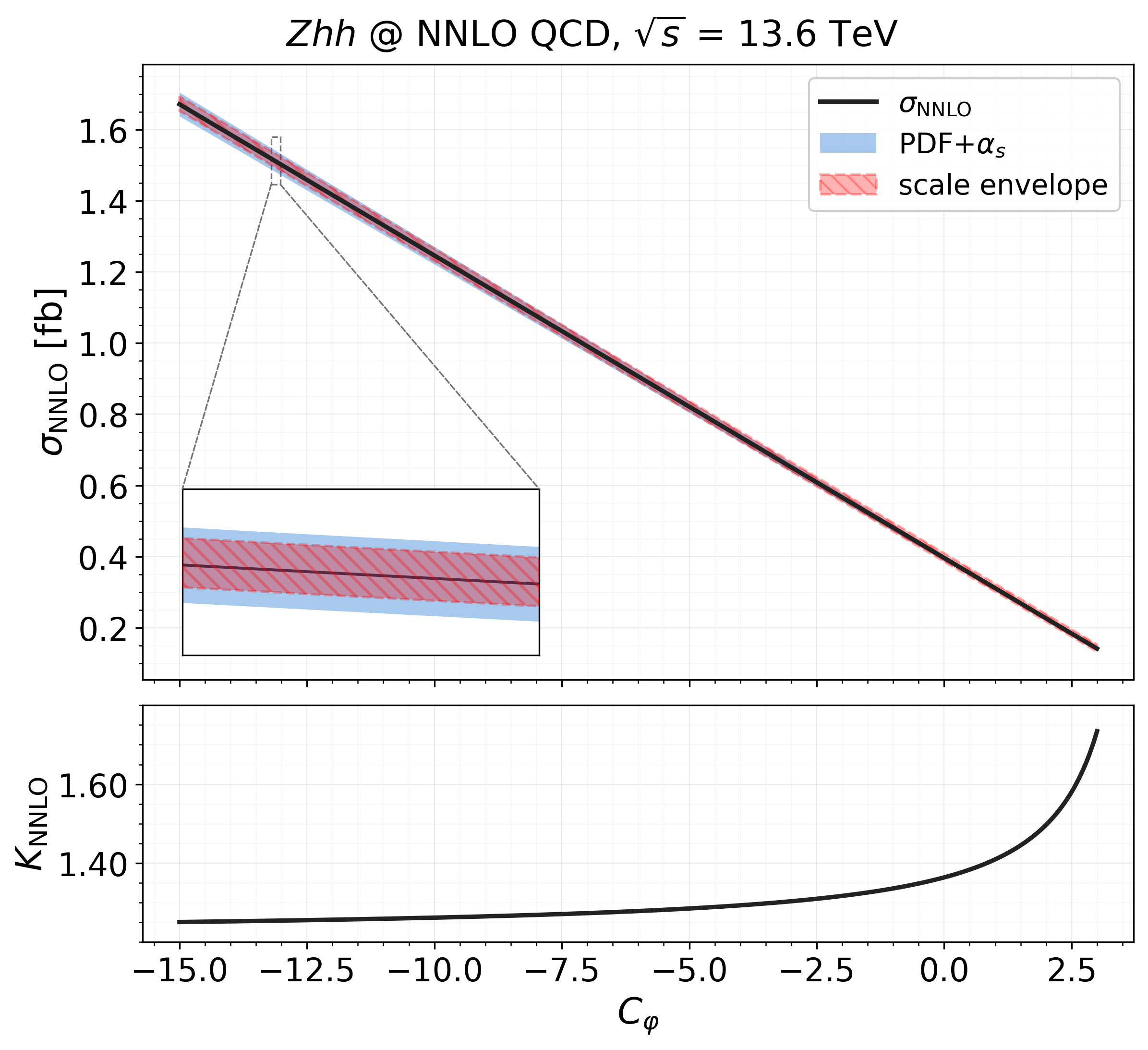}
\includegraphics[width=0.45\textwidth]{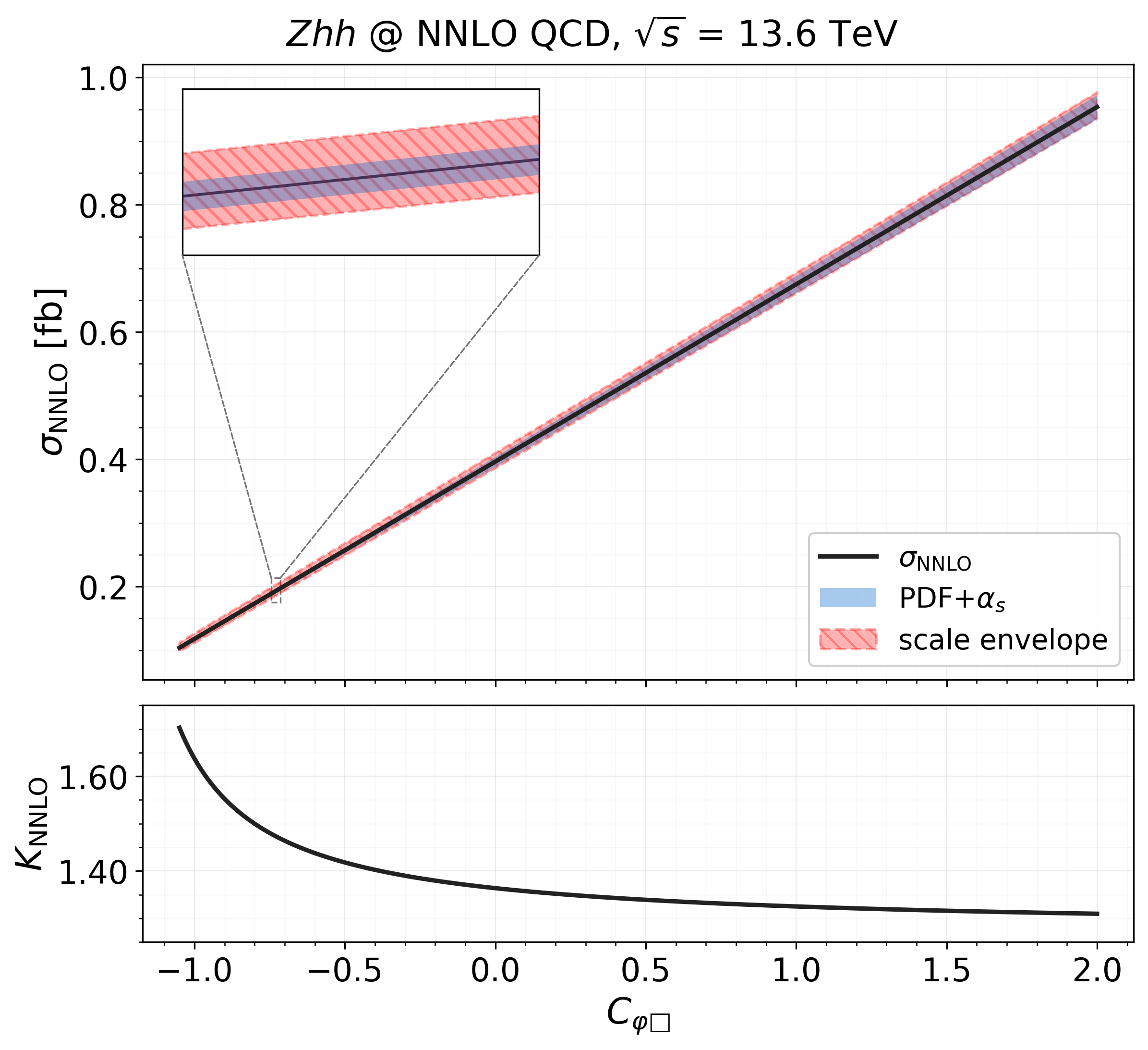}
\includegraphics[width=0.45\textwidth]{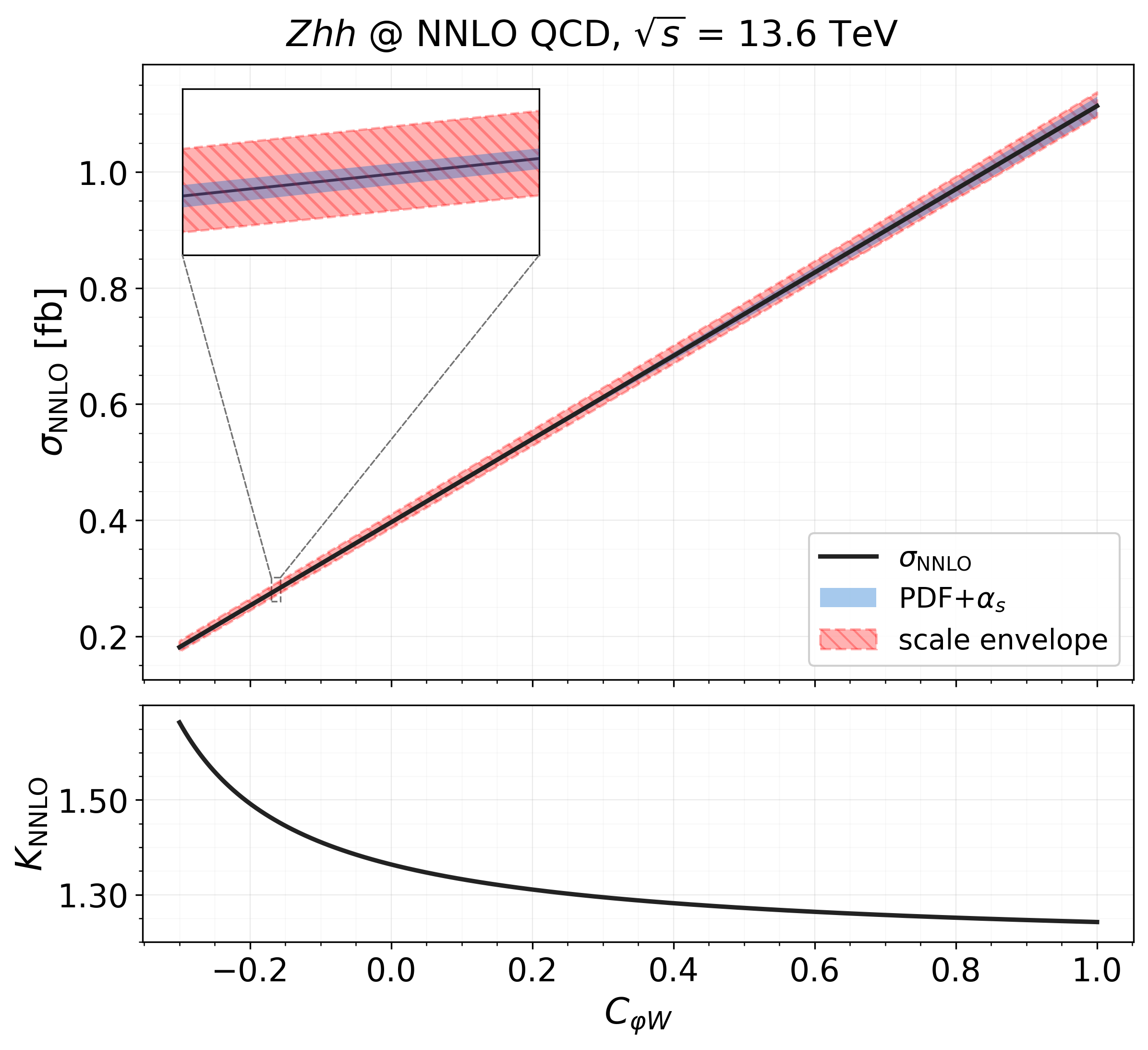}
\includegraphics[width=0.45\textwidth]{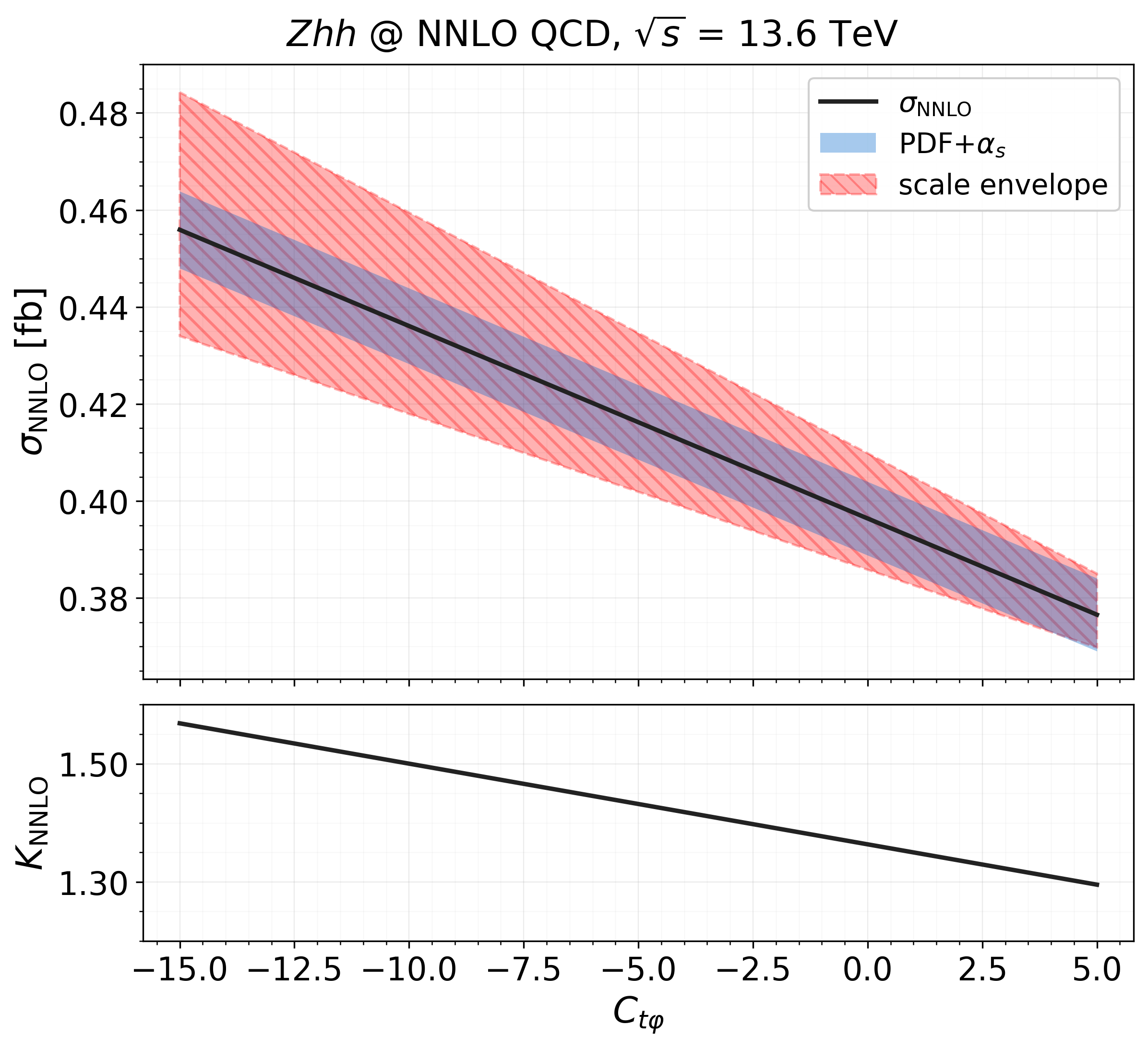}
\caption{Cross section and $K$-factor of  $Z hh$ as a function of $C_{\varphi}$ (left top), $C_{\varphi \Box}$ (right top),  $C_{\varphi W}$ (left bottom) and $C_{t\varphi}$ (right bottom) at $\sqrt{s}=13.6$~TeV.}
\label{fig:smeft:ZHH}
\end{figure}

In this section, we present the results of the SMEFT analysis of the $Vhh$ cross section at NNLO QCD.
We consider single insertions of SMEFT dimension-6 operators from Table~\ref{tab:operators}, as discussed in Sec.~\ref{sec:SMEFT:corr}, and truncate the cross section at $1/\Lambda^2$ order in the EFT expansion. Tables~\ref{tab:B_wphh_combined}, \ref{tab:B_wmhh_combined} and \ref{tab:B_zhh_combined} show the numerical results for the $B_i$ coefficients in Eqs.~\eqref{eq:sigmaWhhSMEFT}, \eqref{eq:sigmaZhhSMEFTLO} and \eqref{eq:sigmaZhhSMEFT}.
The allowed values of the SMEFT Wilson coefficients used for numerical analysis, consistent with the global fits of Refs.~\cite{Celada:2024mcf,terHoeve:2025gey}, are collected in Tables~\ref{tab:SMEFT:coeff:bosonic} and~\ref{tab:SMEFT:coeff:fermionic} .
Maximal cross-section enhancements for each Wilson coefficient, with all others set to their SM values, are provided in Tables~\ref{tab:wplushh_SMEFT}, \ref{tab:wminushh_SMEFT}, \ref{tab:zhh_SMEFT_group1}, and~\ref{tab:zhh_SMEFT_group2}.
Example predictions for the cross section and $K$-factor as a function of a single Wilson coefficient, while keeping all other at zero, are shown in Figs.~\ref{fig:smeft_phi_phiW:WHH}, \ref{fig:smeft_phiBox_phiq:WHH}, and~\ref{fig:smeft:ZHH}.
For brevity, in what follows, we absorb the cut-off scale into the Wilson coefficient definitions by rescaling $C/\Lambda^2 \to C$.

The maximal predicted cross-section enhancements for $W^\pm hh$ production reach $5 \times \sigma_{\rm SM}$ for $C_{\varphi q}^{(3)} = 0.05/\text{TeV}^2$ and $C_{\varphi} = -15.0/\text{TeV}^2$, $4 \times \sigma_{\rm SM}$ for $C_{\varphi W} = 1/\text{TeV}^2$, and lower values for the other operators. For $Zhh$ production, the largest enhancements are again observed for $C_{\varphi} = -15.0/\text{TeV}^2$, leading to a cross-section enhancement of around $4 \times \sigma_{\rm SM}$. A factor of $2 \times \sigma_{\rm SM}$ enhancement is found for $C_{\varphi W} = 1/\text{TeV}^2$ and $C_{\varphi \Box} = 2/\text{TeV}^2$, while $C_{\varphi q}^{(1)} = -3/\text{TeV}^2$ leads to an enhancement of $10 \times \sigma_{\rm SM}$. All other Wilson coefficients result in smaller modifications of the total cross section.

Similarly to HEFT, the predicted theoretical uncertainties can reach at most roughly $10\%$, most of the time being below $4\%$, with lower uncertainty for $W^{\pm}hh$ than for $Zhh$ due to the gluon-induced one-loop contribution relevant for the latter.
One notable exception is the $C_{\varphi q}^{(1)}$ Wilson coefficient from Table~\ref{tab:zhh_SMEFT_group2}, which requires separate discussion.
The cross-section enhancement of order $10 \times \sigma_{\rm SM}$ can be attributed to the comparatively large allowed value of $C_{\varphi q}^{(1)}$, significantly exceeding that of $C_{\varphi q}^{(3)}$ of the same type. The nearly $50\%$ uncertainties originate from the large PDF uncertainties in the high-$x$ region, probed by the sizable values of $C_{\varphi q}^{(1)}$.\footnote{Setting $C_{\varphi q}^{(1)}=0.05/\text{TeV}^2$, the PDF uncertainty becomes $8.7\%$ for $\sqrt{s}=13.6$ TeV, only slightly larger than the one for $C_{\varphi q}^{(3)}$. } This underscores the need for improved PDF determinations.

For the operators that only rescale the couplings already present in the SM, for $W^{\pm}hh$, the effect on the $K$-factor is very small, while for the operator with Wilson coefficient $C_{\varphi W}$ in $W^+ hh$ and $C_{\varphi q}^{(3)}$ in $W^{\pm}hh$, the effect is a bit larger. This can, though, be mostly attributed to a very small LO cross section corresponding to these values of Wilson coefficients, close to 0. The same observation regarding the $K$-factor holds true for $Zhh$. It remains rather flat unless the LO cross section is close to zero, though with the exception  of $C_{t\varphi}$ which plays a role in the gluon-induced component. 

\begin{table}[htbp]
\begin{center}
\resizebox{0.9\textwidth}{!}{%
  \renewcommand{\arraystretch}{1.35}%
  \setlength{\tabcolsep}{6pt}%
  \begin{tabular}{|c|c|c|c|c|c|c|}
    \hline\rule{0pt}{3ex}
    Coefficient $[1/\text{TeV}^2]$ & $C_\varphi$ & $C_{\varphi W}$ & $C_{\varphi B}$ & $C_{\varphi WB}$ & $C_{\varphi D}$ & $C_{\varphi \Box}$ \\[3pt]\hline\rule{0pt}{3ex}
    Uncertainty interval & $[-15,\ 5]$ & $[-1,\ 1]$ & $[-0.5,\ 0.5]$ & $[-1.5,\ 1.5]$ & $[-1.5,\ 1.5]$ & $[-2,\ 2]$ \\ \hline
  \end{tabular}%
}
\end{center}
\caption{Intervals for the bosonic SMEFT Wilson coefficients considered in this work, consistent with the global fits of Refs.~\cite{Celada:2024mcf,terHoeve:2025gey}.}
\label{tab:SMEFT:coeff:bosonic}
\end{table}

\begin{table}[htbp]
\begin{center}
\resizebox{\textwidth}{!}{%
  \renewcommand{\arraystretch}{1.35}%
  \setlength{\tabcolsep}{6pt}%
  \begin{tabular}{|c|c|c|c|c|c|c|c|c|}
    \hline\rule{0pt}{3ex}
    Coefficient $[1/\text{TeV}^2]$ & $C_{\varphi q}^{(3)}$ & $C_{\varphi Q}^{(3)}$ & $C_{\varphi q}^{(1)}$ & $C_{\varphi Q}^{(1)}$ & $C_{\varphi u}$ & $C_{\varphi d}$ & $C_{t\varphi}$ & $C_{\varphi t}$ \\[3pt]\hline\rule{0pt}{3ex}
    Uncertainty interval & $[-0.2,\ 0.05]$ & $[-8,\ 2]$ & $[-3,\ 1]$ & $[-6.5,\ 30.5]$ & $[-3.5,\ 1]$ & $[-4,\ 4]$ & $[-15,\ 5]$ & $[-25,\ 34]$ \\ \hline
  \end{tabular}%
}
\end{center}
\caption{Intervals for the fermionic SMEFT Wilson coefficients considered in this work, consistent with the global fit of Refs.~\cite{Celada:2024mcf,terHoeve:2025gey}.}
\label{tab:SMEFT:coeff:fermionic}
\end{table}

\begin{table}[htbp]
\centering
  \renewcommand{\arraystretch}{1.2}%
\begin{tabular}{c c c c c}
\hline
 & \multicolumn{2}{c}{$\sqrt{s}=13.6$~TeV [fb $\times$ TeV$^{2}$]} & \multicolumn{2}{c}{$\sqrt{s}=14.0$~TeV [fb $\times$ TeV$^{2}$]} \\
\cline{2-5}
 & $B^{W^+}_{i,\mathrm{LO}}$ & $B^{W^+}_{i,\mathrm{NNLO}}$ & $B^{W^+}_{i,\mathrm{LO}}$ & $B^{W^+}_{i,\mathrm{NNLO}}$ \\
\hline
$B^{W^+}_{1}$ & $-0.0746$ & $-0.0897$ & $-0.078$ & $-0.0937$ \\
$B^{W^+}_{2}$ & $0.2319$ & $0.2800$ & $0.2429$ & $0.2928$ \\
$B^{W^+}_{3}$ & $0.0156$ & $0.0189$ & $0.0164$ & $0.0198$ \\
$B^{W^+}_{4}$ & $24.7314$ & $30.3666$ & $26.1576$ & $32.0717$ \\
$B^{W^+}_{5}$ & $0.9314$ & $1.1069$ & $0.9782$ & $1.1614$ \\
\hline
\end{tabular}
\caption{Central values of $B^{W^+}_i$ coefficients for $W^+hh$ at $\sqrt{s}=13.6$~TeV and $14.0$~TeV.}
\label{tab:B_wphh_combined}
\end{table}

\begin{table}[htbp]
\centering
  \renewcommand{\arraystretch}{1.2}%
\begin{tabular}{c c c c c}
\hline
 & \multicolumn{2}{c}{$\sqrt{s}=13.6$~TeV [fb $\times$ TeV$^{2}$]} & \multicolumn{2}{c}{$\sqrt{s}=14.0$~TeV [fb $\times$ TeV$^{2}$]} \\
\cline{2-5}
 & $B^{W^-}_{i,\mathrm{LO}}$ & $B^{W^-}_{i,\mathrm{NNLO}}$ & $B^{W^-}_{i,\mathrm{LO}}$ & $B^{W^-}_{i,\mathrm{NNLO}}$ \\
\hline
$B^{W^-}_{1}$ & $-0.0410$ & $-0.0478$ & $-0.0431$ & $-0.0502$ \\
$B^{W^-}_{2}$ & $0.1242$ & $0.1455$ & $0.1310$ & $0.1532$ \\
$B^{W^-}_{3}$ & $0.0081$ & $0.0096$ & $0.0086$ & $0.0101$ \\
$B^{W^-}_{4}$ & $11.9387$ & $14.2928$ & $12.7052$ & $15.1826$ \\
$B^{W^-}_{5}$ & $0.4820$ & $0.5624$ & $0.5097$ & $0.5940$ \\
\hline
\end{tabular}
\caption{Central values of $B^{W^-}_i$ coefficients for $W^-hh$ at $\sqrt{s}=13.6$ and $14.0$~TeV.}
\label{tab:B_wmhh_combined}
\end{table}

\begin{table}[htbp]
\centering
  \renewcommand{\arraystretch}{1.2}%
\begin{tabular}{c c c c c}
\hline
 & \multicolumn{2}{c}{$\sqrt{s}=13.6$~TeV [fb $\times$ TeV$^{2}$]} & \multicolumn{2}{c}{$\sqrt{s}=14.0$~TeV [fb $\times$ TeV$^{2}$]} \\ 
\cline{2-5}
 & $B^{Z}_{i,\mathrm{LO}}$ & $B^{Z}_{i,\mathrm{NNLO}}$ & $B^{Z}_{i,\mathrm{LO}}$ & $B^{Z}_{i,\mathrm{NNLO}}$ \\
\hline
$B^{Z}_{1}$ & $-0.0697$ & $-0.0850$ & $-0.0732$ & $-0.0891$ \\
$B^{Z}_{2}$ & $0.2188$ & $0.2786$ & $0.2300$ & $0.2930$ \\
$B^{Z}_{3}$ & $0.0135$ & $0.018$ & $0.0142$ & $0.0190$ \\
$B^{Z}_{4}$ & $1.9195$ & $1.8424$ & $1.7508$ & $1.5926$ \\
$B^{Z}_{5}$ & $0.6060$ & $0.7173$ & $0.6389$ & $0.7554$ \\
$B^{Z}_{6}$ & $-1.2537$ & $-1.1325$ & $-1.0932$ & $-0.9025$ \\
$B^{Z}_{7}$ & $0.0826$ & $0.0795$ & $0.0752$ & $0.0687$ \\
$B^{Z}_{8}$ & $-0.0596$ & $-0.0638$ & $-0.0590$ & $-0.0622$ \\
$B^{Z}_{9}$ & $0.0153$ & $0.0181$ & $0.0161$ & $0.0190$ \\
$B^{Z}_{10}$ & $-0.4409$ & $-0.5233$ & $-0.4645$ & $-0.5507$ \\
$B^{Z}_{11}$ & --- & $-0.0040$ & --- & $-0.0042$ \\
$B^{Z}_{12}$ & --- & $0.0038$ & --- & $0.0040$ \\
\hline
\end{tabular}
\caption{Central values of $B^{Z}_i$ coefficients for $Zhh$ at $\sqrt{s}=13.6$ and $14.0$~TeV. Entries marked ``---'' enter only at NNLO.}
\label{tab:B_zhh_combined}
\end{table}

\begin{table}[htbp!]
\centering
\resizebox{\textwidth}{!}{%
  \renewcommand{\arraystretch}{1.45}%
  \setlength{\tabcolsep}{7pt}%
  \begin{tabular}{c  c@{\hspace{10pt}}c@{\hspace{10pt}}c@{\hspace{10pt}}c@{\hspace{10pt}}c@{\hspace{10pt}}c}
    \hline
    \multirow{2}{*}{$\sqrt{s}$} & \multicolumn{6}{c}{$\sigma_{NNLO}$} \\
    \cline{2-7}
     & SM & $C_\varphi$=-15 & $C_{\varphi W}$=1 & $C_{\varphi D}$=1.5 & $C_{\varphi \Box}$=2 & $C_{\varphi q}^{(3)}$=0.05 \\
    \hline
    $13.6$ & $0.358_{+0.36\%}^{-0.40\%} \pm 2.3\%$ & $1.704_{+0.38\%}^{-0.40\%} \pm 2.2\%$ & $1.464_{+1.11\%}^{-0.98\%} \pm 2.4\%$ & $0.386_{+0.36\%}^{-0.40\%} \pm 2.3\%$ & $0.918_{+0.36\%}^{-0.40\%} \pm 2.3\%$ & $1.875_{+0.36\%}^{-0.45\%} \pm 2.8\%$ \\
    $14.0$ & $0.374_{+0.37\%}^{-0.40\%} \pm 2.2\%$ & $1.779_{+0.39\%}^{-0.41\%} \pm 2.2\%$ & $1.535_{+1.10\%}^{-0.97\%} \pm 2.3\%$ & $0.404_{+0.37\%}^{-0.40\%} \pm 2.2\%$ & $0.960_{+0.37\%}^{-0.40\%} \pm 2.2\%$ & $1.977_{+0.36\%}^{-0.44\%} \pm 2.7\%$ \\
    \hline
  \end{tabular}%
}
\caption{Cross-section values $\sigma$ [fb] at NNLO QCD in SMEFT for $W^+hh$ at $\sqrt{s}=13.6,\, 14.0$ TeV, with each Wilson coefficient set to the value maximising $\sigma$ within its scan interval. All other coefficients are set to their SM values.}
\label{tab:wplushh_SMEFT}
\end{table}

\begin{table}[htbp!]
\centering
\resizebox{\textwidth}{!}{%
  \renewcommand{\arraystretch}{1.45}%
  \setlength{\tabcolsep}{7pt}%
  \begin{tabular}{c c@{\hspace{10pt}}c@{\hspace{10pt}}c@{\hspace{10pt}}c@{\hspace{10pt}}c@{\hspace{10pt}}c}
    \hline
    \multirow{2}{*}{$\sqrt{s}$} & \multicolumn{6}{c}{$\sigma_{NNLO}$} \\
    \cline{2-7}
     & SM & $C_\varphi$=-15 & $C_{\varphi W}$=1 & $C_{\varphi D}$=1.5 & $C_{\varphi \Box}$=2 & $C_{\varphi q}^{(3)}$=0.05 \\
    \hline
    $13.6$ & $0.187_{+1.31\%}^{-1.36\%} \pm 2.5\%$ & $0.903_{+1.33\%}^{-1.40\%} \pm 2.4\%$ & $0.749_{+1.91\%}^{-1.67\%} \pm 2.6\%$ & $0.202_{+1.30\%}^{-1.36\%} \pm 2.5\%$ & $0.478_{+1.30\%}^{-1.34\%} \pm 2.5\%$ & $0.901_{+1.16\%}^{-1.25\%} \pm 3.2\%$ \\
    $14.0$ & $0.197_{+1.33\%}^{-1.38\%} \pm 2.4\%$ & $0.949_{+1.35\%}^{-1.42\%} \pm 2.3\%$ & $0.791_{+1.90\%}^{-1.66\%} \pm 2.6\%$ & $0.212_{+1.32\%}^{-1.37\%} \pm 2.4\%$ & $0.503_{+1.31\%}^{-1.36\%} \pm 2.4\%$ & $0.956_{+1.17\%}^{-1.26\%} \pm 3.1\%$ \\
    \hline
  \end{tabular}%
}
\caption{Cross-section values $\sigma$ [fb] at NNLO QCD in SMEFT for $W^-hh$ at $\sqrt{s}=13.6,\, 14.0$ TeV, with each Wilson coefficient set to the value maximising $\sigma$ within its scan interval. All other coefficients are set to their SM values.}
\label{tab:wminushh_SMEFT}
\end{table}

\begin{table}[htbp!]
\centering
\resizebox{\textwidth}{!}{%
  \renewcommand{\arraystretch}{1.45}%
  \setlength{\tabcolsep}{7pt}%
  \begin{tabular}{c c@{\hspace{10pt}}c@{\hspace{10pt}}c@{\hspace{10pt}}c@{\hspace{10pt}}c@{\hspace{10pt}}c}
    \hline
    \multirow{2}{*}{$\sqrt{s}$} & \multicolumn{6}{c}{$\sigma_{NNLO}$} \\
    \cline{2-7}
     & SM & $C_\varphi=-15$ & $C_{\varphi W}=1$ & $C_{\varphi D}=1.5$ & $C_{\varphi \Box}=2$ & $C_{\varphi q}^{(3)}=0.05$ \\
    \hline
    $13.6$ & $0.396_{+3.39\%}^{-2.66\%} \pm 1.9\%$ & $1.671_{+1.32\%}^{-1.14\%} \pm 2.0\%$ & $1.113_{+2.12\%}^{-1.74\%} \pm 2.1\%$ & $0.423_{+3.35\%}^{-2.64\%} \pm 1.9\%$ & $0.954_{+2.42\%}^{-1.95\%} \pm 2.0\%$ & $0.489_{+2.48\%}^{-1.80\%} \pm 5.7\%$ \\
    $14.0$ & $0.417_{+3.42\%}^{-2.69\%} \pm 1.9\%$ & $1.754_{+1.34\%}^{-1.14\%} \pm 2.0\%$ & $1.172_{+2.10\%}^{-1.72\%} \pm 2.1\%$ & $0.446_{+3.38\%}^{-2.66\%} \pm 1.9\%$ & $1.003_{+2.44\%}^{-1.96\%} \pm 8.6\%$ & $0.497_{+2.58\%}^{-1.85\%} \pm 6.0\%$\\
    \hline
  \end{tabular}%
}
\caption{Cross-section values $\sigma$ [fb] at NNLO QCD in SMEFT for $Zhh$ at $\sqrt{s}=13.6,\, 14.0$ TeV, with each Wilson coefficient set to the value maximising $\sigma$ within its scan interval. All other coefficients are set to their SM values.}
\label{tab:zhh_SMEFT_group1}
\end{table}

\begin{table}[htbp!]
\centering
\resizebox{\textwidth}{!}{%
  \renewcommand{\arraystretch}{1.45}%
  \setlength{\tabcolsep}{7pt}%
  \begin{tabular}{c c@{\hspace{10pt}}c@{\hspace{10pt}}c@{\hspace{10pt}}c@{\hspace{10pt}}c}
    \hline
    \multirow{2}{*}{$\sqrt{s}$} & \multicolumn{5}{c}{$\sigma_{NNLO}$} \\
    \cline{2-6}
     & SM & $C_{\varphi q}^{(1)}=-3$ & $C_{\varphi u}=1$ & $C_{\varphi d}=-4$ & $C_{\varphi t} + C_{\varphi Q}^{(3)} - C_{\varphi Q}^{(1)} = 42.5$ \\
    \hline
    $13.6$ & $0.396_{+3.39\%}^{-2.66\%} \pm 1.9\%$ & $3.800_{+5.30\%}^{-2.28\%} \pm 40.7\%$ & $0.476_{+2.58\%}^{-1.90\%} \pm 5.4\%$ & $0.652_{+1.83\%}^{-1.32\%} \pm 4.9\%$ & $0.558_{+10.07\%}^{-7.63\%} \pm 1.6\%$ \\
    $14.0$ & $0.417_{+3.42\%}^{-2.69\%} \pm 1.9\%$ & $3.251_{+4.58\%}^{-5.52\%} \pm 45.0\%$ & $0.486_{+2.67\%}^{-1.95\%} \pm 5.6\%$ & $0.666_{+1.88\%}^{-1.34\%} \pm 5.1\%$ & $0.591_{+10.13\%}^{-7.67\%} \pm 1.4\%$ \\
    \hline
  \end{tabular}%
}
\caption{Cross-section values $\sigma$ [fb] at NNLO QCD in SMEFT for $Zhh$ at $\sqrt{s}=13.6,\, 14.0$ TeV, with each Wilson coefficient set to the value maximising $\sigma$ within its scan interval. All other coefficients are set to their SM values.}
\label{tab:zhh_SMEFT_group2}
\end{table}

\begin{table}[htbp!]
\centering
\resizebox{0.8\textwidth}{!}{%
  \renewcommand{\arraystretch}{1.45}%
  \setlength{\tabcolsep}{7pt}%
  \begin{tabular}{c c@{\hspace{10pt}}c@{\hspace{10pt}}c@{\hspace{10pt}}c}
    \hline
    \multirow{2}{*}{$\sqrt{s}$} & \multicolumn{4}{c}{$\sigma_{NNLO}$} \\
    \cline{2-5}
     & SM & $C_{t\varphi}=-15$ & $C_{\varphi B}=0.5$ & $C_{\varphi WB}=-1$ \\
    \hline
    $13.6$ & $0.396_{+3.39\%}^{-2.66\%} \pm 1.9\%$ & $0.455_{+6.17\%}^{-4.77\%} \pm 1.7\%$ & $0.405_{+3.34\%}^{-2.63\%} \pm 1.9\%$ & $0.920_{+2.09\%}^{-1.72\%} \pm 2.1\%$ \\
    $14.0$ & $0.417_{+3.42\%}^{-2.69\%} \pm 1.9\%$ & $0.480_{+6.21\%}^{-4.78\%} \pm 1.7\%$ & $0.427_{+3.38\%}^{-2.65\%} \pm 1.9\%$ & $0.968_{+2.09\%}^{-1.71\%} \pm 2.1\%$ \\
    \hline
  \end{tabular}%
}
\caption{Cross-section values $\sigma$ [fb] at NNLO QCD in SMEFT for $Zhh$ at $\sqrt{s}=13.6,\, 14.0$ TeV, with each Wilson coefficient set to the value maximising $\sigma$ within its scan interval. All other coefficients are set to their SM values. 
}
\label{tab:zhh_SMEFT_group3}
\end{table}


\section{Conclusion \label{sec:conclusion}}

We have presented NNLO QCD predictions for $Vhh$ production in HEFT and SMEFT, building on the SM NNLO QCD corrections presented in Ref.~\cite{Baglio:2012np}. The calculation exploits the Drell-Yan structure of the quark-induced contribution, in which QCD radiation affects the production of an off-shell vector boson and factorises from the subsequent electroweak decay into $Vhh$. 

For $W^\pm hh$ production, we find that the NNLO QCD corrections in both HEFT and SMEFT largely factorise from the leading-order EFT cross section. As a consequence, the corresponding $K$-factors show only a mild dependence on the EFT parameters, and the residual perturbative uncertainties remain small over the parameter ranges considered. This behaviour makes $W^\pm hh$ a particularly clean channel for interpreting deviations in terms of modified Higgs self-interactions and modified couplings of one or two Higgs bosons to electroweak gauge bosons.

The situation is different for $Zhh$ production. In this channel, the loop-induced $gg \to Zhh$ sub-process enters formally as part of the NNLO QCD correction to the quark-induced Drell-Yan process. Through triangle, box, and pentagon topologies involving heavy-quark loops, this contribution introduces an additional sensitivity to EFT deformations, in particular to coefficients affecting the top Yukawa coupling and the $Zt\bar t$ interaction. It also increases the residual theoretical uncertainty and leads to a more pronounced dependence of the $K$-factor on the chosen EFT point.

In addition to cross sections and $K$-factors for representative benchmark choices of the HEFT and SMEFT parameters, we provide the parametrisation of cross sections in terms of numerical coefficients. These coefficients allow the inclusive cross sections to be reconstructed efficiently for arbitrary values of the EFT parameters within the ranges considered in this work. They are therefore directly useful for phenomenological studies and for future experimental interpretations of associated Higgs pair production with a weak gauge boson.

Our results show that current constraints on EFT coefficients still allow sizeable enhancements of the $Vhh$ rates, while the theoretical uncertainties remain well under control for the level of precision expected in the near future. Present experimental sensitivity to the $Vhh$ cross-section enhancement is still far from our EFT predictions, but this gap is expected to narrow substantially with the upcoming HL-LHC upgrade. Our results thus mark a step towards a systematic EFT programme for sub-leading Higgs pair production channels at the HL-LHC. Natural extensions of this work include fully differential EFT predictions and a more complete treatment of higher-order operators in the HEFT expansion.

\section*{Acknowledgements}
We drew our diagrams using {\tt FeynGame}~\cite{Bundgen:2025utt}.
The work of RG and MR is supported by a STARS@UniPD grant under the acronym “HiggsPairs” and in part by the Italian MUR Departments of Excellence grant 2023-2027 ”Quantum Frontiers”. The authors acknowledge support from the COMETA COST Action CA22130.

\bibliographystyle{utphys}
\bibliography{bibliography}

\end{document}